\documentclass[12pt]{article}
\input{myPreliminary}
\newcommand{\showName}{1}

\graphicspath{{./art/}}
\allowdisplaybreaks

\begin{document}
\title{\bf Trend correlation inference in time series via differencing}
\newcommand*\samethanks[1][\value{footnote}]{\footnotemark[#1]}
\if1\showName{
	\author[1,2]{Hon Kiu To}
    \author[1]{Kin Wai Chan\samethanks\thanks{Corresponding author:
				\href{mailto:kinwaichan@sta.cuhk.edu.hk}
				{\nolinkurl{kinwaichan@sta.cuhk.edu.hk}}}}
	\affil[1]{Department of Statistics and Data Science, The Chinese University of Hong Kong.}
    \affil[2]{Department of Criminology, University of Pennsylvania.}
	\maketitle
}\else{
	\author[]{}
	\maketitle
}\fi
\bigskip
\begin{abstract}
How strongly do latent trends move together when we observe only one noisy multivariate time series?
This paper addresses this issue by introducing trend correlation,
which serves as a measure of structural association.
Formal statistical tests,
including tests for one-pair, two-pair, and all-pair trend correlation, are proposed
via a differencing approach.
The resulting test statistic is computationally fast and straightforward as
it bypasses direct estimation of the unknown trends and change point estimations.
As a natural consequence,
we also obtain new tests on trend stationarity in multivariate time series
based on
trend variance and trend covariance.
The developed difference-based trend tests offer the advantage of being sensitive to small oscillating trend effects.
Finally, the proposed trend correlation test is applied to study the association between
different crime frequencies and weather-related variables in the City of Chicago.
The differences between trend correlation and traditional Pearson correlation are revealed, suggesting a new angle for performing criminological analysis.
\end{abstract}
\noindent
{\it Keywords:}  Difference variate; mean stationarity; non-linear time series; super-efficiency; signal correlation.
\vfill
\newpage
\spacingset{1.6}

\section{Introduction}\label{sec:intro}
\subsection{Background, motivation, and goals}
What is correlation?
Loosely speaking, it is a measure of a statistical association between
variables; see, e.g., \citet{Speed2011}.
Typically,
it is referred to as a stochastic correlation.
For instance, in time series analysis, commonly used measures include
(i) time-domain relationships, such as autocorrelation and cross-correlation for assessing dependence across lags and components; and
(ii) frequency-domain relationships, such as coherence for assessing linear dependence at different frequencies.
These measures focus only on the relationships between stochastic components.

However, the interpretation is
rarely this straightforward for applied scientists.
Consider a bivariate time series:
while a statistician will likely quantify the relationship using one of the aforementioned stochastic measures,
an applied scientist is typically much more interested
in the underlying structural correlation.
To put it in another way,
applied scientists probably care more about the trends
instead of the noises of the data.
An example in climate studies is studying the structural correlation between the temperature trend and sea level trend \citep{VermeerRahmstorf2009}.
Other fields of application include medical studies \citep{Zou2015}, chemistry \citep{Scharnweber2016}, and social science \citep{DonohueWalfers2006}.
Although it is possible to estimate the structural correlations when there are repeated observations \citep{pospisilBair2021},
we lack a method to estimate of structural correlations when there is only one realized time series trajectory.

Consider an $S$-dimensional time series $X_i = (X_{i,1},\ldots,X_{i,S})^{\T} \in \mathbb{R}^S,i = 1,\ldots,n$ generated
from the signal-plus-noise model:
$X_i = \mu_i + Z_i$,
where the deterministic signals $\mu_i = (\mu_{i,1},\ldots,\mu_{i,S})^{\T}$ and the mean-zero stationary noises $Z_i = (Z_{i,1},\ldots,Z_{i,S})^{\T}$ are unobservable.
The structural correlation between $\mu_{j_1}$ and $\mu_{j_2}$, where $j_1,j_2 = 1,\ldots,S$, cannot be consistently estimated by the standard Pearson's correlation of the data; see \S\S\ref{sec:failure} and \ref{sec:TrencCorrTestMC}.
With the abundance of multivariate time series in real life,
it is worthwhile to explore a
formal measure for assessing structural correlations.

On the other hand, the trend correlation can be a useful tool for improving trend stationarity test for multiple time series.
To handle this problem, an intuitive idea would be aggregating univariate test results applied on each component.
However, a key question regarding this approach is how to efficiently aggregate them.
This is non-trivial because
conventional methods for combining $p$-values do not perform well;
see \S\ref{subsubsec:number of means}.
So, it calls for a better procedure to address the multivariate trend stationarity problem.
We conjecture that trend correlation improve trend stationary tests that are based solely on trend variance.

In this paper, we address two main issues: (a) the trend correlation problem and (b) the trend variance problem, with the latter serving as an auxiliary problem derived from the former.
We formally define the notion of trend correlation,
which admits similar interpretations as the conventional correlation coefficient, and can be used to assess the structural correlation between two time series.
We also introduce a new difference-based test for
multivariate trend stationarity in time series
by incorporating the information of
trend correlation into the test.
The proposed test (i) bypasses direct estimation of the underlying mean functions and allows milder conditions on the class of feasible mean functions, and
(ii) is rate-accelerated, i.e., a convergence rate faster than $\surd{n}$, which helps detect
subtle trend effects.

\subsection{Notation}\label{subsec:notations}
Let $\mathbb{N} = \{1,2,\ldots\}$.
Denote the indicator function as $\mathbb{1}(\cdot)$.
For $\left\{a_n \in \mathbb{R}^+\right\}_{n \in \mathbb{N}}$ and $\left\{b_n \in \mathbb{R}^+\right\}_{n \in \mathbb{N}}$,
$a_n \asymp b_n$ means there exists a $C > 0$ such that $1/C \leq a_n/b_n \leq C$;
$a_n \ll b_n$ or $a_n = o\left(b_n\right)$ means $a_n/b_n \rightarrow 0$;
$a_n \lesssim b_n$ or $a_n = O\left(b_n\right)$ means there exists a $C > 0$ such that $a_n/b_n \leq C$ for all large $n$.
Convergence in probability and in distribution are denoted by ``$\inP$'' and ``$\inD$'', respectively.
For $a=(a_1, \ldots, a_S)^{\T}$, denote
$|a|_{\infty} = \max\{|a_1|, \ldots, |a_S|\}$ and
$|a| = (a^{\T}a)^{1/2}$.
Denote Hadamard product by $\odot$ and Kronecker product by $\otimes$.
Denote $I_d$ as the $d\times d$ identity matrix.
For a random variable $X$, denote $\| X \|_p = \E\left(|X|^p\right)^{1/p}$ for $p \geq 1$, and by convention, $\| X \| = \| X \|_2$.
Independent and identically distributed random variables $X_1,\ldots,X_n$ following a distribution $F$ are denoted by $X_i \sim F$ independently.
Denote $\bar{A}_n = \sum_{i = 1}^n A_i/n$ for any $A_1, \ldots, A_n$.
The $i$th element of a vector $a$ is $(a)_i$.
The $(i,j)$th element of a matrix $A$ is $(A)_{i,j}$.
Denote $\diag(a_1,\ldots,a_n)$ as the diagonal matrix with diagonal entries $a_1,\ldots,a_n$.
For matrix $A$,
denote $\vectorized(A)$ as the vectorization of $A$;
$\upper(A)$ and $\upper_0(A)$ as
the vectorization of the upper triangular part with and without the diagonal
of $A$.

\subsection{Trend structure}\label{subsec:mean structure}
Let $\mu_i = \mu(i/n)$, where $\mu(\cdot): [0,1] \rightarrow \mathbb{R}^S$ is a mean function
defined as
$\mu(t) = \mathfrak{c}(t) + \mathfrak{s}(t)$.
Here
$\mathfrak{c}(\cdot)$ is a Lipschitz continuous function, and
$\mathfrak{s}(\cdot)$ is a step discontinuous function:
	\begin{equation}\label{eqt:mean function}
		\mathfrak{s}(t) = \sum^{\mathcal{J}}_{j = 0} \xi_j\mathbb{1}\left(\frac{T_j}{n} \leq t < \frac{T_{j + 1}}{n}\right),
	\end{equation}
	where $\mathcal{J}$ is the number of discontinuities, $1 \equiv T_0 < T_1 < \cdots < T_{\mathcal{J}} < T_{\mathcal{J} + 1} \equiv n + 1$ are times of discontinuities, and $\xi_0, \ldots, \xi_{\mathcal{J}}$ are step sizes such that $|\xi_j - \xi_{j - 1}|_{\infty} \neq 0$ for all $1 \leq j \leq \mathcal{J}$.
It is a general setting asserting that the mean functions $\mu(\cdot)$ is a piecewise Lipschitz continuous function with $\mathcal{J}$ pieces.
Under this setting, we also allow the number of pieces $\mathcal{J}$ to diverge, which further increases the generality of our setting.
Similar settings have been used in other related work; see, e.g., \citet{wu_zhao_2007} and \citet{DallaGiraitisPhillips2015}.

\begin{assumption}\label{assum:jumps}
	The discontinuities of $\mu(\cdot)$ satisfy $\min_{0 \leq j \leq \mathcal{J}} |T_{j + 1} - T_j| \gtrsim \ell$, where $\ell \in \mathbb{N}$ is a bandwidth parameter to be specified.
\end{assumption}
Assumption \ref{assum:jumps} prevents $\mathfrak{s}(\cdot)$ from jumping too frequently.
The parameter $\ell$ may vary in different cases, so it is specified whenever Assumption \ref{assum:jumps} is needed.
The behavior of $\mu(\cdot)$ is assessed by
the smoothness of $\mathfrak{c}(\cdot)$ and the maximum step magnitude of $\mathfrak{s}(\cdot)$:
\begin{equation*}
	\mathcal{C} = \max_{1 \leq s \leq S} \sup_{0 \leq t' < t \leq 1} \left|\frac{\mathfrak{c}_s(t) - \mathfrak{c}_s(t')}{t - t'}\right| = \max_{1 \leq s \leq S} \mathcal{C}_s
	 \qquad \text{and} \qquad
	 \mathcal{S} = \sup_{j = 1,\ldots,\mathcal{J}} |\xi_j - \xi_{j - 1}|_{\infty},
\end{equation*}
respectively.
Specifically, $\mathcal{C}_s = 0$ signifies that the $s$th component series lacks a continuous trend,
and $\mathcal{S} = 0$ or $\mathcal{J} = 0$ implies the absence of step discontinuities.

\subsection{Noise structure}\label{subsec:noise structure}
We adopt the dependence measure framework introduced in \citet{Wu2005}.
Consider a process $Z_i = g(\mathcal{F}_i)$ for some measurable function $g$, where $\mathcal{F}_i = (\ldots,\varepsilon_{i - 1},\varepsilon_i)$ represents the shift process of independent and identically distributed $S$-dimensional innovations $\{\varepsilon_i\}_{i \in \mathbb{Z}}$.
Let $\varepsilon_j'$ be an independent copy of $\varepsilon_j$, and define $\mathcal{F}'_i = (\mathcal{F}_{-1},\varepsilon_0',\varepsilon_1,\ldots,\varepsilon_i)$ and $Z'_i = g(\mathcal{F}'_i)$.
The projection operator is defined as $\mathcal{P}_i \cdot = \E(\cdot \mid \mathcal{F}_i) - \E(\cdot \mid \mathcal{F}_{i - 1})$.
For $p \geq 1$, the physical dependence measure and its aggregated value are given by
$\omega_{i,p} = \left\|  Z_i - Z'_i  \right\|_p$
and
$\Omega_{n,p} = \sum_{i = n}^\infty \omega_{i,p}$,
respectively.
Denote the asymptotic covariance matrix $v = \lim_{n \rightarrow \infty} n\Var(\bar{Z}_n) = \sum_{k = -\infty}^\infty \gamma_k$.
The following weak dependence condition is imposed on the noises $\{Z_i\}_{i \in \mathbb{Z}}$.
\begin{assumption}\label{assum:weak dependence}
	The $S$-dimensional time series $\{Z_i\}_{i \in \mathbb{Z}}$ is mean-zero, strictly stationary, and satisfying
    (i) $\max_{1 \leq j \leq S} \E( | Z_{1j} | ^{4 + \iota}) <\infty$ for some $\iota > 0$;
    (ii) there exist $\lambda_*>0$ and
    $n_*\in\mathbb{N}$ such that
    for all $t\geq 1$ and $n\geq n_*$, we have
    $\lambda_{\min}\{\Var(\sum_{i=t+1}^{t+n}Z_i)\} \geq \lambda_* n$,
    where $\lambda_{\min}(\cdot)$
    denotes the minimum eigenvalue;
    (iii) $\Omega_{n,p} = O\{ n^{-\chi} (\log n)^{-A} \}$
    for $p>2$, $\chi >\{p^2-4+(p-2)(p^2+20p+4)^{1/2}\}/(8p)$
    and,
    $A>0$.
\end{assumption}

Assumption \ref{assum:weak dependence} is needed
to derive
a Gaussian approximation for
$\sum_{i=1}^n Z_i$;
see Theorem 2.2 of
\cite{KarmakarWu2020}.
Other conditions that give a Gaussian approximation is also possible;
see, e.g., \cite{liulin2009}.

\section{Trend covariance and trend correlation}\label{sec:trend correlation test}
\subsection{Definitions}\label{sec:trend correlation}
The celebrated \cite{pearson1896}'s product-moment correlation for $X$ and $Y$ is defined as
$\rho_{\pearson} = \Cov(X,Y)/\{ \Var(X)\Var(Y)\}^{1/2}$ if $\Var(X),\Var(Y) > 0$.
So, a proper definition for the trend correlation
requires proper definitions for the trend covariance and the trend variance.
We propose the trend covariance matrix $\theta$, defined below in (\ref{eqt:TV matrix}),
to interpret the variability of $\{\mu_i\}_{i = 1}^n$ via the standard variance operation.

Let $U_1,\ldots,U_n \simIID \Unif(0,1)$.
Define the randomly permuted trend vectors as
$M_i = \mu(U_i)$,
$i = 1,\ldots,n$.
Then the trend covariance matrix
is given by
\begin{equation}\label{eqt:TV matrix}
	\theta
	= \lim_{n \rightarrow \infty} n\Var\left(\bar{M}_n\right)
	= \int_0^1 \{\mu(t) -\bar{\mu}\}^{\otimes 2} \, \dd t,
\end{equation}
where $\bar{\mu} = \int_0^1 \mu(t)\, \dd t$.
The first expression in (\ref{eqt:TV matrix}) justifies the term covariance matrix.
With (\ref{eqt:TV matrix}), the trend correlation matrix and its $(j_1,j_2)$ entry are
\begin{equation}\label{eqt:trend correlation}
	\rho = \rho_{\trend}
	= (\theta\odot I_S)^{-1/2} \theta (\theta\odot I_S)^{-1/2}
	\quad \text{and} \quad
	\rho_{j_1,j_2}
	= \rho_{j_1,j_2,\trend} =\frac{\theta_{j_1,j_2}}{(\theta_{j_1,j_1}\theta_{j_2,j_2})^{1/2}},
\end{equation}
where $x/0\equiv 0$ for any $x\in\mathbb{R}$ by convention.

The trend correlation shares the usual statistical interpretation of correlation:
(i) when $\rho_{j_1,j_2} = 0$,  $\mu_{j_1}(t)$ and $\mu_{j_2}(t)$ are orthogonal in $\mathcal{L}^2[0,1]$; and
(ii) when $\rho_{j_1,j_2} = \pm 1$, $\mu_{j_1}(t)$ and $\mu_{j_2}(t)$ are perfectly proportional to each other almost everywhere.
One subtle detail about the trend correlation is that it is well-defined when both $\theta_{j_1,j_1},\theta_{j_2,j_2} > 0$.
By convention, we define $\rho_{j_1,j_2} = 0$ when $\theta_{j_1,j_1}\theta_{j_2,j_2} = 0$.
This convention arises from the fact that $\Cov(X,Y) = 0$ if at least one of $X,Y$ is constant.
In the subsequent discussions, we focus on the case that both $\theta_{j_1,j_1},\theta_{j_2,j_2} > 0$.

\subsection{Paradox and failure of commonly used approaches}\label{sec:failure}

\begin{example}[Correlation paradox]\label{eg:paradox}
The trend correlation $\rho_{j_1,j_2, \trend}$ reflects the long-run relationship, whereas the noise correlation
$\rho_{j_1,j_2,\noise} = (\gamma_{0})_{j_1,j_2} (\gamma_{0})_{j_1,j_1}^{-1/2}(\gamma_{0})_{j_2,j_2}^{-1/2}$
reflects the instantaneous relationship, where $(\gamma_0)_{j_1,j_2} = \Cov(X_{i,j_1},X_{i,j_2})$.
Their signs may differ.
It is a form of Simpson's paradox
with time being a lurking variable; see Figure \ref{fig:tcor_K01_paradox}.

For example, in the crime dataset analyzed in
\S\ref{sec:weather and crime},
the estimated trend and noise correlations between theft and offenses involving children are
$\hat{\rho}_{\trend}=-0.34<0$ and
$\hat{\rho}_{\noise}=0.10>0$, respectively,
where both estimators will be formally defined in \S\ref{subsec:differencing}.
From a criminological perspective, these categories reflect distinct generating mechanisms:
theft is an opportunity-driven property offense,
whereas offenses involving children are heterogeneous crimes
that can be detected by families, schools, and child-care facilities.
So, this reveals
divergent long-term trajectories in opportunity-seeking behavior versus incident reporting.
In particular, during 2022--2024,
the average daily number of thefts increased while
the average daily number of recorded offenses involving children decreased.
This divergence may partly reflect post-pandemic changes in retail activity and institutional contact.
The positive $\hat{\rho}_{\noise}$ may be driven by
calendar and recording effects because both series
exhibit a first-of-month effect, which may occur when crimes are backdated to the first day of the month due to
retrospective estimates or prolonged crimes.

Also, the sample Pearson's correlation $\hat{\rho}_{\pearson}$
is neither consistent for $\rho_{j_1,j_2, \trend}$ nor $\rho_{j_1,j_2,\noise}$.
It is not hard to see that
$\hat{\rho}_{\pearson} \rightarrow \rho_{\textsf{mix}}$ in probability,
where
$\rho_{\textsf{mix}}=\{(\gamma_0)_{j_1,j_2}+\theta_{j_1,j_2}\}/[\{(\gamma_0)_{j_1,j_1}+\theta_{j_1,j_1}\}\{(\gamma_0)_{j_2,j_2}+\theta_{j_2,j_2}\}]^{1/2}$.
For example, in the real-data example above,
$\hat{\rho}_{\pearson}=-0.04$,
which is different from $\hat{\rho}_{\trend}$ and $\hat{\rho}_{\noise}$.
The probability limit of $\hat{\rho}_{\pearson}$ can be smaller, equal, or larger
than $\rho_{j_1,j_2, \trend}$ and $\rho_{j_1,j_2,\noise}$.
It gives a warning of using $\hat{\rho}_{\pearson}$ for understanding correlation.

    \begin{figure}
    \captionsetup{font=small}
        \centering
        \includegraphics[width=.55\textwidth]{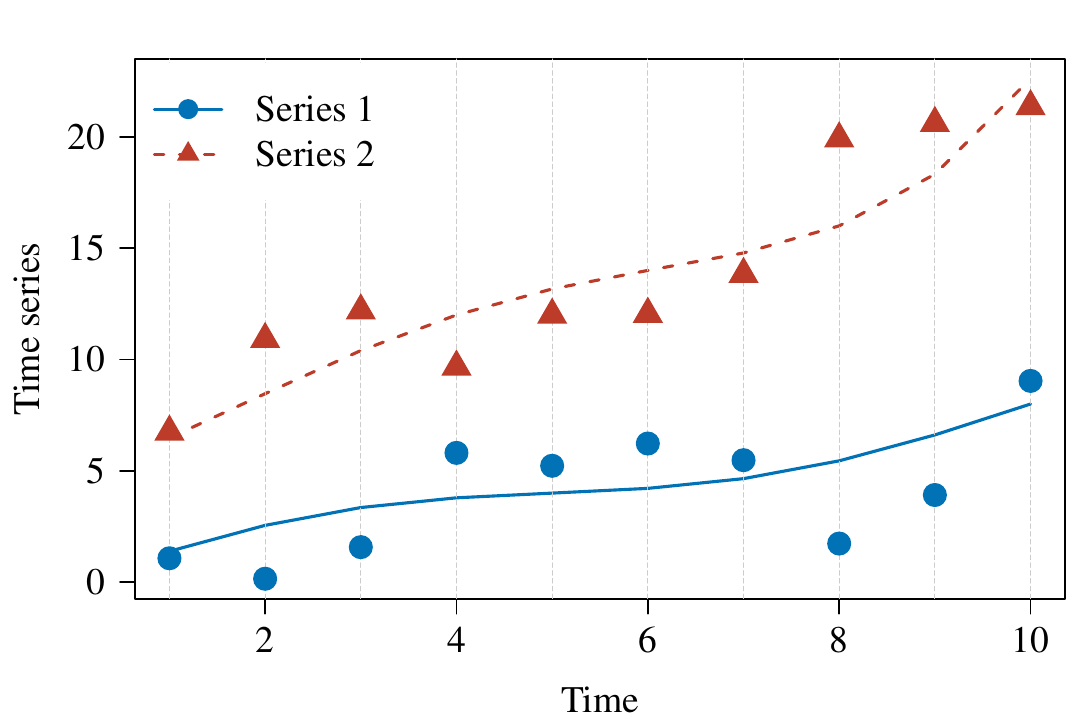}\vspace{-0.2cm}
        \caption{\label{fig:tcor_K01_paradox}Correlation paradox in Example \ref{eg:paradox},
        where  $\rho_{1,2, \trend}>0$ but $\rho_{1,2,\noise}<0$.
        They have a positive trend correlation because they have increasing trends.
        They have a negative noise correlation because
        they tend to move in an opposite direction from their respective trends.
        }
    \end{figure}
\end{example}

\begin{example}[Failure of regression]\label{subsec:TC vs regression}
Suppose we are interested in the trend correlation between $X_{i,j_1}$ and $X_{i,j_2}$, where $j_1,j_2 \in \{1,\ldots,S\}$.
Consider a simple linear regression approach, where we regress $X_{i,j_2}$ on $X_{i,j_1}$, i.e.,
$\E(X_{i,j_2}\mid X_{i,j_1}) = \beta_0 + \beta_1 X_{i,j_1}$
for all $i = 1,\ldots,n$.
    Let $\hat{\beta}_{1}$ be the least-squares estimator of $\beta_1$.
    Then, as $n \rightarrow \infty$,
    \begin{align}\label{eqt:betaWrongLimit}
        \hat{\beta}_{1} &\inP \beta_{1,\mix}
        = \frac{(\gamma_0)_{j_1,j_2} + \theta_{j_1,j_2}}{\{(\gamma_0)_{j_1,j_1} + \theta_{j_1,j_1}\}^{1/2}\{(\gamma_0)_{j_2,j_2} + \theta_{j_2,j_2}\}^{1/2}}
        \left\{\frac{(\gamma_0)_{j_2,j_2} + \theta_{j_2,j_2}}{(\gamma_0)_{j_1,j_1} + \theta_{j_1,j_1}} \right\}^{1/2},
    \end{align}
whose value is jointly controlled by the trend and noise variance.
The proof of (\ref{eqt:betaWrongLimit}) can be found in \S\ref{sec:proofOfReg}.
The sign of $\beta_{1,\mix}$ may disagree with that of $\theta_{j_1,j_2}$.
This phenomenon is similar to the correlation paradox discussed in Example \ref{eg:paradox}.
In other words, we cannot rely on a regression approach to perform trend correlation inferences.
\end{example}

\begin{example}[Coherence at frequency zero]\label{eg:coh}
The coherence between the $j_1$th and $j_2$th components
at frequency $\omega$ is
$\rho^{2}_{j_1, j_2,\coh}(\omega) = |f_{j_1, j_2}(\omega)|^2 / \{f_{j_1, j_1}(\omega)f_{j_2, j_2}(\omega)\}$,
where $f_{j_1, j_2}(\cdot)$ is the cross-spectral density.
It measures the strength of the linear relationship
between the two components at frequency $\omega$.
In particular, at the zero frequency ($\omega=0$),
which is conceptually
the most relevant for analyzing noise dynamics
over the longest possible horizons,
the coherence simplifies to
$\rho^{2}_{j_1, j_2,\coh}(0) = v_{j_1, j_2}^2 / (v_{j_1, j_1} v_{j_2, j_2})$,
where $v = \lim_{n \rightarrow \infty} n \Var(\bar{Z}_n)$ represents the long-run variance matrix.
Consequently, $\rho^{2}_{j_1, j_2,\coh}(0)$
measures the long-run value of $\Corr(\bar{Z}_{n,j_1},\bar{Z}_{n,j_2})$
rather than any trend relationship.
\end{example}

\begin{example}[Cointegration]\label{eg:coint}
A time series $X_i = (X_{i,1}, \ldots, X_{i,S})^{\T}$ is said to be cointegrated
if there exists a non-zero vector $\beta \in \mathbb{R}^S$ such that
$\beta^{\T}X_i$ is stationary; see \citet{Johansen1988}.
Usually,
the cointegration rank $r_{\mathsf{coint}}$, i.e.,
the maximum number of linearly independent $\beta\neq0$ to make $\beta^{\T}X_i$ stationary,
is of interest.
Cointegration characterizes a long-run equilibrium relationship in the sense that
such a linear combination eliminates shared stochastic trends.
In contrast, trend correlation quantifies the degree of co-movement of deterministic trends
rather than just concerning the existence of a linear combination for perfect cancellation.
Also, unlike trend correlation,
cointegration is not a scale-free index in $[-1, 1]$.
\end{example}

\begin{example}[Functional correlation]\label{eg:funCorr}
\cite{ChiouLi2008} studied the functional correlation between two random functions $f_1,f_2:[0,1]\rightarrow\mathbb{R}$,
defined as
$\rho_{\fun} = \langle f_1^\circ, f_2^\circ\rangle/(\|f_1^\circ\|_2\|f_2^\circ\|_2)$,
where $f_j^{\circ}(t) = f_j(t)-\int_0^1 f_j(u)\, \dd u$,
$\langle f, g \rangle=\int_0^1 f(u) g(u)\,\dd u$, $\|f\|_2=\langle f,f \rangle^{1/2}$.
In our setting, the inner product is between the centered latent deterministic trend functions instead of the random function.
More importantly, we observe only one realization of the time series to
study the functional structure instead of $n$ replicates
of functional data.
\end{example}

\begin{example}[Detrended cross-correlation]\label{eg:dcc}
\cite{Zebende2011} considered
the probability limit $\rho_{\dcc}(s)$ of
the sample correlation between
two time series after detrending with a fitted $ r$th-order polynomial over a local window of length $s$;
see also \cite{blytheNikulinMuller2016robust} for the limiting distribution.
It measures only the noise-based relationship if the selected values of $r$ and $s$
adequately remove the deterministic trends.
\end{example}

\begin{example}[Trend synchronism and parallelism]\label{eg:trendSyn}
\citet{LyubchichGel2016} tests for trend synchronism, i.e.,
$H_0:\mu_j(t)=c_j+f(t)$ for all $j=1, \ldots, S$, for some constants $c_j$'s and a common function $f$.
However, it does not quantify the strength and direction of the trend relationship when
the structures are beyond synchronism.
Similar approaches for trend parallelism can be found in, e.g.,
\cite{Degras2011}.
\end{example}

A summary is provided in Table \ref{table:related_trend_correlation_methods}.

\begin{table*}[tp]
\captionsetup{font=small}
\centering
\small
\setlength{\tabcolsep}{3pt}
\renewcommand{\arraystretch}{1.08}
\begin{tabular}{cccccc}
\toprule
Approach
& Reference
& Inferential target
& Noise
& Trend
& Theory \\ \midrule
Trend correlation
& (\ref{eqt:trend correlation})
& $\rho_{\trend}$
& & $\checkmark$ & $\checkmark$ \\

Local-linear regression
& Example~\ref{eg:LLR}
& $\rho_{\trend}$
& & $\checkmark$ & \\

Trend synchronism \& parallelism
& Example \ref{eg:trendSyn}
& $H_0:\mu_j(t)=c_j+f(t)$
& & $\checkmark$ & $\checkmark$\\

Pearson correlation
& Example~\ref{eg:paradox}
& $\rho_{\mix}$
& $\checkmark$ & $\checkmark$ & $\checkmark$\\

Linear regression
& Example~\ref{subsec:TC vs regression}
& $\beta_{1,\mix}$
& $\checkmark$ & $\checkmark$ & $\checkmark$\\

Noise correlation
& Example~\ref{eg:noiseCorr}
& $\rho_{\noise}$
& $\checkmark$ & & $\checkmark$\\

Zero-frequency coherence
& Example~\ref{eg:coh}
& $\rho_{\coh}^2(0)$
& $\checkmark$ & & $\checkmark$\\

Detrended cross-correlation
& Example \ref{eg:dcc}
& $\rho_{\dcc}(s)$
& $\checkmark$ & ?& $\checkmark$\\

Functional-data correlation
& Example \ref{eg:funCorr}
& $\rho_{\fun}$
& -- & -- & $\checkmark$\\

Cointegration
& Example \ref{eg:coint}
& $H_0: r_{\mathsf{coint}}=0$
& -- & -- & $\checkmark$\\ \bottomrule
\end{tabular}

\caption{
Comparison of trend correlation with related approaches.
In columns 4--5,
``$\checkmark$'' indicates that the inferential target,
either the probability limit or the null hypothesis $H_0$,
depends on the stochastic noise $Z_i$ or the deterministic trend $\mu_i$,
where the variables stated in the inferential target can be found in the stated reference;
``--'' means not applicable because the method's
original framework does not fall into or is similar to
our setting defined in \S\S\ref{subsec:mean structure}--\ref{subsec:noise structure};
``?'' means that the results depend on the assumption.
In the last column, ``$\checkmark$'' indicates that
theoretical results are available for performing the inference task.
}
\label{table:related_trend_correlation_methods}
\end{table*}

\section{Method and Theory}\label{sec:theoryMethod}
\subsection{Methods of Differencing and lag-product}\label{subsec:differencing}
Two main principles involved in estimating $\theta$ and $\rho$ are differencing and lag-product.
For any lag $h \in \mathbb{N}$, the $m$th order difference statistics are
\begin{equation}\label{eqt:diffStat}
    D_i^{(m)} = \sum_{j = 0}^m d_j\left(X_{i - jh} - \bar{X}_n\right), \qquad i = mh + 1,\ldots,n,
\end{equation}
where $\{d_j\}_{j = 0}^m$ is a difference sequence satisfying $\sum_{j = 0}^m d_j = 0$ and $\sum_{j = 0}^m d_j^2 = 1$ when $m > 0$,
and $d_0 = 1$ when $m = 0$.
For $k \geq 0$, define lag-$k$ product moment of (\ref{eqt:diffStat}) as
\begin{equation}\label{eqt:dacvf estimator}
	\hat{\gamma}_k^{(m)}
		= \left(\hat{\gamma}_{-k}^{(m)}\right)^{\T}
		= \frac{1}{n}\sum_{i = mh + k + 1}^n D_i^{(m)}\left(D_{i - k}^{(m)}\right)^{\T} .
\end{equation}
The proposed $m$th order difference-based estimators of $\theta$ and $\rho$ are
\begin{equation}\label{eqt:proposed estimator}
	\hat{\theta}^{(m)} = \frac{\sum_{|k| \leq \ell} K(k/\ell)(\hat{\gamma}_k^{(0)} - \hat{\gamma}_k^{(m)})}{\sum_{|k| \leq \ell} K(k/\ell)}
	\quad \text{and} \quad
	\hat{\rho}^{(m)} = (\hat{\theta}^{(m)}\odot I_S)^{-1/2} \hat{\theta}^{(m)} (\hat{\theta}^{(m)}\odot I_S)^{-1/2},
\end{equation}
where $\ell \in \mathbb{N} \cap [1,n)$ is a bandwidth, and $K: \mathbb{R} \rightarrow \mathbb{R}$ is a kernel function such that $K(1) = 0$, $K(t) = K(-t)$ for all $t \in \mathbb{R}$ and $K(t) = 0$ for $|t| > 1$.
Adjusting for the possibility of $\theta=0$,
thresholded estimators of $\theta$ and $\rho$ are
\begin{equation}\label{eqt:proposed estimator_0}
	\bar{\theta}^{(m)} = \hat{H}^{(m)} \hat{\theta}^{(m)} \hat{H}^{(m)}
	\quad \text{and} \quad
	\bar{\rho}^{(m)} = (\bar{\theta}^{(m)}\odot I_S)^{-1/2} \bar{\theta}^{(m)} (\bar{\theta}^{(m)}\odot I_S)^{-1/2},
\end{equation}
where $\hat{H}^{(m)} = \diag\{\mathbb{1}(|\hat{\theta}_{1,1}^{(m)}|\geq\epsilon_1), \ldots, \mathbb{1}(|\hat{\theta}_{S,S}^{(m)}|\geq\epsilon_S) \}$
for some possibly random
$\epsilon_1, \ldots, \epsilon_S$ that satisfy $\max(\epsilon_1, \ldots, \epsilon_S)\inP 0$
and $(n\ell)^{1/2}\min(\epsilon_1, \ldots, \epsilon_S) \inP\infty$.
More discussions of (\ref{eqt:diffStat})--(\ref{eqt:proposed estimator_0}) are in the sequel.

	In (\ref{eqt:diffStat}), we use differencing because
		estimation of $\mu(\cdot)$ is non-trivial
		due to the possibly non-linear trends and possibly divergent number of jumps;
		see Example \ref{eg:LLR}.
		The statistic $D_i^{(m)}$ locally removes $\mu(\cdot)$ when $m>0$
		in the sense that $\E\{D_i^{(m)}\} \approx 0$ due to the constraint $\sum_{j=0}^m d_j=0$.
		Differencing does not require sensitive tuning,
        as the asymptotic properties of $\hat{\theta}^{(m)}$
		and $\hat{\rho}^{(m)}$ are free of all parameters; see Theorem \ref{thm:4.3}.
		It is a commonly used way for removing non-constant trend for time series inference;
		see, e.g., \cite{wu_zhao_2007}, \cite{bai2024difference} and \cite{LeungChan2022}.

In (\ref{eqt:dacvf estimator}), a general lag-product moment is used instead of merely the squared data
		for estimating the second order quantity $\theta$.
		Using multiple lags for estimation leads to two appealing properties:
		(i) a neater asymptotic theory in \S\ref{subsec:pair TC test}
		without resorting to the complicated Bartlett formula (see Theorem 11.2.3 of \cite{brockwellDavis1991}), and
		(ii) accelerated rate of convergence in \S\ref{sec:theory_degenarate}.
		Simple calculations show that
		$\hat{\gamma}_{k}^{(m)} \inP \gamma_k + \theta \mathbb{1}(m = 0)$;
		see Proposition \ref{prop:B3} of the supplement.
		Hence, the contrast of $\hat{\gamma}_k^{(m)}$ (with differencing) and $\hat{\gamma}_k^{(0)}$ (without differencing)
		provides a sensible way to estimate $\theta_{j_1,j_2}$ for any $k$.
		Furthermore, this also allows us to consistently estimate the noise correlation $\rho_{\noise}$
		mentioned in Example \ref{eg:paradox}; see Example \ref{eg:noiseCorr} for details.

 In (\ref{eqt:proposed estimator}), $\hat{\gamma}_k^{(m)}$ are aggregated across $k$ with weights $K(k/\ell)$,
		which is expected to be decreasing, at least roughly, with $|k|$ to reflect
		that $\hat{\gamma}_k^{(m)}$ is a more precise estimate than $\hat{\gamma}_{k'}^{(m)}$
		for $|k|<|k'|$.

 In (\ref{eqt:proposed estimator_0}), $\epsilon$ is a threshold to be derived in Corollary \ref{cor:improvedVersion}.
			We will set $\epsilon_j = \{(\gamma_0)_{j,j}/n\}^{1/2}$ for each $s$, where $\gamma_0$ can be estimated by $\hat{\gamma}_0^{(m)}$.
		Essentially, $\bar{\theta}^{(m)}_{j_1, j_2} = \hat{\theta}^{(m)}_{j_1, j_2} \hat{H}^{(m)}_{j_1, j_1}\hat{H}^{(m)}_{j_2, j_2}$ and
		$\bar{\rho}^{(m)}_{j_1, j_2} = \bar{\theta}^{(m)}_{j_1, j_2}/\{ \bar{\theta}^{(m)}_{j_1, j_1}\bar{\theta}^{(m)}_{j_2, j_2} \}^{1/2} = \hat{\rho}^{(m)}_{j_1,j_2}\hat{H}^{(m)}_{j_1, j_1}\hat{H}^{(m)}_{j_2, j_2}$,
		where
		$\hat{\theta}_{j_1,j_2}^{(m)}$ and
		$\hat{\rho}_{j_1,j_2}^{(m)}$
		are the $(j_1, j_2)$ entries of $\hat{\theta}^{(m)}$ and $\hat{\rho}^{(m)}$.
			The matrix $\hat{\theta}^{(m)}$ may not be positive definite.
			In such case, we bound the eigenvalues of $\hat{\theta}^{(m)}$ from below by an order of $1/n^{1/2}$.
			This may happen when the $\mu(t)$ is approximately constant.
Some special cases and other related estimators are presented below:

\begin{example}[Default estimator]\label{eg:default}
The proposal is parametrized by $m, \{d_j\}_{j=0}^m, K, h$ and $\ell$.
The first four can be chosen without tuning.
By default, we set
$m=1$, $(d_0,d_1)=(1/\surd{2},-1/\surd{2})$,
$K(t) = (1-|t|^2)^+$, and $h=2\ell$.
While $\ell$ should diverge to infinity at a suitable rate,
it does not affect the asymptotic performance according to Theorem \ref{thm:4.3}.
By default, we set $\ell = \lfloor 2n^{1/5} \rfloor$,
where neither the coefficient $2$ nor the power $1/5$ affects the asymptotic distribution of $\hat{\theta}$
in Theorem \ref{thm:4.3}.
Under this default setup, we have
\begin{align}\label{eqt:default}
	\hat{\theta}^{(1)}
	= \frac{3}{4\ell^3-\ell}\sum_{|k| \leq \ell} (\ell^2-k^2) (\hat{\gamma}_k^{(0)} - \hat{\gamma}_k^{(1)})
	\approx \frac{3}{2(4\ell^3-\ell)}\sum_{|k| \leq \ell} (\ell^2-k^2) (\hat{\gamma}_{k+2\ell}^{(0)}+\hat{\gamma}_{k-2\ell}^{(0)}),
\end{align}
where $\hat{\gamma}_k^{(0)}$ is the standard sample autocovariance function.
The approximation in (\ref{eqt:default}) is due to omitting $X_i$ at boundary indices $i$.
			Besides, replacing $\ell \kappa_{\ell} = \sum_{|k| \leq \ell} K(k/\ell)$
			by $\ell \kappa = \ell \int_{-1}^1 K(t) \, \dd t$ in (\ref{eqt:proposed estimator})
			does not affect its asymptotic properties.
			A comprehensive sensitivity analysis of these parameters
            is performed in \S\ref{sec:sensitivity},
			which shows that the proposed estimator performs well and stably under the default parameters.
\end{example}

\begin{example}[Independent noises]
If $\gamma_k=0$ for $k\neq 0$, we have $\hat{\gamma}_{k}^{(m)}\inP 0$ for $k\neq 0$ and $m>0$.
An alternative estimator of $\theta$ is
$\hat{\theta}^{(m)}_{\text{ind}} = \{\hat{\gamma}_0^{(0)} - \hat{\gamma}_0^{(m)} + \sum_{0<|k| \leq \ell} K(k/\ell)\hat{\gamma}_k^{(0)}\}/\sum_{|k| \leq \ell} K(k/\ell)$.
For trend problem,
the data are more naturally assumed to be temporally dependent.
So, we suggest using $\hat{\theta}^{(m)}$ by default.
\end{example}

\begin{example}[Plug-in estimator via nonparametric regression]\label{eg:LLR}
Suppose $\hat{\mu}_{\text{LLR}}(t)$ is local linear regression of $\mu(t)$.
A plug-in estimator of $\theta$ is
$\hat{\theta}_{\text{LLR}} = \int_0^1 \{\hat{\mu}_{\text{LLR}}(t)- \bar{\mu}_{\text{LLR}} \}^2 \, \dd t$,
where $\bar{\mu}_{\text{LLR}} = \int_0^1 \hat{\mu}_{\text{LLR}}(t) \, \dd t$;
see \S\ref{sec:TrencCorrTestMC}.
This estimator has several drawback despite being intuitive.
The estimator $\hat{\mu}_{\text{LLR}}(\cdot)$ needs to be $\mathcal{L}^2$-consistent for $\mu(\cdot)$
in the presence of trend and jumps.
It is not trivial.
Also, it involves sensitive tuning of parameters and requires numerical optimization,
which sometimes makes the estimate unstable or even undefined when the optimization is numerically singular.
\end{example}

\begin{example}[Noise correlation]\label{eg:noiseCorr}
Note that $\hat{\gamma}_{k}^{(m)} \inP \gamma_k + \theta \mathbb{1}(m = 0)$; see Proposition \ref{prop:B3} of the supplement.
A consistent estimator of the noise correlation $\rho_{\noise}$ is
$\hat{\rho}^{(m)}_{\noise} = (\hat{\gamma}_0^{(m)}\odot I_S)^{-1/2} \hat{\gamma}_0^{(m)} (\hat{\gamma}_0^{(m)}\odot I_S)^{-1/2}$
for any $m>0$.
\end{example}

\subsection{Theory of non-degenerate sample trend covariance}\label{subsec:pair TC test}
To perform inference on the trend correlation $\rho$,
we need the asymptotic distribution of $\hat{\rho}^{(m)}$, or more fundamentally of $\hat{\theta}^{(m)}$.
It requires regularizing the kernel $K$ as follows:
\begin{assumption}\label{assum:kernel}
	(i) $t \mapsto K(t)$ is Lipschitz continuous on $(-1,1)$,
	(ii) $A = \int_{0}^1 K^2(t) \, \dd t <\infty$ exists, and
	(iii) there exist $q \in \mathbb{N}$ and $B \in\mathbb{R}\setminus\{0\}$ such that $\lim_{t \downarrow 0} \{K(t) - K(0)\}/|t|^q = B$.
\end{assumption}

We need the following notation.
Let $u_q = \max_{j = 1,\ldots,S} \sum_{k \in \mathbb{Z}} |k|^q |(\gamma_k)_{j,j}|$ for $q\in\mathbb{N}_0$.
Let $K_{S^2}$ and $M_{S^2}$ be the $S^2 \times S^2$ commutation matrix and diagonal selection matrix
defined as
\begin{align}\label{eqt:KM}
	K_{S^2} = \mathop{\sum\sum}_{j_1,j_2=1}^S E_{j_1, j_2} \otimes E_{j_2, j_1}
	\quad \text{and} \quad
	M_{S^2} = \sum_{j=1}^S E_{j, j} \otimes E_{j, j} ,
\end{align}
where $E_{j_1,j_2}\in\{0,1\}^{S\times S}$ whose
$(a,b)$th entry is $\mathbb{1}_{\{(a,b)=(j_1,j_2)\}}$;
see \S\ref{sec:communicationMatrix} for details.

\begin{theorem}[Non-degenerate case: $\theta\neq0$]\label{thm:4.3}
	Let $m \in\mathbb{N}$, $u_q < \infty$ for some $q \in \mathbb{N}$, $h=2\ell$
	and
	$n^{1/(2 + 2q)} \lesssim \ell \ll n^{1/2}$.
	Suppose Assumptions \ref{assum:jumps}, \ref{assum:weak dependence} and \ref{assum:kernel} hold.
	Assume
	$\mathcal{C} +\mathcal{S}\mathcal{J} = o(  n^{1/4}/\ell^{1/2} )$ and
	$\mathcal{S}^2\mathcal{J} = o( n^{3/2}/\ell^2  )$.
    (i) If $\theta_{j,j} > 0$ for all $j \in\{ 1,\ldots,S\}$, then as $n \rightarrow \infty$,
    \begin{align}\label{eqt:CLT_thetaHat}
        n^{1/2} \vectorized \{\hat{\theta}^{(m)}-\theta \}  \inD \Normal\left(0, \Omega_{\theta}\right),
        \quad \text{where} \quad
        \Omega_{\theta} = \left(I_{S^2} + K_{S^2}\right)(v \otimes \theta + \theta \otimes v).
    \end{align}
    (ii) If, in addition $|\rho_{j_1,j_2}|<1$ for all unequal $j_1,j_2 \in \{1,\ldots,S\}$, then
		as $n \rightarrow \infty$,
        \begin{gather}
    		n^{1/2} \vectorized\{\hat{\rho}^{(m)}-\rho\} \inD \Normal\left(0,\Omega_{\rho}\right),
			\quad \text{where} \quad
			\Omega_{\rho} = G \Omega_{\theta} G^{\T}
			\quad \text{and} \quad \nonumber\\
			G = \left\{ I_{S^2}-\frac{1}{2}(\rho \otimes I_S +  I_S \otimes \rho ) M_{S^2} \right\} \{ (\theta\odot I_S)^{-1/2}\otimes (\theta\odot I_S)^{-1/2} \}.
            \label{eqt:def_G}
    	\end{gather}
\end{theorem}

We discuss the assumptions and the interpretations of Theorem \ref{thm:4.3}:
First, Theorem \ref{thm:4.3} requires the lag $h = 2\ell$ for doing differencing so that
			$h$ is long enough to ensure
			the autocovariance of $D_i^{(m)}$
			is approximately the same as that of $Z_i$
			up to lags $|k|\leq \ell$, otherwise $\hat{\gamma}_k^{(0)}-\hat{\gamma}_k^{(m)}$ in (\ref{eqt:proposed estimator})
			may not be a consistent estimate of $\theta$; see \cite{wangchan2026} for more discussions.
			Indeed, the theory supports $h\geq 2\ell$,
            but it is not suggested to use large $h$ as it weakens the differencing effect.
			Thus, we only present the result with the smallest possible $h$, i.e,. $h=2\ell$.
			The assumption of
		$u_q < \infty$ for some $q$ regularizes the strength of serial dependence.
		The larger the value of $q$, the weaker the dependence. Thus, a smaller bandwidth $\ell$ is allowed.
		Neither $q$, $\ell$ nor $h$ affects the limiting distribution (\ref{eqt:CLT_thetaHat}).

Second,
the conditions
$\mathcal{C} +\mathcal{S}\mathcal{J} = o(  n^{1/4}/\ell^{1/2} )$ and
	$\mathcal{S}^2\mathcal{J} = o( n^{3/2}/\ell^2  )$
			come from the de-trending errors due to the differencing operations in Proposition \ref{prop:B2} and Lemma \ref{lem:B8}.
			Differencing asymptotically removes arbitrary trends if
			$\mathcal{C}$, $\mathcal{S}$ and $\mathcal{J}$
			do not diverge to infinity too quickly.
			For example, suppose the recommended $\ell \asymp n^{1/(1+2q)}$ is used
			with $q=2$,
			it covers trend functions with a Lipschitz constant $\mathcal{C}= o(n^{3/20})$
			and $\mathcal{J} = o(n^{3/20})$ jump discontinuities of finite magnitude.

Third, the limiting variance of $\hat{\theta}^{(m)}$ in
			Theorem \ref{thm:4.3} (i) admits a different form than the
			limiting variance of the
			standard sample variance $\hat{\gamma}_0^{(0)} = \sum_{i=1}^n (X_i-\bar{X}_n)^{\otimes 2}/n$ in the classical results \citep{nel1985},
			where $n\Var(\vectorized \hat{\gamma}_0^{(0)}) \rightarrow  (I_{S^2}+K_{S^2}) (\gamma_0 \otimes \gamma_0)$
			for independent and identically distributed~$X_1, \ldots, X_n$.
			It is also different from the limiting variance of the kernel long-run variance estimator
			$\hat{v}^{(0)} = \sum_{|k|\leq \ell} K(k/\ell) \hat{\gamma}_k^{(0)}$,
			which satisfies
			$(n/\ell)\Var(\vectorized \hat{v}^{(0)}) \rightarrow  2 A(I_{S^2}+K_{S^2})(v \otimes v)$
			for stationary $X_1, \ldots, X_n$;
			see, e.g., \cite{andrews1991}.
			We note that $\hat{\theta}^{(m)}$ is $n^{1/2}$-consistent and
			$n\Var\{ \vectorized \hat{\theta}^{(m)}\}$ depends on the more involved quantity $v\otimes \theta + \theta \otimes v$
			reflecting the influence of the trend variability.
			It is worth mentioning that using the sum of lag-products
			in $\hat{\theta}^{(m)}$
			leads to a neater form of asymptotic variance $v\otimes \theta + \theta \otimes v$.
			If only one individual lag-product is used, the asymptotic variance will admit a more complicated form similar to
			the Bartlett formula; see \cite{brockwellDavis1991} and \cite{Wu2009ACVF}.
			Part (ii) of Theorem \ref{thm:4.3} follows from the Delta method,
			that is similar to \cite{SteigerHakstian1982} and \cite{NeudeckerWesselman1990}.

Fourth, kernels that satisfy Assumption \ref{assum:kernel} include the truncated polynomial kernel
		$K(t) = (1 - |t|^q)\mathbb{1}(|t| \leq 1)$ \citep{Parzen1957} and
		Tukey--Hanning kernel $K(t) = \{1 + \cos(\pi t)\}\mathbb{1}(|t| \leq 1)/2$ \citep{andrews1991}.
		See Example \ref{eg:default} for the default choice.

Fifth,
the entries of $\Omega_{\theta}$ and $\Omega_{\rho}$ corresponding to
the asymptotic values of
$n\Cov(\hat{\theta}_{j_1,j_2}^{(m)},\hat{\theta}_{j_3,j_4}^{(m)})$
and $n\Cov(\hat{\rho}_{j_1,j_2}^{(m)},\hat{\rho}_{j_3,j_4}^{(m)})$ are, respectively,
        \begin{align}
        	\Omega_{\theta,(j_1,j_2),(j_3,j_4)}
			&= v_{j_1,j_3}\theta_{j_2,j_4} + v_{j_1,j_4}\theta_{j_2,j_3}
				+ \theta_{j_1,j_3}v_{j_2,j_4} + \theta_{j_1,j_4}v_{j_2,j_3} ,
				\label{eqt:Omegathetaj1234}\\
			\Omega_{\rho,(j_1,j_2),(j_3,j_4)}
			 &=\bigg\{ \widetilde{\Omega}_{\theta,(j_1,j_2),(j_3,j_4)}
                + \frac{1}{4}\rho_{j_1,j_2}\rho_{j_3,j_4}  \sum_{a\in\{j_1,j_2\}}\sum_{b\in\{j_3,j_4\}}\widetilde{\Omega}_{\theta,(a,a),(b,b)}
			 \bigg\}
            	\nonumber \\
			& \qquad	- \frac{1}{2}\rho_{j_1,j_2} \sum_{a\in\{j_1,j_2\}}\widetilde{\Omega}_{\theta,(a,a),(j_3,j_4)}
				- \frac{1}{2}\rho_{j_3,j_4} \sum_{a\in\{j_3,j_4\}}\widetilde{\Omega}_{\theta,(j_1,j_2),(a,a)}
            ,
			 \nonumber
        \end{align}
        where $\widetilde{\Omega}_{\theta,(j_1,j_2),(j_3,j_4)} = \Omega_{\theta,(j_1,j_2),(j_3,j_4)}/ (\theta_{j_1,j_1}\theta_{j_2,j_2}\theta_{j_3,j_3}\theta_{j_4,j_4})^{1/2}$ for any $j_1, j_2, j_3, j_4$.

\subsection{Theory of degenerate sample trend covariance}\label{sec:theory_degenarate}
When $\theta_{j,j}=0$ or $\theta_{j',j'}=0$ for $j\neq j'$, the trend correlation $\rho_{j,j'}$ is defined as zero.
In this degenerate case, the asymptotic behavior of $\hat{\theta}^{(m)}$ changes substantially;
see Theorem \ref{thm:5.3} below.
This case will be shown useful for inference of the trend variance $\theta$.
Let
\begin{align}\label{eqt:Delta}
	\Delta_m = \sum_{|s|\leq m} \delta_s^2\qquad \text{and} \qquad \delta_s = \sum_{j=|s|}^m d_j d_{j-|s|}.
\end{align}

\begin{theorem}[Degenerate case: $\theta=0$]\label{thm:5.3}
    Let $m \in\mathbb{N}$, $u_q < \infty$ for some $q \in \mathbb{N}$,
    $h =2\ell$, and $n^{1/(1 + 2q)} \lesssim \ell \ll n^{1/2}$.
    Suppose Assumptions \ref{assum:jumps}, \ref{assum:weak dependence} and \ref{assum:kernel} hold.
    If $\theta=0$, then as $n \rightarrow \infty$
    \begin{align}\label{eqt:limitDist_thetaHat_H0}
        (n\ell \kappa_\ell^2)^{1/2}\vectorized \hat{\theta}^{(m)} \inD \Normal(0,\Omega^{\circ}_{\theta}),
        \quad \text{where} \quad
        \Omega^{\circ}_{\theta} = 2A\left(\Delta_m - \Delta_0\right)(I_{S^2}+K_{S^2})(v\otimes v).
    \end{align}
\end{theorem}

Discussions of Theorem \ref{thm:5.3} are in the sequel:
First, the assumptions needed for Theorem \ref{thm:5.3} are the same as Theorem \ref{thm:4.3},
			except that $\theta=0$ is needed and the range of feasible $\ell$ is different.
			The shortest possible $\ell$ is of order $n^{1/(1+2q)}$,
			which is larger than $n^{1/(2+2q)}$ required for Theorem \ref{thm:4.3}.
			However, the results are substantially different.
			From (\ref{eqt:limitDist_thetaHat_H0}), $\hat{\theta}^{(m)}\inP\theta=0$ at the rate of $(n\ell)^{1/2}$ instead of the standard $n^{1/2}$-rate.
			The accelerated rate is achieved by aggregating $\ell+1$ sample lag-products in (\ref{eqt:proposed estimator}).
			We will show in \S\ref{sec:inference_tCorr} that this faster rate
			leads to better test statistics.

Second, the rate of convergence $(n\ell)^{1/2}$ is increasing with $\ell$.
			However, we recommend using the minimal possible $\ell$ in order to be conservative
			because longer $\ell$ leads to a poorer differencing effect.
			Unless $n$ is very large, we do not recommend $\ell\gg n^{1/(1+2q)}$.

Third, for any $j_1,j_2,j_3,j_4 \in\{ 1,\ldots,S\}$,
		the entry of $\Omega^{\circ}_{\theta}$ corresponding to the asymptotic value of
		$n\ell\kappa_{\ell}^2\Cov(\hat{\theta}_{j_1,j_2}^{(m)},\hat{\theta}_{j_3,j_4}^{(m)})$ is
		$
			\Omega^{\circ}_{\theta,(j_1,j_2),(j_3,j_4)} = 2A\left(\Delta_m - \Delta_0\right)(v_{j_1,j_3}v_{j_2,j_4} + v_{j_1,j_4}v_{j_2,j_3}).
		$
		A special case of Theorem \ref{thm:5.3} is
		$(n\ell\kappa_\ell^2)^{1/2} \diag \hat{\theta}^{(m)} \inD \Normal\{ 0,4A\left(\Delta_m - \Delta_0\right)(v \odot v)\}$,
        which further reduces to Theorem 4.1 of \cite{ToChan2024} when $S=1$.

Fourth,
the optimal difference sequence $\{d_j\}_{j = 0}^m$ for a fixed $m \in \mathbb{N}$ can be found by minimizing $\Delta_m - \Delta_0 = 1/(2m)$ subject to the conditions $\sum_{j = 0}^m d_j = 0$ and $\delta_0 = \sum_{j = 0}^m d_j^2 = 1$; see, e.g., \citet{hall1990}, including the optimal solution $\{d_j^\star\}_{j = 0}^m$.
For example, when $m=2$, the optimal difference sequence is $(d_0,d_1,d_2)=[(5^{1/2}+1)/4, -1/2, -(5^{1/2}-1)/4]$.
While $m$ should ideally be as large as possible in theory, excessively large values of $m$ can lead to suboptimal finite-sample performance as a larger $m$ weakens the detrending effect.
In practice, we recommend $m = 1$ by default and suggest $m=2$ when $n$ is large.

Owing to the accelerated convergence rate of $\hat{\theta}^{(m)}$ under $\theta=0$,
we can improve $\hat{\theta}^{(m)}$
to $\bar{\theta}^{(m)}$ by thresholding, as stated in (\ref{eqt:proposed estimator_0}).
We have the following general properties.
\begin{corollary}[Thresholded version]\label{cor:improvedVersion}
    Let $m \in\mathbb{N}$, $u_q < \infty$ for some $q \in \mathbb{N}$,
    $h=2\ell$, and $n^{1/(1 + 2q)} \lesssim \ell \ll n^{1/2}$.
    Suppose Assumptions \ref{assum:jumps}, \ref{assum:weak dependence} and \ref{assum:kernel} hold.
    Let $H = \diag\{\mathbb{1}(\theta_{1,1}\neq0) , \ldots, \mathbb{1}(\theta_{S,S}\neq0)\}$.
    If $\max(\epsilon_1, \ldots, \epsilon_S) \inP 0$
    and
    $(n\ell)^{1/2}\min(\epsilon_1, \ldots, \epsilon_S) \inP \infty$,
    then, no matter $\theta=0$ or not, as $n \rightarrow \infty$,
    \begin{align}
        n^{1/2} \vectorized (\bar{\theta}^{(m)} -\theta)
        	&\inD \Normal\{ 0, (H\otimes H)\Omega_{\theta}(H\otimes H) \},  \label{eqt:limitDist_general_theta} \\
        n^{1/2} \vectorized (\bar{\rho}^{(m)} - \rho)
        	&\inD \Normal\{ 0, (H\otimes H)\Omega_{\rho}(H\otimes H)\}.  \label{eqt:limitDist_general_rho}
    \end{align}
\end{corollary}

In (\ref{eqt:limitDist_general_rho}),
the computation of $\Omega_{\rho}$ involves
the diagonal matrix $(\theta \odot I_S)^{-1/2}$; see (\ref{eqt:def_G}).
If some components of $\theta$ are zero,
then we interpret it as $(\theta \odot I_S)^{-1/2} = \diag\{ \theta_{j,j}^{-1/2} \mathbb{1}(\theta_{j,j}\neq 0) \}_{j=1}^S$
by convention.
The estimators $\bar{\theta}^{(m)}$ and $\bar{\rho}^{(m)}$ can be regarded as shrinkage estimators.
The accelerated convergence rate achieved in Theorem \ref{thm:5.3} is similar to the Hodges estimator;
see Chapter 8 of \cite{vanDerVaart2000} for a review.
This faster-rate of convergence is useful for handling the degenerate case of $\theta=0$.
The limiting covariance in (\ref{eqt:limitDist_general_theta}) corresponding to $(j_1, j_2)$ and $(j_3, j_4)$ entries is
$\Omega_{\theta, (j_1, j_2), (j_3, j_4)} \mathbb{1}(\theta_{j_1,j_1}\theta_{j_2,j_2}\theta_{j_3,j_3}\theta_{j_4,j_4}\neq 0)$.
In particular,
\[
	n^{1/2}(\bar{\theta}^{(m)}_{j_1, j_2} - \theta_{j_1, j_2})
		\left\{
		\begin{array}{ll}
		\inD \Normal(0, \Omega_{\theta, (j_1, j_2), (j_1,j_2)}) & \text{if $\theta_{j_1,j_1}\theta_{j_2,j_2}\neq 0$}; \\
		\inP 0 & \text{if $\theta_{j_1,j_1}\theta_{j_2,j_2}= 0$}.
		\end{array}
		\right.
\]

\subsection{Implementation}
According to (\ref{eqt:CLT_thetaHat}) and (\ref{eqt:limitDist_thetaHat_H0}),
inference of $\theta$ and $\rho$ requires
the knowledge of $v = \sum_{k=-\infty}^{\infty} \gamma_k$.
In practice, it should be estimated robustly in the presence of trend and jumps,
otherwise it may lead to incorrect quantification of the precision.
One parametric method and one nonparametric method are provided;
see Examples \ref{eg:para_v}--\ref{eg:nonpara_v}.
The other unknown $\theta$ in the asymptotic variance of (\ref{eqt:CLT_thetaHat}) can be estimated by
plugging in $\hat{\theta}^{(m)}$.
Hence, we can define
$\hat{\Omega}_{\theta}$
by
$\Omega_{\theta}$
with $(v, \theta)$
replaced by  $(\hat{v},\hat{\theta})$.
Similarly, we  define
$\bar{\Omega}_{\theta}$
by
$\Omega_{\theta}$
with $(v, \theta)$
replaced by  $(\hat{v},\bar{\theta})$.
Other asymptotic covariance matrices are defined similarly.

\begin{example}[Parametric plug-in]\label{eg:para_v}
Assume a parametric model on $\{Z_i\}$.
For illustration, we consider an $\textsc{ar}(1)$ model:
$Z_i = \Phi Z_{i-1} + \varepsilon_i$, where $\Phi\in\mathbb{R}^{S\times S}$ and
$\varepsilon_i$'s are independent and identically distributed with $\E(\varepsilon_i)=0$.
We estimate $\Phi$ by the difference-based estimator $\hat{\Phi}^{(m)} = \hat{\gamma}_1^{(m)}(\hat{\gamma}_0^{(m)})^{-1}$,
where $\hat{\gamma}_0^{(m)}$ and $\hat{\gamma}_1^{(m)}$ are defined in (\ref{eqt:dacvf estimator})
with $h=\lfloor n^{1/2}\rfloor$ and $m=1$.
Then an estimator of $v$
is $\hat{v}_{\textsc{para}}^{(m)} = \hat{\gamma}_{0}^{(m)} + \hat{C}^{(m)} + (\hat{C}^{(m)})^{\T}$, where
$\hat{C}^{(m)} = \hat{\gamma}_0^{(m)} (\hat{\Phi}^{(m)})^{\T} \{I_S-(\hat{\Phi}^{(m)})^{\T}\}^{-1}$.
\end{example}

\begin{example}[Nonparametric plug-in]\label{eg:nonpara_v}
There are many possible nonparametric estimators of $v$; see, e.g.,
\cite{wu2001}, \cite{wu_zhao_2007},  \cite{DetteWu2019}, \cite{chan2022_mac,Chan2020}, etc.
Similar to Example \ref{eg:para_v},
we may use $\hat{\gamma}_k^{(m)}$ to estimate $\gamma_k$.
It gives an estimator of $v$ as $\hat{v}^{(m)}_{\textsc{nonp}} = \sum_{|k| \leq \ell} K(k/\ell)\hat{\gamma}_k^{(m)}$
with $\ell = \lfloor 2 n^{1/5} \rfloor$, $h=2\ell$, $m=1$ and $K(t)= (1-t^2)^+$;
see \citet{ChanYau2022} and \citet{LiuChan2022} for a model-free way for optimally selecting $\ell$.
By default,
we will use this pilot estimator throughout the article.
\end{example}

\section{Applications}\label{sec:applications}
\subsection{Inference of trend correlation}\label{sec:inference_tCorr}
We consider three inference problems for $\rho$:
(i) one-pair, (ii) two-pair, and (iii) all-pair:
\[
	H_0^{\one}: \rho_{j_1,j_2}=\rho_0, \quad
	H_0^{\two}: \rho_{j_1,j_2}=\rho_{j_3,j_4}, \quad
	H_0^{\all}: \forall \, 1\leq j_1 < j_2\leq S,\, \rho_{j_1,j_2} = 0.
\]
where $\rho_0\in(-1,1)$ is a specific null value; see
Example \ref{eg:onePairTest},
Example \ref{eg:twoPairTest}, and
Example \ref{eg:allPairTest1}, respectively.
More examples are deferred to \S\ref{sec:additionInference_tCor}.
The thresholded $\bar{\theta}^{(m)}$ and $\bar{\rho}^{(m)}$
are used in the trend correlation tests to handle the degenerate case of $\theta=0$.

\begin{example}[One-pair trend correlation test]\label{eg:onePairTest}
We test $H_0^{\one}$ against $H_1^{\one}: \rho_{j_1,j_2}>\rho_0$
for a particular pair $j_1\neq j_2$.
In particular, setting $\rho_0 = 0$
tests for uncorrelated trends,
meaning that their deterministic components have no linear relationship.
This is particularly useful when the direction and strength of the association are of primary interest,
such as determining whether the crime rate trend increases with the temperature trend; see \S\ref{sec:weather and crime}.
The test statistic is
\[
	\hat{T}^{(m)}_{\one,\rho} = \frac{\bar{\rho}^{(m)}_{j_1, j_2} - \rho_0}{\{\bar{\Omega}_{\rho,(j_1,j_2),(j_1,j_2)}/n\}^{1/2}}.
\]
If $\rho_0=0$, the null hypothesis $H_0^{\one}:\rho_{j_1,j_2}=0$ is equivalent to
$\theta_{j_1,j_2}=0$. In this case, another possible test statistic is
$
	\hat{T}^{(m)}_{\one,\theta}
		=  n^{1/2}\bar{\theta}^{(m)}_{j_1, j_2} / \bar{\Omega}_{\theta,(j_1,j_2),(j_1,j_2)}^{1/2} .
$
Suppose $\hat{T} \in \{\hat{T}^{(m)}_{\one,\rho}, \hat{T}^{(m)}_{\one,\theta}  \}$ is used.
Then we reject $H_0^{\one}$ at size $\alpha$ asymptotically
if $\hat{T}> \Phi^{-1}(1-\alpha)$, where $\alpha\in(0,1)$ and $\Phi^{-1}$ is the quantile function of $\Normal(0,1)$.
Two-sided tests can be derived similarly.
\end{example}

\begin{example}[Two-pair trend correlation test]\label{eg:twoPairTest}
We test $H_0^{\two}$ against
$H_1^{\two}: \rho_{j_1,j_2} > \rho_{j_3,j_4}$ for some
$j_1,j_2,j_3,j_4 \in \{1,\ldots,S\}$ such that $j_1 \neq j_2$, $j_3 \neq j_4$, and $j_1 \neq j_3$.
It leads to two cases: $j_2=j_4$ and $j_2\neq j_4$.
The first case compares two trend correlations with one component in common.
We can interpret $H_1^{\two}$ as the trend of component $j_2$ moving more in sync with the trend of component $j_1$ than with the trend of component $j_3$.
In \S\ref{sec:weather and crime}, we will discuss an application of the equal trend correlation test where we compare the trend correlations between air temperature and two different types of crimes.

The second case aims to test
whether one pair of time series is more strongly correlated
than another pair.
By using indices, it provides a unit-free comparison for any time series data.
In this sense, it is more convenient and objective than
comparing two regression coefficients.
It is worth mentioning that
the asymptotic variance $\Omega_{\rho,(j_1,j_2),(j_1,j_2)}$ admits different forms in these two cases.
In both cases, the test statistic is
\begin{align*}
	\hat{T}^{(m)}_{\two,\rho} &= \frac{\bar{\rho}^{(m)}_{j_1, j_2} - \bar{\rho}^{(m)}_{j_3, j_4}}{\{( \bar{\Omega}_{\rho,(j_1,j_2),(j_1,j_2)}+\bar{\Omega}_{\rho,(j_3,j_4),(j_3,j_4)}-2\bar{\Omega}_{\rho,(j_1,j_2),(j_3,j_4)} )/n\}^{1/2}}  .
\end{align*}
Then we reject $H_0^{\two}$ at size $\alpha$ asymptotically
if $\hat{T}_{\two, \rho}^{(m)} > \Phi^{-1}(1-\alpha)$.
\end{example}

\begin{example}[All-pair trend correlation test]\label{eg:allPairTest1}
We test $H_0^{\all}$
against
$H_1^{\all}: \,\rho_{j_1,j_2} \neq 0$ for some $j_1, j_2\in\{1, \ldots, S\}$ with $j_1\neq j_2$.
We call it a portmanteau trend test.
Assume $\theta_{j,j}>0$ for all $j$.
Instead of performing the one-pair test for all $S(S-1)/2$ pairs and then combining them,
we do it after adjusting for the intercorrelation.
This test is particularly effective for
identifying significant directional trend relationships
within a set of time series.
Recall that $\upper_0$ is the vectorization of the upper triangular part without the diagonal.
The test statistic is
\begin{align*}
	\hat{T}_{\all,\rho}^{(m)}
		&= n \upper_0(\bar{\rho}^{(m)})^{\T}\bar{\Omega}_{\rho,\upper_0}^{-1} \upper_0(\bar{\rho}^{(m)}).
\end{align*}
Define $\hat{T}_{\all,\theta}^{(m)}$ similarly.
Both statistics are of the Hotelling's $T^2$ type and are based on the Mahalanobis distance.
Under $H_0^{\all}$, we have $\hat{T}_{\all,\bullet}^{(m)} \inD \chi^2_{S(S-1)/2}$ for $\bullet\in\{\rho, \theta\}$.
So, the asymptotic $p$-value is
$\hat{p}_{\all,\bullet}^{(m)} = 1-F_{S(S-1)/2}(\hat{T}_{\all, \bullet}^{(m)})$,
where $F_k(\cdot)$ is the distribution function of $\chi^2_k$.

Note that in practice, when $\theta_{j,j}=0$ is possible for some $j$,
one may perform the test only on the components
$\hat{\mathcal{A}}=\{ 1\leq j\leq S : \hat{\theta}_{j,j}^{(m)} \geq \epsilon_j\}$, i.e.,
the estimated set of components with non-constant trends.
Let  $\hat{A} = |\hat{\mathcal{A}}|$, so
$\hat{A} \inP A := |\mathcal{A}|$, where $\mathcal{A}=\{ 1\leq j\leq S : \theta_{j,j}>0\}$.
Based on the modified statistic $\hat{T}_{\all,\rho}^{(m)}$ based on components in $\hat{\mathcal{A}}$,
the asymptotic $p$-value is
$\hat{p}_{\all,\bullet}^{(m)} = 1-F_{\hat{A}(\hat{A}-1)/2}(\hat{T}_{\all, \bullet}^{(m)})$.
\end{example}

\subsection{Inference of trend variance}\label{sec:trendVar}
This section presents
auxiliary results derived from our main methodology
for trend correlation.
The goal is to test $H_0^{\trend}$: $\mu_j(\cdot)$ is a constant function for $j=1, \ldots, S$,
or equivalently,
\begin{align}\label{eqt:H0trend}
	H_0^{\trend}: \theta_{1,1}=\cdots=\theta_{S,S}=0
\end{align}
against $H_1^{\trend}: \theta_{j,j} > 0$ for some $j = 1,\ldots,S$.
The mean stationarity problem addressed here
builds upon the frameworks of \citet{HeinrichsDette2021}, \citet{ToChan2024}, and other approaches utilizing the integrated squared deviation criterion.
We contribute by
(i) allowing for a general $S \geq 1$,
(ii) using a difference-based approach to bypass trend estimation, and
(iii) assessing the utility of trend covariances in trend stationarity test.

Two trend tests are proposed: one based on the trend variances, i.e., the diagonal of $\theta$,
and another incorporating the full trend covariance matrix;
see Examples \ref{eg:trendTest_var}--\ref{eg:trendTest_covar}, respectively.
Example \ref{remark 4.2} shows a procedure for selecting components that have non-constant trends via
false-discovery rate control.
Note that the non-thresholded versions
${\hat{\theta}}^{(m)}$ and ${\hat{\rho}}^{(m)}$ are used in the tests
because their continuous asymptotic distributions under the degenerate case are well-established.

\begin{example}[Trend variance test]\label{eg:trendTest_var}
In view of (\ref{eqt:H0trend}), a sensible test statistic is built on an estimator of
$\diag(\theta) = (\theta_{1,1},\ldots, \theta_{S,S})$. Define the trend variance test statistic as
\begin{equation}\label{eqt:TS}
	\hat{T}_{\trend,\diag(\theta)}^{(m)}
		= \frac{n\ell \kappa_\ell^2}{4A(\Delta_m - \Delta_0)}\left\vert (\hat{v}^{(m)} \odot \hat{v}^{(m)})^{-1/2}\diag(\hat{\theta}^{(m)})\right\vert^2.
\end{equation}
By Theorem \ref{thm:5.3}, the continuous mapping theorem, and the Slutsky's Theorem,
$\hat{T}_{\trend,\diag(\theta)}^{(m)}\inD \chi^2_{S}$ under $H_0^{\trend}$.
On the other hand, under  $H_1^{\trend}$,
$\hat{T}_{\trend,\diag(\theta)}^{(m)} = C n\ell  |\hat{\theta}^{(m)}|^2  = Cn\ell \{ |\theta|^2  +o_p(1)\}\inP \infty$
for some $C\asymp 1$.
Hence, using
the asymptotic $p$-value
$\hat{p}_{\trend,\diag(\theta)}^{(m)} = 1- F_S(\hat{T}_{\trend,\diag(\theta)}^{(m)})$
leads to a valid and consistent test.
\end{example}

\begin{example}[Trend covariance test]\label{eg:trendTest_covar}
The trend variance test $\hat{T}_{\trend,\diag(\theta)}^{(m)}$
only utilizes the diagonal of $\hat{\theta}^{(m)}$.
However,
$\theta_{1,1}= \cdots = \theta_{S,S} = 0$
implies that $\theta_{j_1,j_2} = 0$ for all $j_1,j_2 \in \{1,2,\ldots,S\}$.
This observation prompts us to define a test using the upper triangular sub-matrix of $\hat{\theta}^{(m)}$.
Similar to (\ref{eqt:TS}), we define the trend covariance test statistic as
\begin{equation}\label{eqt:22}
    \hat{T}_{\trend,\theta}^{(m)}
    = n\ell \kappa_\ell^2 (\upper \hat{\theta}^{(m)})^{\T}
    		\left(\hat{\Omega}^{\circ}_{\theta,\upper}\right)^{-1}(\upper \hat{\theta}^{(m)}),
\end{equation}
where $\hat{\Omega}^{\circ}_{\theta,\upper}$ is the submatrix of $\hat{\Omega}^{\circ}_{\theta}$
that corresponds to the components in $\upper(\theta)$.
Similarly, Theorem \ref{thm:5.3} implies that $\hat{T}_{\trend,\theta}^{(m)} \inD \chi^2_{S(S+1)/2}$.
The asymptotic $p$-value
$\hat{p}_{\trend,\theta}^{(m)} = 1- F_{S(S+1)/2}(\hat{T}_{\trend,\theta}^{(m)})$ gives a valid and consistent test.
This trend covariance test $\hat{T}_{\trend,\theta}^{(m)}$ is more powerful
than the trend variance test $\hat{T}_{\trend,\diag(\theta)}^{(m)}$ when $\theta_{j_1, j_2}\neq 0$ for all
$j_1, j_2\in\{1, \ldots, S\}$; see
\S\ref{subsubsec:number of means} for a simulation demonstration.
\end{example}

\begin{example}[Trend component selection]\label{remark 4.2}
	The goal is to extract the components that have non-constant trends.
	For any $\mathcal{S}\subseteq\{1, \ldots, S\}$,
	denote $H_0^{\mathcal{S}} : \theta_j = 0$ for $j\in\mathcal{S}$.
	Hence, the global null hypothesis is $H_0^{\trend} = H_0^{\{1,\ldots,S\}}$.
	Also, denote the $p$-value for testing $H_0^{\mathcal{S}}$ as $\hat{p}_{\mathcal{S}}$,
	which can be found according to Example \ref{eg:trendTest_var} or \ref{eg:trendTest_covar}.
	We set $\hat{p}_{\emptyset}=0$.
	Suppose that users have a pre-defined testing order of the components,
	which may depend on the context.
	Let the target overall false discovery rate be $\alpha$.
	Without loss of generality,
	we test $H_0^{\{1, \ldots, S\}}, H_0^{\{1, \ldots, S-1\}}, \ldots, H_0^{\{1\}}$ sequentially.
	Let $\mathcal{S}_j = \{1, \ldots, j\}$ for $j=S, \ldots, 0$, where
	$\mathcal{S}_0=\emptyset$.
	The selection procedure is as follows:
	For $j = S,\ldots, 0$,
	if $\hat{p}_{\mathcal{S}_j} \leq \alpha_j \equiv \min\{S\alpha/(S - j + 1),1\}$ and $j>0$,
	then proceed to the next $j$; otherwise stop and report
	the estimated set of components with non-constant trend: $\hat{\mathcal{S}}= \{1, \ldots, S\}\setminus \mathcal{S}_j$.

	By Theorem 3.1 of \citet{Lynch_etal2017}, this procedure strongly controls the false discovery rate at level $\alpha$ under arbitrary dependence of the $p$-values; see also Theorem 2 of \citet{LynchGuo2016}.
	Section \ref{subsubsec:number of means} provides a simulation study on this procedure.
\end{example}

\section{Simulation experiments}\label{sec:simulation}

\subsection{Trend correlation tests}\label{sec:TrencCorrTestMC}
We assess the performance of trend correlation tests for
$H_0^{\one}:\rho_{1,2}=0$, $H_0^{\two}:\rho_{1,2}=\rho_{1,3}$, and $H_0^{\all}:\rho_{1,2}=\rho_{1,3}=\rho_{2,3}=0$
against their respective two-sided alternative hypotheses.

We compare our proposed tests with different competing tests
based on
sample Pearson's correlation,
linear regression, and
local linear regression:
\begin{itemize}[noitemsep,topsep=0pt]
	\item Let $\hat{\rho}_{\pearson, j_1 ,j_2}$ be the sample Pearson's correlation of $\{X_{i,j_1}\}_{i=1}^n$
			and $\{X_{i,j_2}\}_{i=1}^n$.
	\item Let $(\hat{\beta}_{\LR,j_1,j_2},\hat{\rho}_{\LR, j_1, j_2})$ be the
the least-squares estimator of $(\beta_{j_1,j_2},\rho_{\LR,j_1,j_2})$ in the
linear regression model: $\E(X_{i, j_1} \mid X_{i, j_2} )  = \beta_{j_1,j_2} + \rho_{\LR,j_1,j_2} X_{i,j_2}$.
	\item Let $\hat{\rho}_{\LLR,j_1,j_2}$ be
the sample Pearson's correlation of the series $\{\hat{\mu}_{\LLR,j_1}(i/n)\}_{i=1}^n$ and $\{\hat{\mu}_{\LLR,j_2}(i/n)\}_{i=1}^n$,
where $\hat{\mu}_{\LLR,j}(\cdot)$ is the local linear regression estimator of $\mu_j(\cdot)$
by fitting $\{ X_{i,j} \}_{i=1}^n$ on $\{ i/n \}_{i=1}^n$ with the {R} function
{``KernSmooth::locpoly''} and bandwidth selected by the {R} function {``KernSmooth::dpill''}.
\end{itemize}
Let the corresponding one-pair, two-pair and all-pair trend correlation test statistics be
\begin{gather*}
	\hat{T}_{\one,\bullet} = \hat{\rho}_{\bullet,j_1,j_2}^2, \qquad
	\hat{T}_{\two,\bullet} = (\hat{\rho}_{\bullet,j_1,j_2}-\hat{\rho}_{\bullet,j_3,j_4})^2,  \qquad
	\hat{T}_{\all,\bullet} = \sum_{j_1=1}^{S-1}\sum_{j_2=j_1+1}^S \hat{\rho}_{\bullet,j_1,j_2}^2,
\end{gather*}
for $\bullet \in \{\text{\pearson}, \LR, \LLR\}$.
To the best of our knowledge, there are no direct theoretical results for testing trend correlation using those devices.
For each $\mathsf{problem}\in\{\one,\two,\all\}$ and $\bullet \in \{\text{\pearson}, \LR, \LLR\}$,
we reject $H_0^{\mathsf{problem}}$ if $\hat{T}_{\mathsf{problem},\bullet}>C_{\mathsf{problem},\bullet}$, where
$C_{\mathsf{problem},\bullet}$
is the 95\% null quantiles of $\hat{T}_{\mathsf{problem},\bullet}$ computed
by simulation under each specific mean and noise structure and sample size $n$ described below, so that the tests
are calibrated to have exact 5\% size and are regarded as oracle tests.
The proposed tests $\hat{T}_{\mathsf{problem}, \theta}^{(m)}$ and $\hat{T}_{\mathsf{problem}, \rho}^{(m)}$
are computed with $m=1$ and $q=2$; see Examples \ref{eg:onePairTest}, \ref{eg:twoPairTest}, and \ref{eg:allPairTest1}
for the definitions.

\begin{table}[tp]
\captionsetup{font=small}
\small
\setlength{\tabcolsep}{3.5pt}
\begin{center}
\begin{tabular}{ccccc}
\toprule
\multicolumn{2}{c}{Case} & One-pair test & Two-pair test  & All-pair test \\ \midrule
$\mu(t)$ & D 	& $\begin{bmatrix}4t  + \delta \mathbb{1}_{(t>1/2)} \\8(t-1/2)^2  + \delta \mathbb{1}_{(t>1/2)} \end{bmatrix}$
						& $\begin{bmatrix}15|t-0.3| \\ 15|t-0.7+0.5\delta\mathbb{1}_{(t>1/2)}| \\ 15|t-0.7| \end{bmatrix} $
						& $\begin{bmatrix}4\sqw(t) + \delta\sqw(3t/2)  \\
									2\sqw(2t) + \delta\sqw(3t/2)  \\
									\sqw(4t) + \delta\sqw(3t/2)  \end{bmatrix}$ \\
 \cmidrule(r){2-5}
    	& C 	& $\begin{bmatrix}4t  + 2\delta \surd{t} \\ 8(t-1/2)^2  + 2\delta \surd{t} \end{bmatrix}$
					& $\begin{bmatrix}15|t-0.3| \\ 15|t-0.7+0.5\delta\surd{t}| \\ 15|t-0.7| \end{bmatrix} $
					& $\begin{bmatrix}2(2t-1) + \delta \surd{t} \\
									2(6t^2-6t+1) + \delta \surd{t}\\
									2(20t^3-30t^2+12t-1) +\delta \surd{t} \end{bmatrix}  $ \\
 \cmidrule(r){1-5}
$\Phi$ 	 	& $+$ 	& $\begin{bmatrix} 0.5 & 0.25 \\ 0 & 0.5 \end{bmatrix}$
 				& $\begin{bmatrix} 0.5 & 0.25 & 0 \\ 0 & 0.5 & 0.25 \\ 0.25 & 0 & 0.5 \end{bmatrix}$
 				& $\begin{bmatrix} 0.5 & 0.25 & 0 \\ 0 & 0.5 & 0.25 \\ 0.25 & 0 & 0.5 \end{bmatrix}$  \\
  \cmidrule(r){2-5}
 	 	& $-$ 	& $\begin{bmatrix} -0.5 & -0.25 \\ 0 & -0.5 \end{bmatrix}$
 				& $\begin{bmatrix} -0.5 & -0.25 & 0 \\ 0 & -0.5 & -0.25 \\ -0.25 & 0 & -0.5 \end{bmatrix}$
 				& $\begin{bmatrix} -0.5 & -0.25 & 0 \\ 0 & -0.5 & -0.25 \\ -0.25 & 0 & -0.5 \end{bmatrix}$  \\
			 \bottomrule
\end{tabular}
\caption{\label{table:MC_mean_noise_cases}
For each $\mathsf{problem}\in\{\one,\two,\all\}$,
the mean function $\mu(t)$ has two cases: discontinuous (D) and continuous (C), while
the $\AR$ coefficient matrix $\Phi$ has two cases, denoted ($+$) and ($-$), corresponding to positive
and negative dependence.
Let the square-wave function be $\sqw(t) = \sgn\{\sin(2\pi t)\}$, where $\sgn(\cdot)$ is the signum function.
If $\delta=0$, then $H_0^{\one}$, $H_0^{\two}$, and $H_0^{\all}$ hold.
When the magnitude of $\delta$ increases,
the data generating process deviates more from $H_0^{\mathsf{problem}}$, i.e.,
$|\rho_{1,2}|$, $|\rho_{1,2}-\rho_{1,3}|$, and $|\rho_{1,2}|+|\rho_{1,3}|+|\rho_{2,3}|$ are further away from zero.
}
\end{center}
\end{table}

Assume the noise $Z_i$ is generated as $Z_i = (v')^{-1/2}Z'_i$.
Here $\{Z'_i\}$
is a vector $\AR(1)$ process:
$Z'_i = \Phi Z'_{i-1} + \varepsilon_i$, where $\varepsilon_i\sim\Normal_2(0,\Sigma_{\varepsilon})$ independently;
the $(j_1,j_2)$ entry of $\Sigma_{\varepsilon}$ is
$(\Sigma_{\varepsilon})_{j_1,j_2}=\mathbb{1}_{(j_1=j_2)} + 0.5\mathbb{1}_{\{\max(j_1,j_2)>\min(j_1,j_2)=1\}}$; and
$v' = \lim_{n\rightarrow \infty}n\Var(\bar{Z}'_n)$.
So, the long-run variance matrix $v$ for $\bar{Z}_n$ is standardized to an identity matrix for any $\Phi$.
For each $\mathsf{problem}\in\{\one,\two,\all\}$,
we consider two mean structures $\mu(t)$ and two $\AR$ structures $\Phi$,
summarized in Table \ref{table:MC_mean_noise_cases}.
We set the nominal size $\alpha=5\%$.
We consider $n\in\{200, 400, 800\}$.
Table \ref{table:tcor_G17_tcorTest_n200_m1_adjust1} shows the size and
the smallest power across $\pm\delta$ when
$n=200$ under Cases $\{\text{D}, \text{C}\}\times \{+,-\}$.
All simulations are replicated 1000 times.

\begin{table*}[t]
\captionsetup{font=small}
\centering
 \setlength{\tabcolsep}{4pt}
\small
\begin{tabular}{ccc rrrrr rrrrr}
 \toprule
	&& & \multicolumn{5}{c}{Case D (discontinuous $\mu$)}& \multicolumn{5}{c}{Case C (continuous $\mu$)}\\
	 \cmidrule(r){4-8}	\cmidrule(r){9-13}
$\mathsf{problem}$	& Case & $H_{0/1}^\mathsf{problem}$ & $\hat{T}_{\pearson}$ & $\hat{T}_{\LR}$ & $\hat{T}_{\LLR}$ & $\hat{T}_{\theta}$ & $\hat{T}_{\rho}$ & $\hat{T}_{\pearson}$ & $\hat{T}_{\LR}$ & $\hat{T}_{\LLR}$ & $\hat{T}_{\theta}$ & $\hat{T}_{\rho}$  \\
 \cmidrule(r){1-13}
	$\mathsf{one}$ & $+$ & 	$\delta=0$	&	$ 5.0$	&	$ 5.0$	&	$ 5.0$	&	$ 6.9$	&	$ 7.4$	&	$ 5.0$	&	$ 5.0$	&	$ 5.0$	&	$ 6.9$	&	$ 7.4$	\\
	& & $\delta=\pm0.5$	&	$51.3$	&	$32.4$	&	$64.5$	&	$62.7$	&	$64.9$	&	$50.0$	&	$27.7$	&	$59.3$	&	$59.1$	&	$61.2$	\\
	& & $\delta=\pm1$	&	$83.0$	&	$37.5$	&	$94.0$	&	$95.3$	&	$95.3$	&	$88.0$	&	$35.5$	&	$93.9$	&	$95.4$	&	$95.9$	\\
	 \cmidrule(r){2-13}
		 & $-$ & 	$\delta=0$	&	$ 5.0$	&	$ 5.0$	&	$ 5.0$	&	$ 3.9$	&	$ 4.3$	&	$ 5.0$	&	$ 5.0$	&	$ 5.0$	&	$ 3.9$	&	$ 4.3$	\\
	& & $\delta=\pm0.5$	&	$11.7$	&	$ 8.8$	&	$72.1$	&	$60.8$	&	$63.7$	&	$12.1$	&	$ 8.4$	&	$67.9$	&	$57.2$	&	$60.7$	\\
	& & $\delta=\pm1$	&	$17.3$	&	$11.1$	&	$98.5$	&	$94.6$	&	$95.8$	&	$18.6$	&	$10.5$	&	$96.8$	&	$93.6$	&	$94.7$	\\
	 \cmidrule(r){1-13}
	$\mathsf{two}$ & $+$ & 	$\delta=0$	&	$ 5.0$	&	$ 5.0$	&	$ 5.0$	&	$ 5.4$	&	$ 4.7$	&	$ 5.0$	&	$ 5.0$	&	$ 5.0$	&	$ 5.4$	&	$ 4.7$	\\
	& & $\delta=\pm0.5$	&	$12.2$	&	$14.4$	&	$34.6$	&	$27.1$	&	$32.7$	&	$27.0$	&	$21.5$	&	$51.3$	&	$53.0$	&	$49.3$	\\
	& & $\delta=\pm1$	&	$68.9$	&	$66.2$	&	$86.3$	&	$67.8$	&	$82.2$	&	$93.3$	&	$78.7$	&	$97.3$	&	$97.8$	&	$97.2$	\\
	 \cmidrule(r){2-13}
		 & $-$ & 	$\delta=0$	&	$ 5.0$	&	$ 5.0$	&	$ 5.0$	&	$ 4.4$	&	$ 4.5$	&	$ 5.0$	&	$ 5.0$	&	$ 5.0$	&	$ 4.4$	&	$ 4.5$	\\
	& & $\delta=\pm0.5$	&	$ 1.1$	&	$ 1.2$	&	$30.8$	&	$23.5$	&	$30.4$	&	$ 0.3$	&	$ 0.5$	&	$48.5$	&	$49.9$	&	$48.2$	\\
	& & $\delta=\pm1$	&	$ 0.2$	&	$ 0.2$	&	$83.5$	&	$64.1$	&	$80.3$	&	$ 0.0$	&	$ 0.1$	&	$96.7$	&	$97.6$	&	$96.6$	\\
	 \cmidrule(r){1-13}
	$\mathsf{all}$ & $+$ & 	$\delta=0$	&	$ 5.0$	&	$ 5.0$	&	$ 5.0$	&	$ 3.5$	&	$ 1.2$	&	$ 5.0$	&	$ 5.0$	&	$ 5.0$	&	$ 3.1$	&	$ 5.0$	\\
	& & $\delta=\pm0.5$	&	$ 0.0$	&	$ 3.0$	&	$ 0.0$	&	$13.6$	&	$ 9.2$	&	$ 0.3$	&	$ 3.0$	&	$ 0.3$	&	$23.4$	&	$26.0$	\\
	& & $\delta=\pm1$	&	$ 0.0$	&	$60.5$	&	$ 0.0$	&	$88.7$	&	$91.8$	&	$ 0.0$	&	$20.6$	&	$ 0.0$	&	$76.1$	&	$83.9$	\\
	 \cmidrule(r){2-13}
		 & $-$ & 	$\delta=0$	&	$ 5.0$	&	$ 5.0$	&	$ 5.0$	&	$ 2.4$	&	$ 8.1$	&	$ 5.0$	&	$ 5.0$	&	$ 5.0$	&	$ 2.3$	&	$ 3.8$	\\
	& & $\delta=\pm0.5$	&	$ 1.5$	&	$ 9.1$	&	$ 0.2$	&	$10.4$	&	$ 8.9$	&	$ 3.2$	&	$ 3.2$	&	$ 0.2$	&	$15.5$	&	$20.9$	\\
	& & $\delta=\pm1$	&	$ 0.0$	&	$35.6$	&	$ 0.0$	&	$82.3$	&	$90.4$	&	$ 1.6$	&	$ 2.6$	&	$ 0.0$	&	$63.3$	&	$69.4$	\\
 \bottomrule
\end{tabular}
\caption{
\label{table:tcor_G17_tcorTest_n200_m1_adjust1}
For $\mathsf{problem}\in\{\one,\two,\all\}$, we show
	(i) the size under the null hypotheses (i.e., $\delta=0$), and
	(ii) the smallest power across $\pm\delta$ under the alternative hypotheses.
	The oracle tests $\hat{T}_{\pearson}$, $\hat{T}_{\LR}$, and $\hat{T}_{\LLR}$
	are calibrated to have exact 5\% size, which is unachievable in practice.
    The mean structure (Case D and C) and the noise structure (Cases $+$ and $-$)
    are defined in Table \ref{table:MC_mean_noise_cases}.
	The sample size is $n=200$ and nominal size is $5\%$}
\end{table*}

From Table \ref{table:tcor_G17_tcorTest_n200_m1_adjust1}, we observe that the tests
$\hat{T}_{\pearson}$ and $\hat{T}_{\LR}$ are typically
inconsistent because they incorrectly count the noise variance as part of the
trend variance; see \S\ref{sec:failure}.
It is remarked that we also implemented $\hat{T}_{\pearson}$ with Fisher's $z$-test for dependent correlations
via the ``diffcor'' package \citep{diffcor}.
A similar phenomenon of inconsistency was observed.
The test $\hat{T}_{\LLR}$ works in most cases.
Although it may have power comparable to our proposed tests
$\hat{T}_{\theta}$ and $\hat{T}_{\rho}$
in the one-pair and two-pair tests,
its power is substantially reduced, even to zero, in the all-pair test.
We emphasize that the oracle tests $\hat{T}_{\pearson}$, $\hat{T}_{\LR}$, and $\hat{T}_{\LLR}$ are not
implementable in practice because theoretical results on their
limiting null distributions are lacking.

\begin{figure}
    \captionsetup{font=small}
        \centering
        \includegraphics[width=\textwidth]{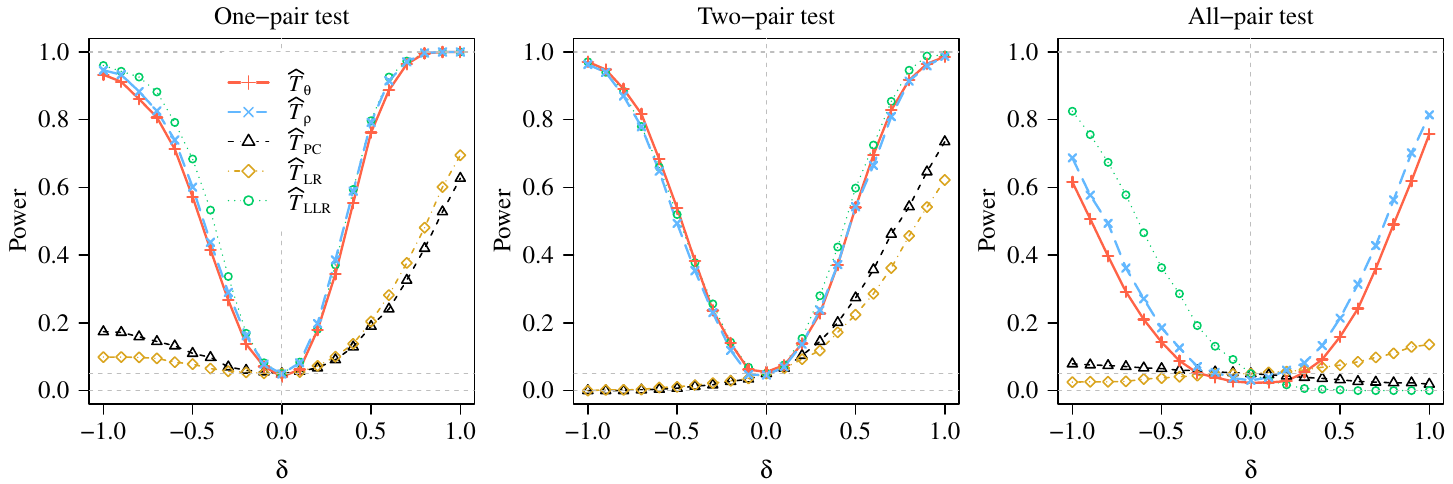}
        \caption{\label{fig:tcor_G17_tcorTest_n200_m1_adjust1_simplePlot}Power curves of one-pair, two-pair, and all-pair trend correlation tests
		under Case C$-$ when $n=200$; see Table \ref{table:MC_mean_noise_cases} for the definitions.
        The nominal size 5\% is indicated by the horizontal gray dashed line.}
    \end{figure}

Our proposed tests $\hat{T}_{\theta}$ and $\hat{T}_{\rho}$
exhibit monotonic power against $\pm\delta$ in all cases, and demonstrate fairly good size control even in finite-samples with $n=200$.
We remark that $\hat{T}_{\theta}$ may not be suitable
for the two-pair trend correlation test when the trend variances differ across components.
In most cases, $\hat{T}_{\rho}$ concerns a unit free measure, shows
slightly better control of the type-I error rate, and more stable power in our experiments.
Hence we recommend $\hat{T}_{\rho}$ as the default
choice.
Figure \ref{fig:tcor_G17_tcorTest_n200_m1_adjust1_simplePlot} shows the full power curves against $\delta\in[-1,1]$
when $n=200$ under Case C$-$.
It visualizes
that
$\hat{T}_{\pearson}$, $\hat{T}_{\LR}$, and
$\hat{T}_{\LLR}$ can have no power in some cases.
This serves as a warning to us against using them in practice.
Additional simulation results are presented in \S\ref{sec:TrencCorrTestMC_supp}.

\subsection{Different numbers of non-constant mean components}\label{subsubsec:number of means}
We investigate whether the covariance trend test
$\hat{T}_{\theta}$
is more powerful than the variance trend test
$\hat{T}_{\diag(\theta)}$
or tests based on combining univariate test results.

We test $H_0^{\trend}$ using $\hat{T}_{\theta}$ and $\hat{T}_{\diag(\theta)}$ with $m=1$ and $q=2$.
For reference, we also compute the tests based on naively applying the univariate test $\hat{T}_{\theta}$
on each component, then combining them via the Bonferroni correction and the harmonic mean of the $p$-values \citep{Wilson2019}.
We denote them by $\hat{T}_{\theta\text{-B}}$ and $\hat{T}_{\theta\text{-W}}$, respectively.

Let $Z_i = A W_i$, where $A\in\mathbb{R}^{5\times 5}$ whose $(j_1,j_2)$th entry is $(-0.8)^{|j_1-j_2|}$ and
$W_i$ is a 5-dimensional independent bilinear time series:
$W_{i,s} = (0.2+0.4\varepsilon_{i,s})W_{i-1,s} + \varepsilon_{i,s}$
and $\varepsilon_{i,s}\sim\Normal(0,1)$ independently.
We consider $\mu(t) = \delta a \odot \eta(t)$, where
$\delta\in\mathbb{R}$ controls the jump magnitude,
$a\in\{ (1,0,0,0,0)^{\T} , (1,1,1,0,0)^{\T}, (1,1,1,1,1)^{\T}\}$ determines the non-zero components, and
$\eta(t)$ defines the trend correlation structure.
We consider two cases of $\eta(t)$:
{\small
\[
    \mu(t)=
    \begin{bmatrix}
    1 &  -1 &  1 & -1 \\
    \frac{1}{2} & -1 & 1 & -1 \\
    1 & -\frac{1}{2} & 1 & -1 \\
    1 & -1 & \frac{1}{2} & -1 \\
    1 & -1 & 1 & -\frac{1}{2}
    \end{bmatrix}
    \begin{bmatrix}
    \mathbb{1}_{(t > 0.2)}\\
    \mathbb{1}_{(t > 0.4)}\\
    \mathbb{1}_{(t > 0.6)}\\
    \mathbb{1}_{(t > 0.8)}
    \end{bmatrix}
    \;\; \text{and} \;\;
    \mu(t)=
    \begin{bmatrix}
    0 & 1 & 0 & -1 & 0 & 1 & 0 & -1 \\
    -1 & \frac{1}{2} & 0 & 0 & 0 & 0 & 0 & 0 \\
    0 & \frac{1}{2} & -1 & \frac{1}{2} & 0 & 0 & 0 & 0 \\
    0 & 0 & 0 & \frac{1}{2} & -1 & \frac{1}{2} & 0 & 0 \\
    0 & 0 & 0 & 0 & 0 & \frac{1}{2} & -1 & \frac{1}{2}
    \end{bmatrix}
    \begin{bmatrix}
    \mathbb{1}_{(t > 0.1)} \\
    \mathbb{1}_{(t > 0.2)} \\
    \vdots\\
    \mathbb{1}_{(t > 0.7)} \\
    \mathbb{1}_{(t > 0.8)}
    \end{bmatrix},
\]}
which lead to strong and zero trend correlations, respectively.

\begin{figure}[t]
    \captionsetup{font=small}
        \centering
        \includegraphics[width=\textwidth]{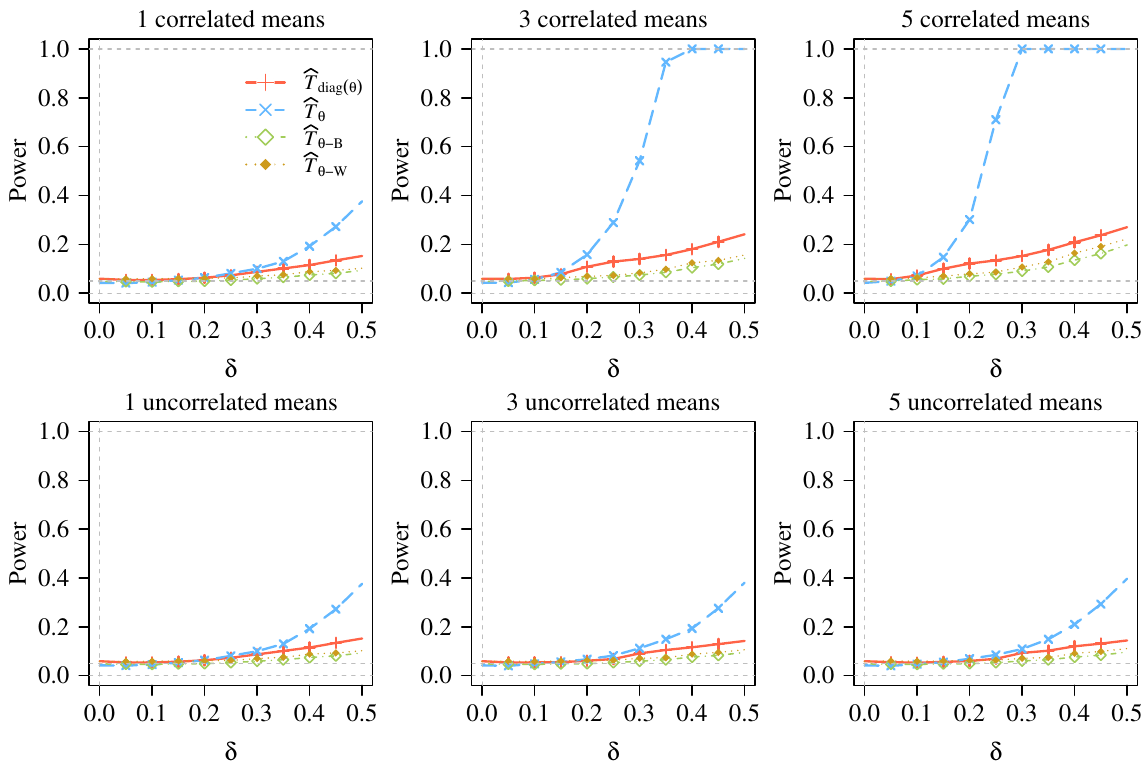}
        \caption{\label{fig:tcor_I02v1_trendCombine_n200_m1_adjust0}
		The power curves of $\hat{T}_{\theta}$, $\hat{T}_{\diag(\theta)}$, $\hat{T}_{\theta\text{-B}}$,
		and $\hat{T}_{\theta\text{-W}}$; see \S\ref{subsubsec:number of means} for details.
        Upper and lower panels correspond to correlated and uncorrelated means, respectively.
		The sample size is $n=200$.
        The nominal size 5\% is indicated by the horizontal gray dashed line.}
    \end{figure}

The power curves are shown in
Figure \ref{fig:tcor_I02v1_trendCombine_n200_m1_adjust0}.
When $|\rho_{j_1,j_2}|>0$ for $j_1\neq j_2$,
the power of $\hat{T}_{\theta}$ is significantly higher than that of $\hat{T}_{\diag(\theta)}$.
Also, power increases with the number of trend-correlated components.
Furthermore, we observe that univariate combined tests
($\hat{T}_{\theta\text{-B}}$ and $\hat{T}_{\theta\text{-W}}$)
are generally less powerful than the multivariate combined test ($\hat{T}_{\diag(\theta)}$).
This occurs because information is lost during the inefficient $p$-value combining step.
In \S\ref{subsubsec:number of means_selection},
we also evaluate the selection procedure described in Example \ref{remark 4.2}.
In \S\ref{sec:trendVarTestMC}, the proposed trend test
is compared with other existing tests.

\section{Real-data application: Weather and crime}\label{sec:weather and crime}
Studies of the correlation between weather and crime have been longstanding in the criminology literature, with several criminological theories such as the situational approach to crime and the routine activities theory suggesting potential influences of weather on crime \citep{Cohn1990}.
In this section, we examine the relationship between the frequencies of different crimes and weather-related variables.

Prior studies generally find that weather is associated with crime, particularly offenses linked to routine activities and outdoor exposure.
Warmer temperatures are often associated with higher levels of violent crime and, in many settings, property crime, although the magnitude of the association varies across offense types and context.
Evidence regarding precipitation is less consistent, with studies reporting positive, negligible, or negative associations depending on the crime category, location, and modeling framework \citep{Defronzo1984,Cohn1990,HorrocksMenclova2011,Jacob_etal2007,Choi_etal2024,CohnRotton2000}.

We consider the daily number of reported incidents of crime in the City of Chicago from 17 May 2021 to 1 April 2026 ($n=1781$).
We consider all types of crimes that have at least 100 incidents during the concerned period,
giving a total of $S_{\text{crime}}=26$ types of crime and $1,189,867$ reported incidents.
We study $S_{\text{weather}}=11$ daily weather-related quantitative variables recorded at the
Oak Street Weather Station.
The datasets were downloaded from Chicago data portal (\url{https://data.cityofchicago.org}).
Across the study period, data are missing for 68 days, corresponding to $3.8\%$ of the sample.
Since the proportion of missing records was small, missing values were linearly interpolated using the nearest observations.
For missing initial or terminal points, backward or forward imputation was applied, respectively.
This is expected to have a small impact on the analysis.
We transform the variables
`wind speed', `maximum wind speed', `rain intensity', and `interval rain' by $x \mapsto \log(1+x)$ due to right-skewness and
high-frequency of zero,
whereas we replace far outliers for other variables by the boundaries of the Tukey's fences.

\begin{figure}[t]
\captionsetup{font=small}
    \centering
    \hspace{-.2cm}
    \begin{subfigure}{.5\textwidth}
		\centering
		\includegraphics[width=1.0\textwidth]{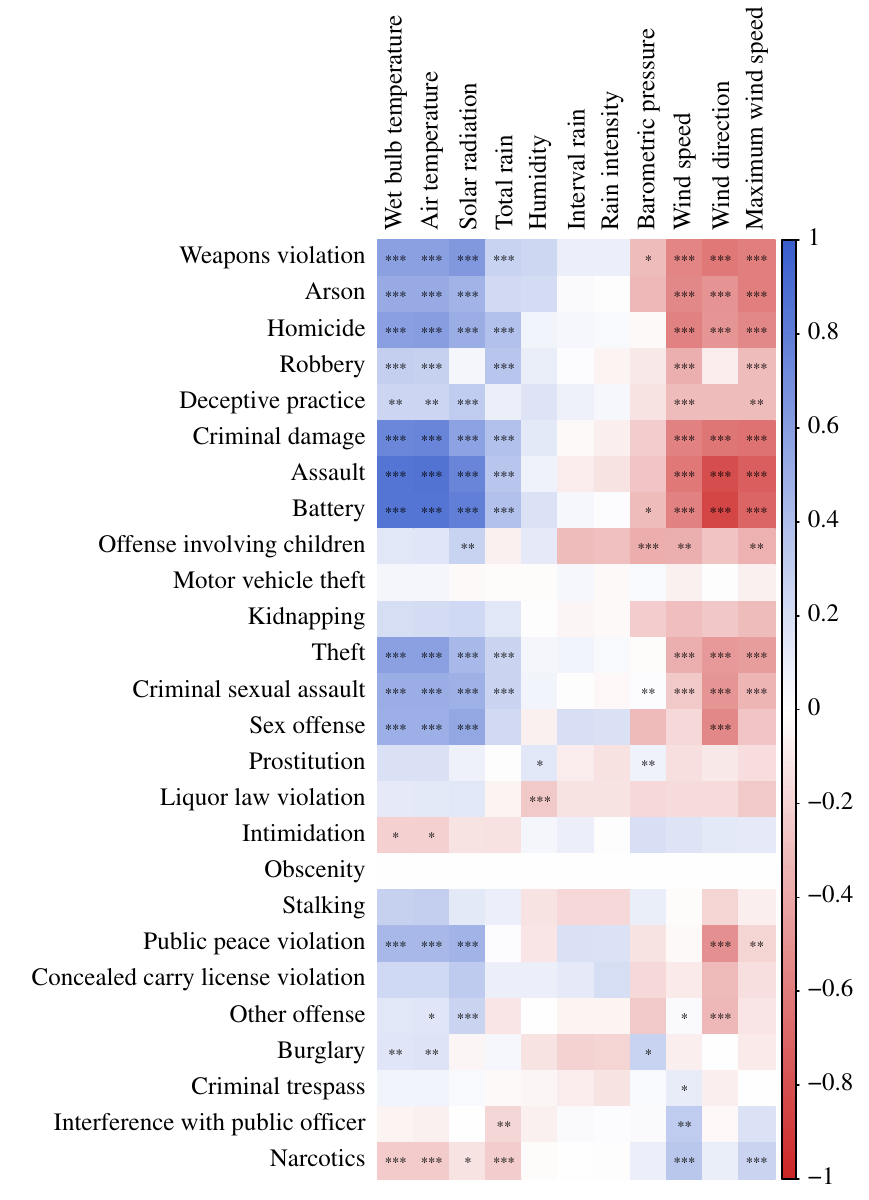}
		\caption{\label{fig:corr_crime_weather_TC_submatrix}Trend correlation matrix.}
	\end{subfigure}
	\hspace{-0.2cm}
	\begin{subfigure}{.5\textwidth}
		\centering
		\includegraphics[width=1.0\textwidth]{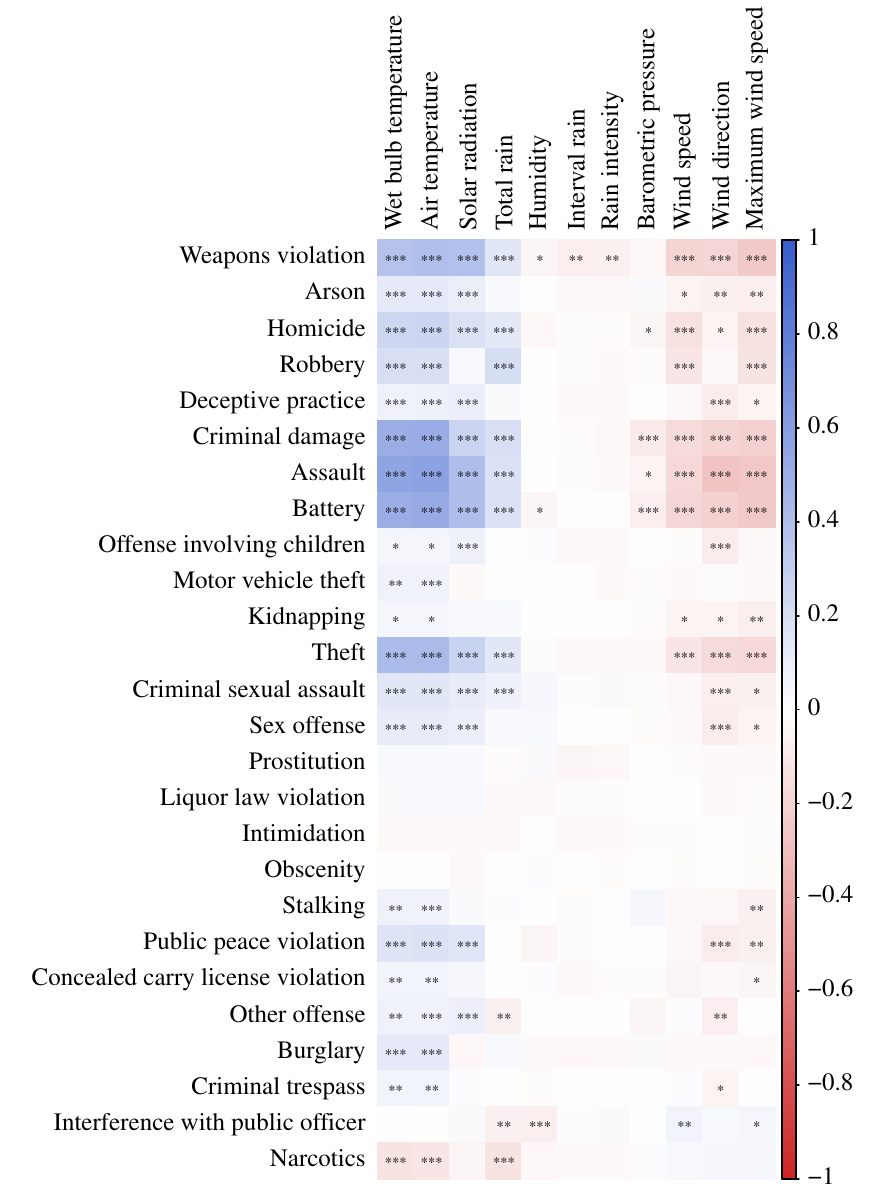}
		\caption{\label{fig:corr_crime_weather_PC_submatrix}Pearson correlation matrix.}
	\end{subfigure}
    \caption{Two types of estimated correlation matrices.
    The first $S_{\text{crime}}=26$ variables
    are crime-related, whereas the last $S_{\text{weather}}=11$ are weather-related variables.
    The ordering of the variables within each of these two categories is determined by the first principal components of
    the trend correlation matrix.
    The proposed trend correlation test and the Pearson's correlation test based on the R-function ``{cor.test}''
    are performed.
    The cells are marked with `\texttt{*}', `\texttt{**}',
    and `\texttt{***}' if the corresponding $p$-value is less than $5\%$, $1\%$, and $0.1\%$,
    respectively.
    The full correlation matrices, including
    the weather-to-weather and crime-to-crime relationships,
    are shown in Figure \ref{fig:corr_crime_weather}.
    }
    \label{fig:corr_crime_weather_submatrix}
\end{figure}

First, we perform one-pair trend correlation tests on
all possible pairs of the $S = S_{\text{crime}} + S_{\text{weather}} = 37$ variables
based on our proposed test with $m=1$ and $q=2$.
The estimated trend correlations and the $p$-values for a two-sided alternative of zero trend correlation are computed
and reported in Figure  \ref{fig:corr_crime_weather_TC_submatrix}.
For reference, we also compute the Pearson's correlation and the corresponding $p$-values;
see Figure  \ref{fig:corr_crime_weather_PC_submatrix}.
Note that regressing one variable on another yields a $p$-value effectively identical to that of the Pearson correlation test.
There are several key observations.
Pearson correlation coefficients are typically much smaller in magnitude than the trend correlation because trend correlation is correctly standardized by the trend variance without being inflated by the noise variance.
Also, the Pearson correlation test is typically more aggressive, since the variances of the estimators are incorrect when there is a non-constant trend component.
We note that the temperature-related variables, i.e., wet-bulb temperature, air temperature, and solar radiation,
are usually significantly trend-correlated with crimes against persons (e.g., homicide, assault, robbery, and battery),
property crimes (e.g., theft, burglary, and criminal damage), and crimes against public safety (e.g., weapons violation and public peace violation), but negatively correlated with narcotics.
In particular, we study air temperature, weapons violation, and narcotics;
see the time series plot in Figure \ref{fig:tcor_J01_crime_tsplot}.
The trend correlation between air temperature and weapons violation is $0.590$ with a 95\% confidence interval $(0.520, 0.660)$,
but the Pearson's correlation is just $0.395$.
On the other hand,
the trend correlation between air temperature and narcotics is $-0.279$ with a 95\% confidence interval $(-0.400, -0.159)$,
but the Pearson's correlation is just $-0.118$.
This is consistent with prior work showing that crime exhibits substantial temporal structure and that temperature is informative for explaining and predicting variation in some crime series, particularly violent or aggressive offenses \citep{Jacob_etal2007,Choi_etal2024,Towers_etal2018}.
The wind- and pressure-related variables, i.e., barometric pressure, wind speed, wind direction, and maximum wind speed,
typically show negative trend correlation with crimes against persons (e.g., homicide, battery, and criminal sexual assault) and property crimes (e.g., arson, criminal damage, and theft).

\begin{table*}[t]
\captionsetup{font=small}
\centering
\setlength{\tabcolsep}{2pt}
\small
\begin{tabular}{cc rcr rcr}
\toprule
	&& \multicolumn{3}{c}{Trend correlation} & \multicolumn{3}{c}{Linear regression}\\
	 \cmidrule(r){3-5}\cmidrule(r){6-8}
	Crime 1& Crime 2 & Difference  & 95\% CI & $p$-value
					& Difference  & 95\% CI & $p$-value \\
	 \cmidrule(r){1-8}
	Arson & Weapons violation	&	$ 0.03$	&	$(-0.16, 0.22)$	&	$72.4\%$	&	$-0.20$	&	$(-0.39, 0.00)$	&	$31.5\%$	\\
	Homicide & Weapons violation	&	$-0.08$	&	$(-0.20, 0.03)$	&	$16.4\%$	&	$-0.60$	&	$(-0.76,-0.44)$	&	$ 0.0\%$	\\
	Homicide & Arson	&	$-0.12$	&	$(-0.32, 0.09)$	&	$27.0\%$	&	$-0.71$	&	$(-0.96,-0.45)$	&	$ 0.9\%$	\\
	Robbery & Weapons violation	&	$ 0.30$	&	$( 0.20, 0.39)$	&	$ 0.0\%$	&	$ 0.36$	&	$( 0.31, 0.40)$	&	$ 0.0\%$	\\
	Robbery & Arson	&	$ 0.26$	&	$( 0.07, 0.46)$	&	$ 0.8\%$	&	$ 0.75$	&	$( 0.54, 0.96)$	&	$ 0.0\%$	\\
	Robbery & Homicide	&	$ 0.38$	&	$( 0.25, 0.50)$	&	$ 0.0\%$	&	$ 1.35$	&	$( 1.20, 1.51)$	&	$ 0.0\%$	\\
     \bottomrule
\end{tabular}
\caption{Results of the two-pair trend correlation test and linear regression between two crime variables (including arson, weapons violations, homicide, and robbery) and air temperature. The point estimates of the differences in trend correlation and linear regression coefficients, their 95\% CIs, and the corresponding $p$-values are also reported. Differences are calculated as the measure related to crime 2 minus the measure related to crime 1.
}
\label{table:2pairTest_crime}
\end{table*}

Second, we examine whether temperature sensitivity differs across selected crime types, focusing on robbery, homicide, arson, and weapons violation.
The difference in trend correlations between two of these crime variables and air temperature is computed,
together with the 95\% confidence interval (CI).
The $p$-value of the two-pair trend correlation test is also reported.
The results are shown in Table \ref{table:2pairTest_crime}.
For reference, we also fit separate linear regression models
with air temperature as the independent variable and each crime variable as the dependent variable.
The difference between the coefficients is computed.
We also report the 95\% CI of the difference and the $p$-value for testing whether the difference is zero, after adjusting for time-series error using the variance estimator in \cite{Chan2020}.
We find that linear regression tests are more aggressive and reject the null hypothesis in most cases.
In contrast, the two-pair trend correlation test rejects
equal trend correlations only for the cases that involve robbery.
These apparent contrasts are likely due to
the inconsistency of linear regression discussed in Example \ref{subsec:TC vs regression}.
Note that both tests agree in rejecting the null hypothesis of equal correlations only for the
comparison robbery and weapons violations, with weapons violations exhibiting a stronger positive association with temperature.

Overall, the results indicate heterogeneous weather sensitivity across offense types.
The temperature-related variable trends co-move positively with several violent, property, and public-order offenses, but
co-move negatively associated with narcotics.
This pattern is consistent with routine activity and situational opportunity perspectives, under which weather may alter guardianship, street visibility, and enforcement opportunities rather than directly changing underlying offending behavior.
We remark, however, that these associations include shared annual seasonality.
It may be interesting to analyze whether these associations remain even after removing calendar effects.

\section*{Funding}
Chan was partially supported by grants GRF-14306421, 14307922, and 14307925 provided by the
Research Grants Council of HKSAR.

\section*{Statement on AI use}
The authors used ChatGPT to assist with proofreading, language editing, and checking mathematical proofs.
The authors thoroughly reviewed and verified all material retained in the article and take
full responsibility for the final work.

\appendix

\numberwithin{table}{section}
\numberwithin{figure}{section}
\numberwithin{equation}{section}
\numberwithin{assumption}{section}

\renewcommand{\thetheorem}{\thesection.\arabic{theorem}}
\renewcommand{\theproposition}{\thesection.\arabic{proposition}}
\renewcommand{\thelemma}{\thesection.\arabic{lemma}}
\renewcommand{\thecorollary}{\thesection.\arabic{corollary}}
\renewcommand{\theequation}{\thesection.\arabic{equation}}
\renewcommand{\thefigure}{\thesection.\arabic{figure}}
\renewcommand{\thetable}{\thesection.\arabic{table}}
\renewcommand{\thesection}{\Alph{section}}

\section{Additional results and discussions}
\subsection{Details of $K_{S^2}$ and $M_{S^2}$}\label{sec:communicationMatrix}
The commutation matrix $K_{S^2}$ in (\ref{eqt:KM})
transforms $\vectorized(A)$ into $\vectorized(A^{\T})$, where $A\in\mathbb{R}^{S\times S}$,
in the sense that $K_{S^2}\vectorized(A)=\vectorized(A^{\T})$.
When $S=2$, the $4\times 4$ matrix $K_4$ has zeros everywhere except for the
$(1,1)$, $(2,3)$, $(3,2)$, and $(4,4)$ entries, which are $1$.
Similarly, the diagonal selection matrix $M_{S^2}$ ensures that $M_{S^2} \vectorized(A) = \vectorized(A\odot I_S)$.
When $S=2$, all entries of the $4\times 4$ matrix $M_4$ are $0$,
except that the $(1,1)$ and $(4,4)$ entries are $1$.

\subsection{Sensitivity analysis}\label{sec:sensitivity}
The proposed trend correlation estimators have some parameters
designed for flexibly handling various situations.
Table \ref{tab:summaryParameters} summarizes
the definitions, theoretical properties, and applications of the proposals,
alongside the default parameters used for implementation.
The rationale for these default values is discussed in the main text,
with a brief summary provided below:
\begin{itemize}[noitemsep,topsep=0pt]
	\item Threshold level:
            The factor $(\widehat{\gamma}_0^{(m)})_{j,j}^{1/2}$
            in the threshold level $\epsilon_j = \{(\widehat{\gamma}_0^{(m)})_{j,j}/n\}^{1/2}$
            ensures
            the estimators are scale-invariant.
            The $n^{1/2}$ rate of $\epsilon_j$ does not affect the asymptotic properties
            according to Corollary \ref{cor:improvedVersion}.
	\item Differencing order:
            The lowest order, $m=1$, is used to achieve trend-robustness,
            as it provides optimal de-meaning effects in finite sample.
	\item Differencing sequence:
            For $m=1$, the sequence $(d_0,d_1) = (2^{-1/2},-2^{-1/2})$
            is unique up to a reversal of order.
            For higher orders of $m$,
            we utilize the optimal values established in Theorem \ref{thm:5.3}.
	\item Kernel order:
            This is determined based on the serial dependence assumption $u_q<\infty$
            (see Theorems \ref{thm:4.3} and \ref{thm:5.3}).
            If $\gamma_k \lesssim k^a$ for $a<-q-1$ for all sufficiently large $k$,
            then $u_q<\infty$. Requring $q=2$ is a standard assumption (see, e.g., \cite{TBB01}).
            We employ the polynomial kernel $K(t) = (1-|t|^q)^+$ in all numerical studies.
	\item Bandwidth:
            We set $\ell \asymp  n^{1/(1+2q)}$
            to achieve the optimal rate for estimating the long-run variance $v$.
            This order is further supported by
            Theorems \ref{thm:4.3} and \ref{thm:5.3}.
            Thus, the same bandwidth is used in all estimators for implementation consistency.
\end{itemize}

\begin{table}[t!]
\footnotesize
\begin{center}
\begin{tabular}{l|ll}
\toprule
& $\bar{\theta}^{(m)}$ and $\bar{\rho}^{(m)}$ 	& $\widehat{\theta}^{(m)}$ and $\widehat{\rho}^{(m)}$ \\
\midrule
Definition & (\ref{eqt:proposed estimator_0}) & (\ref{eqt:proposed estimator}) \\
Applicable cases & General case ($\theta\neq 0$ and $\theta=0$) & Degenerate case ($\theta = 0$) \\
Asymptotic properties & Corollary \ref{cor:improvedVersion} & Theorem \ref{thm:5.3} \\
Inference applications & Trend correlation test (\S\ref{sec:inference_tCorr}) \& estimation & Trend test (\S\ref{sec:trendVar}) \\
\cmidrule(r){1-3}
Threshold level $\epsilon_1, \ldots, \epsilon_S$ & $\epsilon_j = \{(\widehat{\gamma}_0^{(m)})_{j,j}/n\}^{1/2}$ for all $j$ & --\\
Differencing order $m$ & $m=1$ & $m=1$ \\
Difference sequence $d_0, \ldots, d_m$ & $(d_0,d_1)=(2^{-1/2},-2^{-1/2})$ & $(d_0,d_1)=(2^{-1/2},-2^{-1/2})$ \\
Kernel order $q$ & $q=2$ & $q=2$\\
Bandwidth $\ell$ & $\ell = \lfloor 2 n^{1/5}\rfloor$ & $\ell = \lfloor 2 n^{1/5}\rfloor$ \\
\bottomrule
\end{tabular}
\caption{\label{tab:summaryParameters}The upper part of the table summarizes
the definitions, theoretical properties, and usages of the proposed estimators,
whereas the lower part lists all the parameters needed for implementation
alongside their default values.}
\end{center}
\end{table}

\begin{figure}
        \centering
        \includegraphics[width=\textwidth]{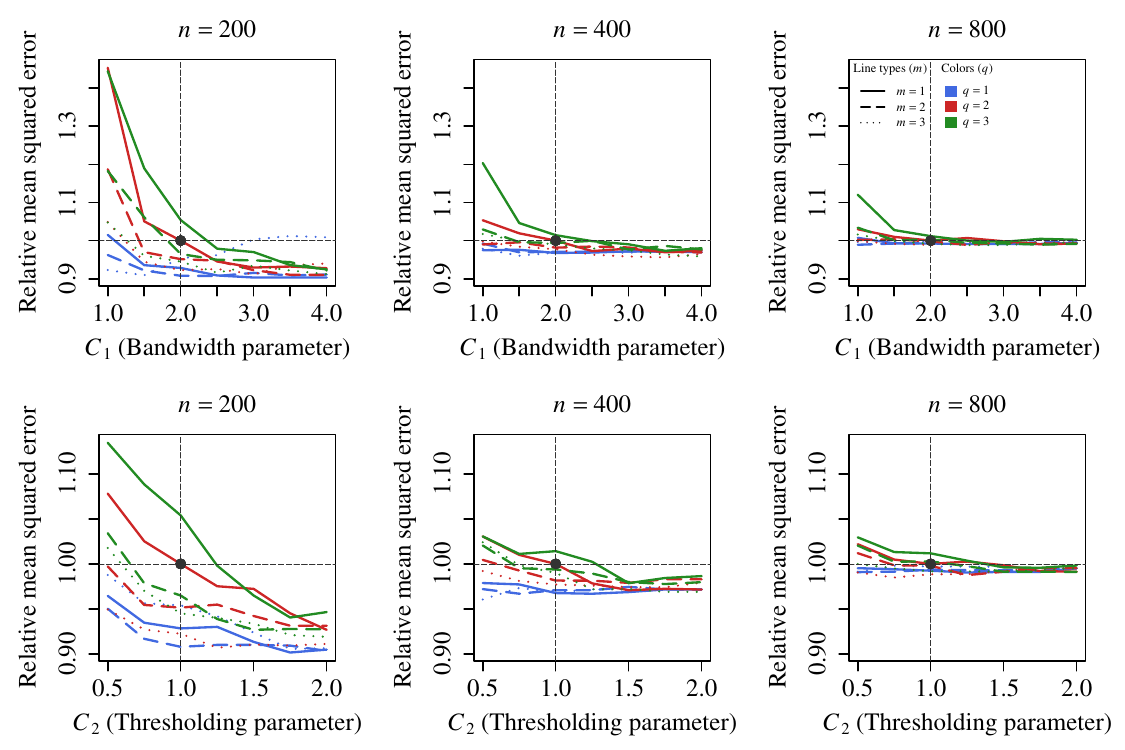}
        \caption{\label{fig:tcor_L16_sensitivity_epsilon_idelta_1}
        When $\delta = 0$,
        the relative mean-squared errors $\MSE(\widehat{\theta}^{(m)}_{1,2})/\MSE(\widehat{\theta}^{\text{default}}_{1,2})$
        are shown above,
        where $\widehat{\theta}^{\text{default}}_{1,2}$ denote the proposed estimator under default parameters.
        The first and second rows of graphs are plotted against $C_1$ and $C_2$, respectively.
        In each plot,
        different line types denote different values of $m\in\{1,2,3\}$, whereas
        different colors denote different values of $q\in\{1,2,3\}$.
        The default estimator is indicated by a dark gray dot.
        }
    \end{figure}

\begin{figure}
        \centering
        \includegraphics[width=\textwidth]{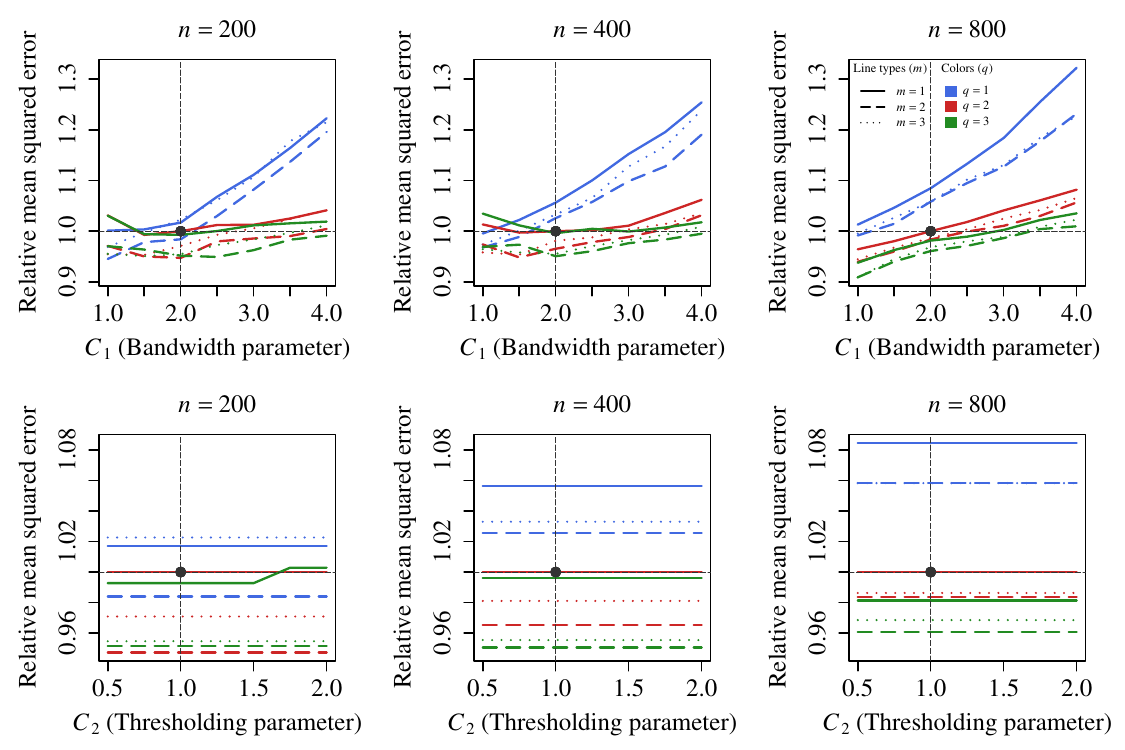}
        \caption{\label{fig:tcor_L16_sensitivity_epsilon_idelta_3}
        The caption in Figure \ref{fig:tcor_L16_sensitivity_epsilon_idelta_1} also applies here,
        except that $\delta = 1$ is used in the above plots. }
    \end{figure}

We perform a sensitivity analysis for the proposed estimator $\widehat{\theta}^{(m)}_{1,2}$ across the following parameters:
\begin{align}\label{eqt:sensitivity}
	m\in\{1,2,3\}, \quad
	q\in\{1,2,3\}, \quad
	\ell = \lfloor C_1 n^{1/(1+2q)}\rfloor , \quad
	\epsilon_j = \{C_2 ( \widehat{\gamma}_0^{(m)})_{j,j}/n\}^{1/2},
\end{align}
where $C_1 \in\{1, 1.5, 2, \ldots, 4\}$ and $C_2 \in\{0.5, 0.75, 1, \ldots, 2\}$.
The noise is generated from a vector $\AR(1)$ model:
$Z_i = \Phi Z_{i-1} + \varepsilon_i$,
where $\varepsilon_i\simIID\Normal_2(0,I_2)$ and
$\vectorized(\Phi) = (0.2, 0.1, 0.1, 0.2)^{\T}$.
The mean function is
$\mu(t) = 4\left[ \delta\sqrt{t},\,  (t-0.5)^2+\delta\sqrt{t} \right]^{\T}$,
where $\delta \in\{0 , 1\}$.
The sample size is $n\in\{200,400,800\}$.
We compute the relative mean-squared errors of $\widehat{\theta}^{(m)}_{1,2}$
with respect to the estimator under the default parameters stated in Table \ref{tab:summaryParameters}.
The results for $\delta=0$ and $\delta = 1$ are shown in
Figures \ref{fig:tcor_L16_sensitivity_epsilon_idelta_1} and
\ref{fig:tcor_L16_sensitivity_epsilon_idelta_3}, respectively.

When $\delta=0$,
the first component of the time series has a constant mean,
representing a degenerate case (i.e., $\theta_{1,1}=0$).
According to Corollary \ref{cor:improvedVersion},
the asymptotic normality result does not depend on the parameters
$m$, $q$, $C_1$, and $C_2$ outlined in (\ref{eqt:sensitivity}).
This is confirmed by the simulation results in
Figure \ref{fig:tcor_L16_sensitivity_epsilon_idelta_1},
which demonstrate that all estimators, regardless
of the chosen parameters,
converge to the same mean-squared error as the sample size $n\rightarrow \infty$.

When $\delta=1$, all components are non-constant,
representing a non-degenerate case (i.e., $\theta\neq 0$).
From Theorem \ref{thm:5.3},
the asymptotic variance of $\widehat{\theta}_{1,2}^{(m)}$ decreases with $m$ and $\ell$.
Because we set $\ell = \lfloor C_1 n^{1/(1+2q)}\rfloor$,
the asymptotic variance also decreases as $C_1$ and $q$ increase.
In contrast, $C_2$ has no asymptotic effect on $\widehat{\theta}_{1,2}^{(m)}$.
These theoretical findings are corroborated by the results in
Figure \ref{fig:tcor_L16_sensitivity_epsilon_idelta_3};
however, the efficiency gains from using a larger $m$ or $q$
only become apparent when the sample size $n$ is sufficiently large.

Overall, the recommended default parameters perform robustly,
striking an effective balance between theoretical efficiency and finite-sample performance.

\subsection{Supplements to Section \ref{sec:inference_tCorr}}\label{sec:additionInference_tCor}
Example \ref{eg:allPairTest2} provides an alternative all-pair trend correlation test in Example \ref{eg:allPairTest1}.
Example \ref{eg:SCI} details the construction of confidence regions for the trend correlation.

\begin{example}[All-pair trend test -- weighted Euclidean distance]\label{eg:allPairTest2}
Assume $\theta_{jj}>0$ for all $j$.
Alternative tests to those in Example \ref{eg:allPairTest1} are
\begin{gather*}
	\widehat{T}_{\all,\rho}^{(m)}
		= n \mathop{\sum\sum}_{ \substack{j_1, j_2\in \{1,\ldots, S\} \\ j_1<j_2} } \frac{(\bar{\rho}_{j_1,j_2}^{(m)})^2}{\bar{\Omega}_{\rho,(j_1,j_2),(j_1,j_2)}},
\end{gather*}
whereas $\widehat{T}_{\all,\theta}^{(m)}$ is defined similarly.
For any covariance matrix $M\in\mathbb{R}^{k\times k}$, let $\lambda_1, \ldots, \lambda_k$
be the eigenvalues of $M^{1/2}(I_{k}\odot M)^{-1}M^{1/2}$.
Denote $\chi^2(M)$ as the distribution of $\sum_{j=1}^{k} \lambda_j U_j$, where $U_1, \ldots, U_k\simIID \chi^2_1$.
Also, denote the distribution function of $\chi^2(M)$ by $F(\cdot, M)$,
which can be computed by standard statistical software.
Then under $H_0^{\all}$, we have
$\widehat{T}_{\all,\bullet}^{(m)} \inD \chi^2(\Omega_{\bullet})$ for $\bullet\in\{\rho, \theta\}$.
So, the asymptotic $p$-value is
$\widehat{p}_{\all,\bullet}^{(m)} = 1-F(\widehat{T}_{\all}^{(m)},\bar{\Omega}_{\bullet})$.
\end{example}

\begin{example}\label{eg:SCI}
Assume $\theta_{jj}>0$ for all $j$.
An asymptotic $100(1 - \alpha)\%$ confidence region for $\upper_0\rho$ is
\begin{equation}\label{eqt:confidence region}
    \widehat{R} = \left\{\rho_0 \in \mathbb{R}^{S(S-1)/2}:
    		n \upper_0\left(\bar{\rho}^{(m)}-\rho_0\right)^{\T} \bar{\Sigma}^{-1} \upper_0\left(\bar{\rho}^{(m)}-\rho_0\right)
		 \leq \chi^2_{S(S-1)/2,1 - \alpha}\right\},
\end{equation}
where $\chi^2_{k,1-\alpha}$ is the $1-\alpha$ quantile of $\chi^2_k$.
To facilitate presentation, it is sometimes preferred to use
the simultaneous confidence intervals via Bonferroni correction:
\begin{equation*}
	\widehat{R}^{\square} = \bigtimes_{j_1=1}^{S-1}\bigtimes_{j_2=j_1+1}^S \widehat{I}_{j_1,j_2},
	\quad \text{where} \quad
	\widehat{I}_{j_1,j_2} = \bar{\rho}_{j_1,j_2}^{(m)} \pm \Phi^{-1}\left(1-\frac{\alpha}{S(S-1)/2}\right)\sqrt{\bar{\Sigma}_{(j_1,j_2),(j_1,j_2)}/n}.
\end{equation*}
Note that $\pr(\upper_0\rho \in \widehat{R}^{\square}) \rightarrow 1-\alpha'$, where $\alpha'\leq\alpha$.
We will illustrate and compare different types of confidence intervals through the application in Section \ref{sec:weather and crime}.
\end{example}

\subsection{Supplements to Section \ref{sec:TrencCorrTestMC}}\label{sec:TrencCorrTestMC_supp}
\begin{figure}[t]
        \centering
        \includegraphics[width=\textwidth]{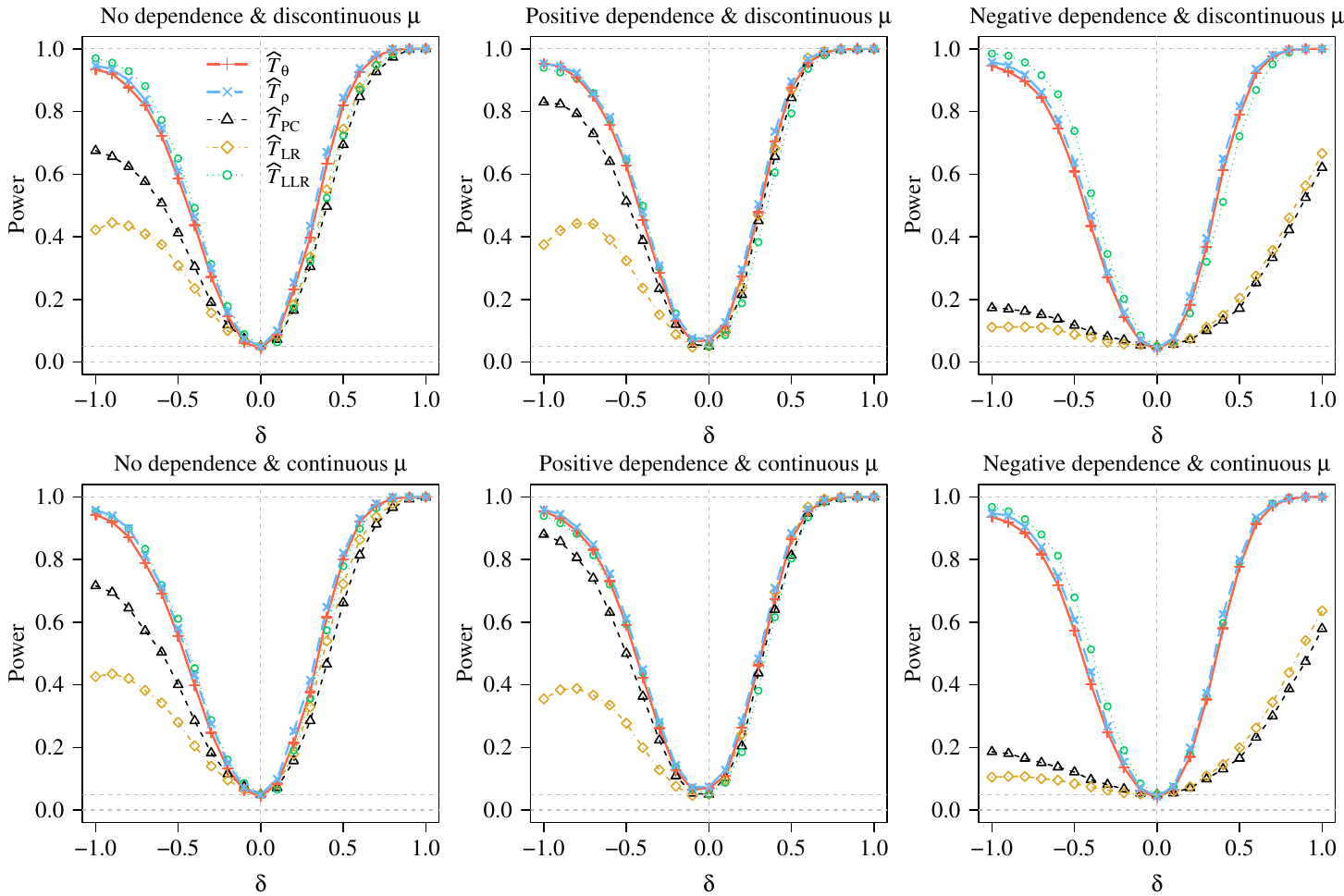}
        \caption{\label{fig:tcor_G17_tcorTest_n200_m1_adjust1_one.pdf}
        Power curves of one-pair trend correlation tests
		under Cases $\{\text{D}, \text{C}\}\times \{0, +,-\}$
		when $n=200$; see Table \ref{table:MC_mean_noise_cases} for the definitions.
        The nominal size 5\% is indicated by the horizontal gray dashed line.}
    \end{figure}

\begin{figure}[t]
        \centering
        \includegraphics[width=\textwidth]{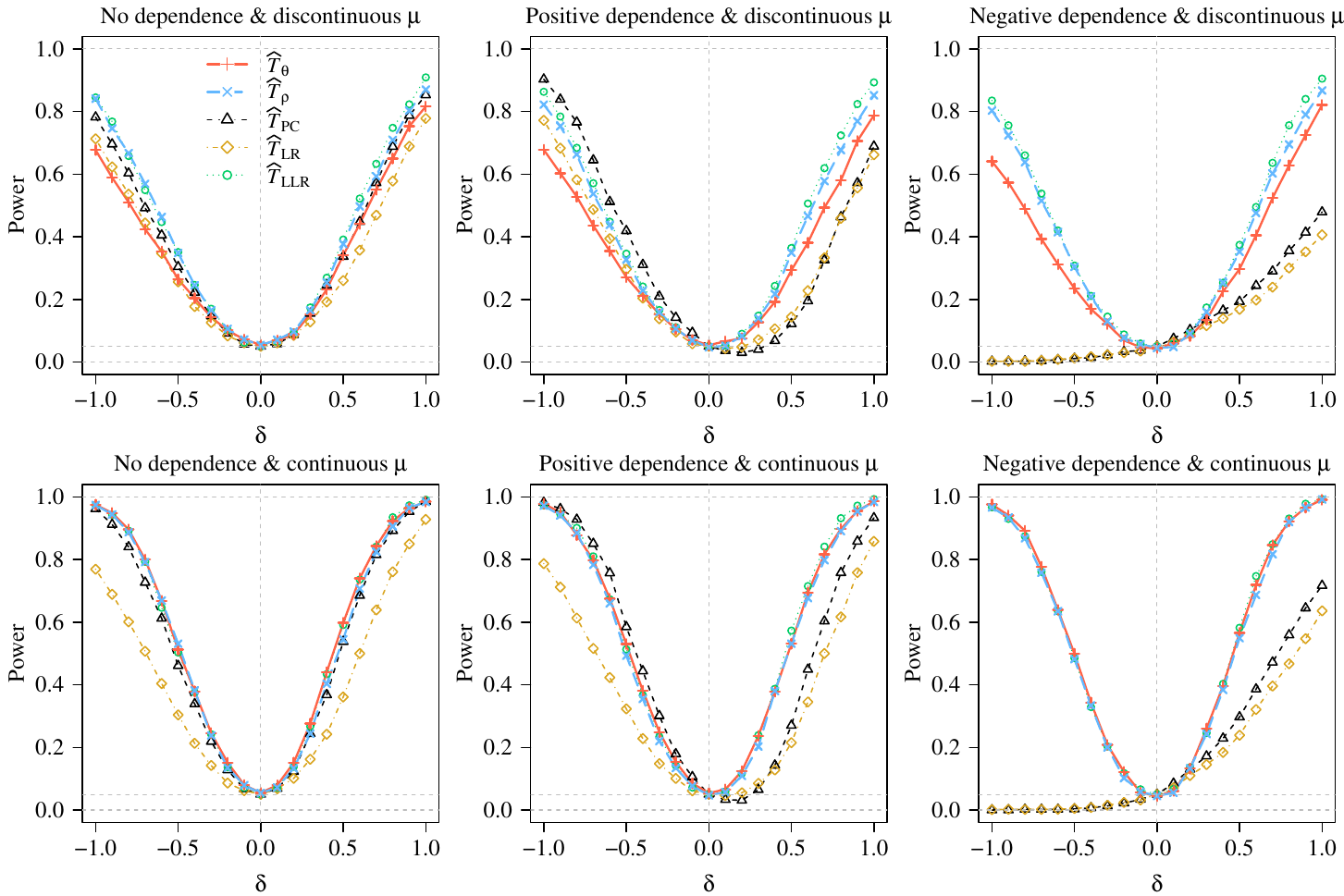}
        \caption{\label{fig:tcor_G17_tcorTest_n200_m1_adjust1_two.pdf}
        Power curves of two-pair trend correlation tests
		under Cases $\{\text{D}, \text{C}\}\times \{0, +,-\}$
		when $n=200$; see Table \ref{table:MC_mean_noise_cases} for the definitions.
        The nominal size 5\% is indicated by the horizontal gray dashed line.}
    \end{figure}

\begin{figure}[t]
        \centering
        \includegraphics[width=\textwidth]{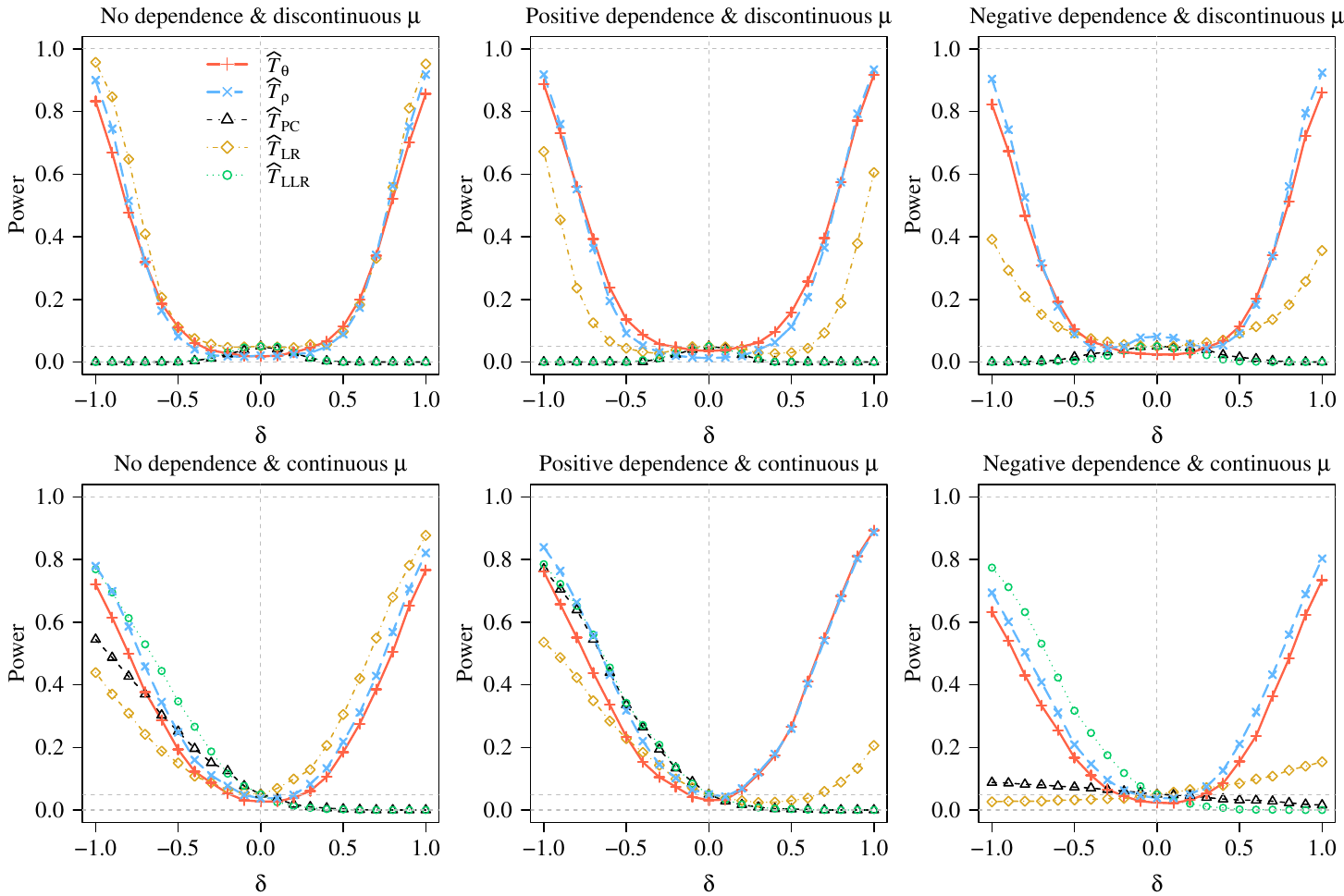}
        \caption{\label{fig:tcor_G17_tcorTest_n200_m1_adjust1_all.pdf}
        Power curves of all-pair trend correlation tests
		under Cases $\{\text{D}, \text{C}\}\times \{0, +,-\}$
		when $n=200$; see Table \ref{table:MC_mean_noise_cases} for the definitions.
        The nominal size 5\% is indicated by the horizontal gray dashed line.}
    \end{figure}

We present additional simulation results for the experiments detailed in Section \ref{sec:TrencCorrTestMC}.
In addition to Cases $+$ and $-$ stated in Table \ref{table:MC_mean_noise_cases},
we consider an additional Case $0$ for the dependence structure: $\Phi=0$,
i.e., the independent case.
The full versions of the power curves displayed in Figure \ref{fig:tcor_G17_tcorTest_n200_m1_adjust1_simplePlot}
are shown in Figures
\ref{fig:tcor_G17_tcorTest_n200_m1_adjust1_one.pdf},
\ref{fig:tcor_G17_tcorTest_n200_m1_adjust1_two.pdf}, and
\ref{fig:tcor_G17_tcorTest_n200_m1_adjust1_all.pdf}
for the one-pair, two-pair, and all-pair trend correlation tests, respectively.
It can be seen that $\widehat{T}_{\pearson}$, $\widehat{T}_{\LR}$, and $\widehat{T}_{\LLR}$
are not necessarily powerful,
even when their critical values are calibrated using the simulation results.
It is also worth noting that this phenomenon holds true even in the independent case.
On the other hand, the proposed tests $\widehat{T}_{\theta}$ and $\widehat{T}_{\rho}$
demonstrate good size control and power performance across all cases.

\subsection{Simulation experiments regarding trend variance test}\label{sec:trendVarTestMC}
We test $H_0^{\trend}: \theta = 0$ against $H_1^{\trend}: \theta \neq 0$.
The following tests are compared:
\begin{itemize}[noitemsep,topsep=0pt]
	\item (WZ) The mean stationarity test \citep{wu_zhao_2007} applied to each component series,
				with their proposed non-overlapping batch-means long-run variance estimator.
				The tests are aggregated across components via
				the Bonferroni correction $\hat{T}_{\trend,\text{WZ-B}}$ and
				the harmonic mean $p$-value $\hat{T}_{\trend,\text{WZ-W}}$ \citep{Wilson2019}.
	\item (KPSS) The stationarity test $\hat{T}_{\trend,\text{KPSS-C}}$ proposed by \cite{Kwiatkowski1992}
				 with the trend-robust long-run variance estimator \citep{Chan2020}
				 based on Bartlett kernel, first order differencing, and nonparametric plug-in bandwidth.
				 The critical values for $S> 1$ can be found in \cite{Nyblom1989}.

	\item (WS) We include the self-normalized $U$-statistic test $\hat{T}_{\trend,\text{WS}}$ \citep{WangShao2020}
			with a trimming parameter $m = 10$.
			It works for high-dimensional data.

	\item (KS) The multivariate Kolmogorov--Smirnov change point test $\hat{T}_{\trend,\text{KS-C}}$ \citep{Horvath_etal1999}
					is evaluated using the long-run variance estimator \citep{Chan2020} as in $\hat{T}_{\trend,\text{KPSS-C}}$.
	\item (SZ) The self-normalized KS change point test $\hat{T}_{\trend,\text{SZ}}$ \citep{shao_zhang_2010}.
\end{itemize}
The proposed
trend variance test $\hat{T}_{\trend, \diag(\theta)}^{(m)}$ and
trend covariance test $\hat{T}_{\trend, \theta}^{(m)}$
are computed with $m=1$ and $q=2$; see Examples \ref{eg:trendTest_var}--\ref{eg:trendTest_covar}.

Assume the noises are generated from a linear combination of bilinear models:
$Z_i = A W_i$,
where
$\vectorized(A) = (1,-0.8, -0.8, 1)^{\T}$,
$W_{i,s} = (a+b\varepsilon_{i,s})W_{i-1,s} + \varepsilon_{i,s}$
and $\varepsilon_{i,s}\simIID\Normal(0,1)$.
We consider three sets of $(a,b)\in\{(0.1, 0.2), (0.2, 0.4), (0.3, 0.6)\}$
corresponding to weaker, mild, and stronger dependence.
Four mean functions are considered:
\begin{gather*}
\small
	\mu_{\text{oneCP}}(t) = \delta\begin{bmatrix}
			\mathbb{1}_{(t>0.5)} \\
			-\mathbb{1}_{(t>0.5)}
		\end{bmatrix},
	\qquad
	\mu_{\text{fourCP}}(t) = \delta
	\begin{bmatrix}
			\mathbb{1}_{(t>0.2)}-\mathbb{1}_{(t>0.4)}+\mathbb{1}_{(t>0.6)}-\mathbb{1}_{(t>0.8)} \\
			-\mathbb{1}_{(t>0.2)}-\mathbb{1}_{(t>0.4)}+\mathbb{1}_{(t>0.6)}-\mathbb{1}_{(t>0.8)}
		\end{bmatrix},	\\
	\mu_{\text{cos}}(t) = \delta\begin{bmatrix}
			\cos(4\pi t) \\
			-\cos(4\pi t)
		\end{bmatrix},
	\qquad
	\mu_{\text{ramp}}(t) =\delta
	\begin{bmatrix}
			\max(t-0.5,0) \\
			-\max(t-0.5,0)
		\end{bmatrix},
\end{gather*}
corresponding to one changepoint, four changepoints, cosine wave, and ramp function.

    \begin{figure}
    \captionsetup{font=small}
        \centering
        \includegraphics[width=\textwidth]{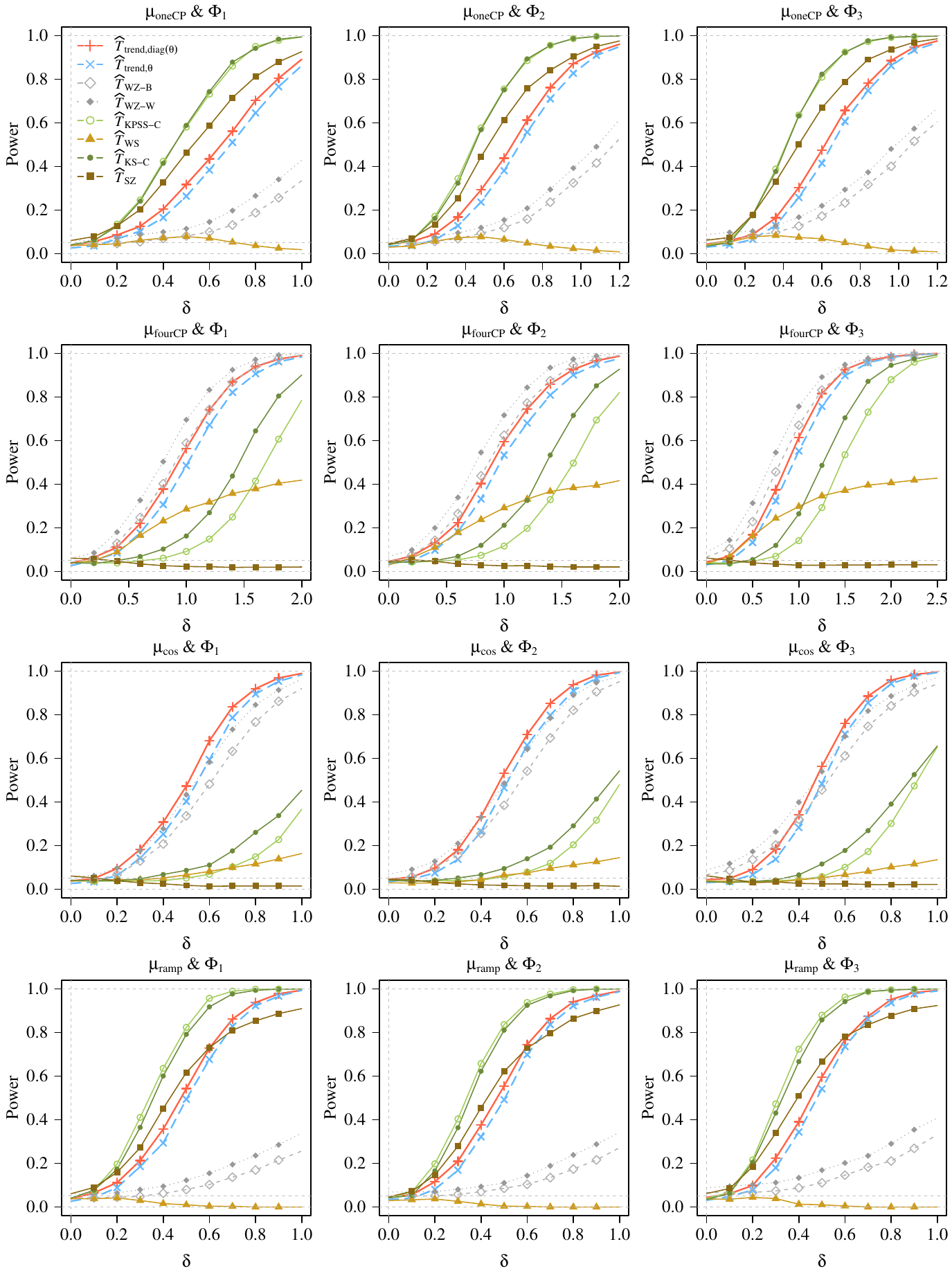}
        \caption{\label{fig:tcor_F13v2_trendTest_n200_m1_adjust0}
        Power curves for testing $H_0^{\trend}$
		under Cases $\{\text{oneCP}, \text{fourCP}, \text{cos}, \text{ramp}\}\times\{1,2,3\}$
		when $n=200$; see \S\ref{sec:trendVarTestMC}.
        Horizontal gray dashed lines indicate the nominal size 5\%.}
    \end{figure}

The power curves for $n=200$ are shown in Figure \ref{fig:tcor_F13v2_trendTest_n200_m1_adjust0}.
We observe that all tests admit mild size inflation under weaker to mild dependence (i.e., Cases 1--2),
while non-self-normalized tests exhibit slightly more severe size inflation under stronger dependence (i.e., Case 3).
The cumulative sum–type tests $\hat{T}_{\text{KPSS-C}}$ and $\hat{T}_{\text{KS-C}}$ perform well under
the one-CP and ramp mean functions; however, their power is severely affected under more complicated mean structures
of the four-CP and cosine mean structures.
The test $\hat{T}_{\text{SZ}}$ is a self-normalized version of $\hat{T}_{\text{KS-C}}$.
Although it controls size very well and shows good power under the one-CP and ramp cases,
it loses all power in the other two cases.
The trend tests $\hat{T}_{\text{WZ-B}}$ and $\hat{T}_{\text{WZ-W}}$ are maximum-type tests.
They can detect sudden mean changes that appear in the four-CP and cosine cases,
but
they lose a substantial amount of power in the one-CP and ramp cases.
The power loss is also partly due to information loss in the $p$-value combination step.
The $U$-statistic–based test $\hat{T}_{\text{WS}}$ is designed for high dimensions.
It manages to show some power in the four-CP and cosine cases;
however, the power can be less encouraging in other cases.
The proposed tests $\hat{T}_{\theta}$ and $\hat{T}_{\diag(\theta)}$
show monotone power in all cases.

\begin{figure}
        \centering
        \includegraphics[width=\textwidth]{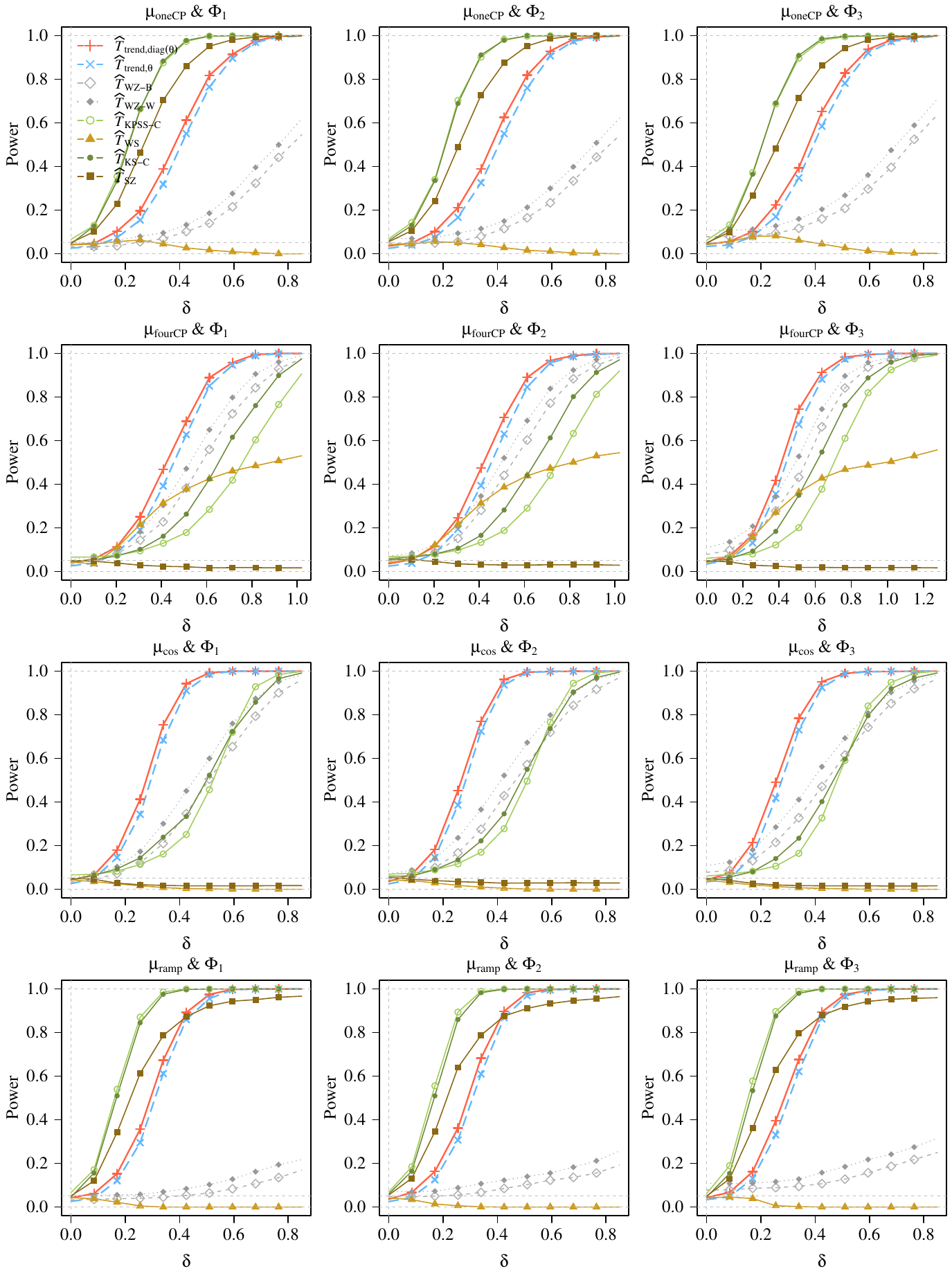}
        \caption{\label{fig:tcor_F13v2_trendTest_n800_m1_adjust0}
        Power curves for testing $H_0^{\trend}$
		under Cases $\{\text{oneCP}, \text{fourCP}, \text{cos}, \text{ramp}\}\times\{1,2,3\}$
		when $n=800$; see Section \ref{sec:trendVarTestMC}.
        Here $m=1$ is used.
        Horizontal gray dashed lines indicate the nominal size 5\%.}
    \end{figure}

\begin{figure}
        \centering
        \includegraphics[width=\textwidth]{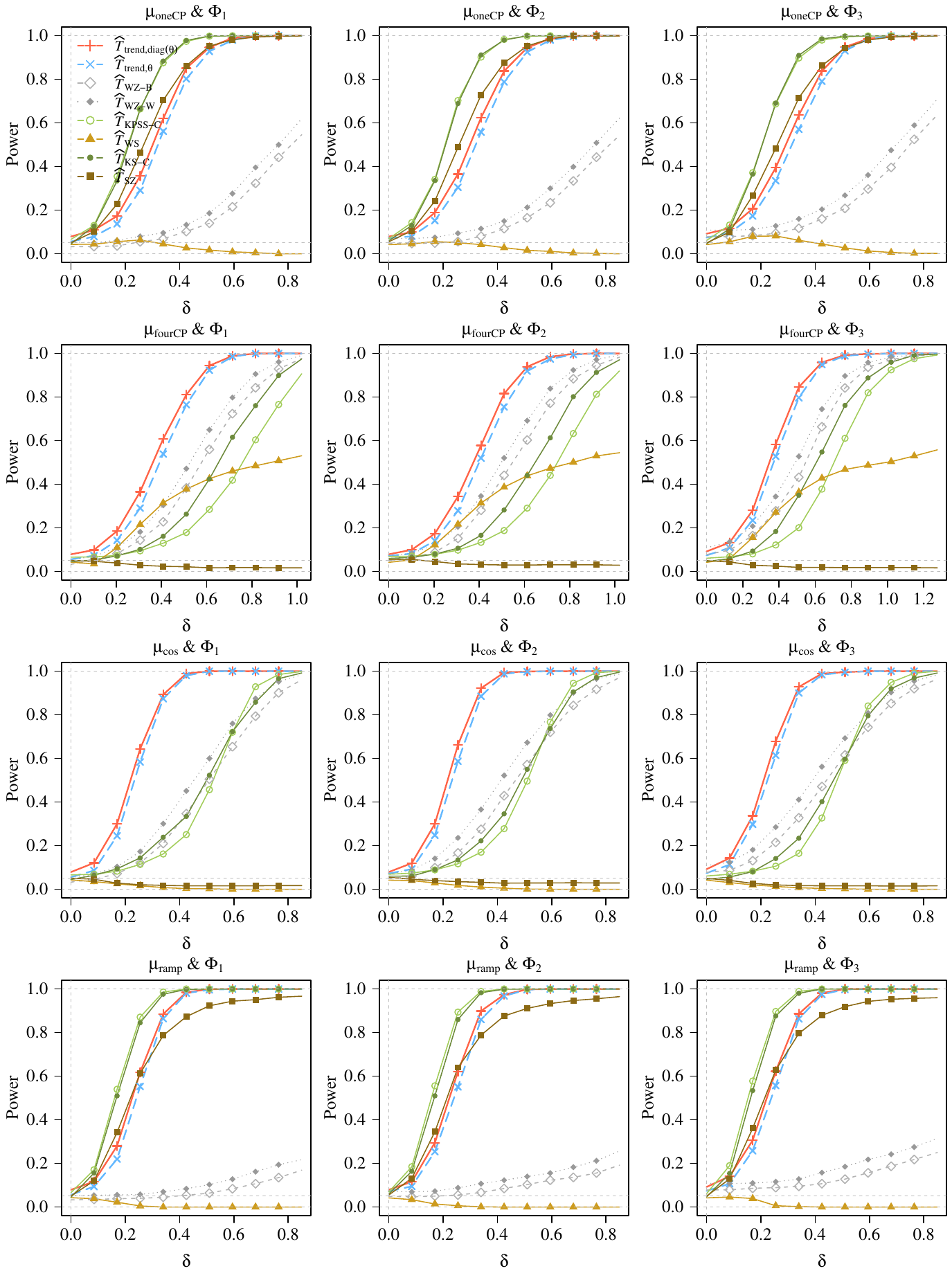}
        \caption{\label{fig:tcor_F13v2_trendTest_n800_m3_adjust0}
        Power curves for testing $H_0^{\trend}$
		under Cases $\{\text{oneCP}, \text{fourCP}, \text{cos}, \text{ramp}\}\times\{1,2,3\}$
		when $n=800$; see Section \ref{sec:trendVarTestMC}.
        Here $m=3$ is used.
        Horizontal gray dashed lines indicate the nominal size 5\%.}
    \end{figure}

For reference,
we also present the results for
$n=800$ in
Figure \ref{fig:tcor_F13v2_trendTest_n800_m1_adjust0}.
Compared to the results for $n=200$,
which is shown in Figure \ref{fig:tcor_F13v2_trendTest_n200_m1_adjust0},
we observe a clear increase in power,
confirming the consistency of the test.
Furthermore, Theorem \ref{thm:5.3} states that
the power of the test increases with $m$.
This property is verified by the power curves for $m=3$,
presented in Figure \ref{fig:tcor_F13v2_trendTest_n800_m3_adjust0}.
We emphasize, however, that this enhanced power is
only realized when the sample size $n$ is sufficiently large (e.g., $n=800$)
for asymptotic properties to apply.
Therefore, in practice, we recommend $m=1$ as a safer choice.

Hence, the proposed tests $\hat{T}_{\theta}$ and $\hat{T}_{\diag(\theta)}$
can serve as good general-purpose mean tests.

\subsection{Supplements to Section \ref{subsubsec:number of means}}\label{subsubsec:number of means_selection}
We revisit the simulation experiments in Section \ref{subsubsec:number of means}.
Here, we evaluate the selection procedure described in Example \ref{remark 4.2}.
In this study, we sequentially test the null hypotheses
\[
	H_0^{\{1,2,3,4,5\}}, H_0^{\{1,2,3,4\}}, \ldots, H_0^{\{1\}}.
\]
Let $N$ denote the true number of components exhibiting a non-constant trend,
and let $\widehat{N}$ represent the number of components identified
as having a non-constant trend.
Consequently, the selected set of non-constant components
is defined as $\{1, \ldots, \widehat{N}\}$ if $\widehat{N}>0$,
and as $\emptyset$ if $\widehat{N}=0$.
For these experiments,
we employ the correlated mean structure
$\eta(t)$ and set $a \in \{(0,0,0,0,0)^{\T}, (1,0,0,0,0)^{\T}, \ldots, (1,1,1,1,1)^{\T}\}$
to correspond with the cases $N=0, 1, \ldots, 5$.
We use $n=200$ and set the nominal size at $5\%$.

The results are presented in Table \ref{table:tcor_H01v2_trendSelect}.
Under the global null $H_0^{\{1,2,3,4,5\}}$, i.e., $N = 0$,
the proportion of selecting no component, i.e., $\widehat{N}=0$, is around 95\%, meaning that the size is around $5\%$.
When $N>0$, the algorithm can report a correct $\widehat{N}$ with increasing probability when $\delta$ increases.
It verifies the consistency of the selection algorithm.

\begin{table*}[t]
\centering
\setlength{\tabcolsep}{3pt}
\renewcommand{\arraystretch}{0.8}
\small
\begin{tabular}{c cccccc cccccc}\toprule
& \multicolumn{6}{c}{$\delta=0.5$} &\multicolumn{6}{c}{$\delta=1$} \\[0.5ex]
\cmidrule(r){2-7}\cmidrule(r){8-13}
	 & $N=0$ & $N=1$ & $N=2$ & $N=3$ & $N=4$ & $N=5$ & $N=0$ & $N=1$ & $N=2$ & $N=3$ & $N=4$ & $N=5$  \\[0.5ex]
	\cmidrule(r){1-13}
	$\widehat{N}=0$	&	\fbox{$95.4$}	&	\fbox{$63.7$}	&	$ 0.0$	&	$ 0.0$	&	$ 0.0$	&	$ 0.0$	&	\fbox{$95.4$}	&	$ 0.1$	&	$ 0.0$	&	$ 0.0$	&	$ 0.0$	&	$ 0.0$	\\[0.5ex]
	$\widehat{N}=1$	&	$ 0.8$	&	$31.0$	&	\fbox{$63.5$}	&	$ 0.0$	&	$ 0.0$	&	$ 0.0$	&	$ 0.8$	&	\fbox{$93.8$}	&	$ 0.1$	&	$ 0.0$	&	$ 0.0$	&	$ 0.0$	\\[0.5ex]
	$\widehat{N}=2$	&	$ 1.4$	&	$ 2.4$	&	$32.3$	&	$21.2$	&	$ 0.0$	&	$ 0.0$	&	$ 1.4$	&	$ 2.8$	&	\fbox{$95.3$}	&	$ 0.0$	&	$ 0.0$	&	$ 0.0$	\\[0.5ex]
	$\widehat{N}=3$	&	$ 0.8$	&	$ 1.2$	&	$ 1.6$	&	\fbox{$73.6$}	&	$16.1$	&	$ 0.0$	&	$ 0.8$	&	$ 1.3$	&	$ 1.8$	&	\fbox{$94.7$}	&	$ 0.0$	&	$ 0.0$	\\[0.5ex]
	$\widehat{N}=4$	&	$ 0.3$	&	$ 0.3$	&	$ 0.6$	&	$ 0.8$	&	\fbox{$71.7$}	&	\fbox{$72.1$}	&	$ 0.3$	&	$ 0.5$	&	$ 0.6$	&	$ 0.8$	&	\fbox{$87.3$}	&	$38.5$	\\[0.5ex]
	$\widehat{N}=5$	&	$ 1.3$	&	$ 1.4$	&	$ 2.0$	&	$ 4.4$	&	$12.2$	&	$27.9$	&	$ 1.3$	&	$ 1.5$	&	$ 2.2$	&	$ 4.5$	&	$12.7$	&	\fbox{$61.5$}	\\[0.5ex]
	\bottomrule
\end{tabular}
\caption{\label{table:tcor_H01v2_trendSelect}
Percentage of replications resulting in which component(s) is/are selected
as having non-constant trends based on 1,000 replications.
The overall false discovery rate is set to be $5\%$.
Boxed cells denote the category with the highest proportion.}
\end{table*}

\subsection{Supplements to Section \ref{sec:weather and crime}}\label{sec:RealData_plot}

\begin{figure}[t]
    \centering
    \hspace{-.2cm}
    \begin{subfigure}{.5\textwidth}
		\centering
		\includegraphics[width=1.0\textwidth]{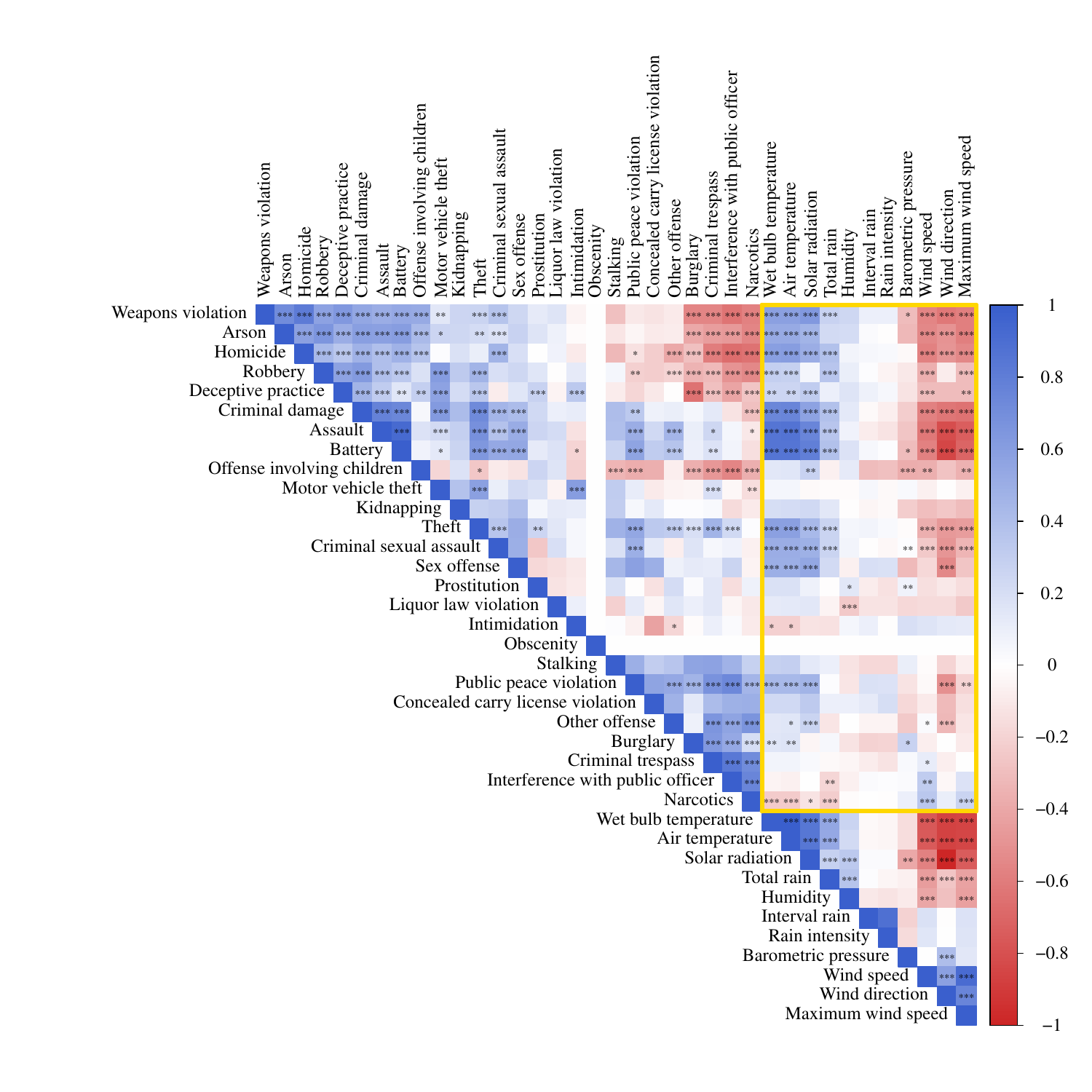}
		\caption{\label{fig:corr_crime_weather_TC}Trend correlation matrix.}
	\end{subfigure}
	\hspace{-0.2cm}
	\begin{subfigure}{.5\textwidth}
		\centering
		\includegraphics[width=1.0\textwidth]{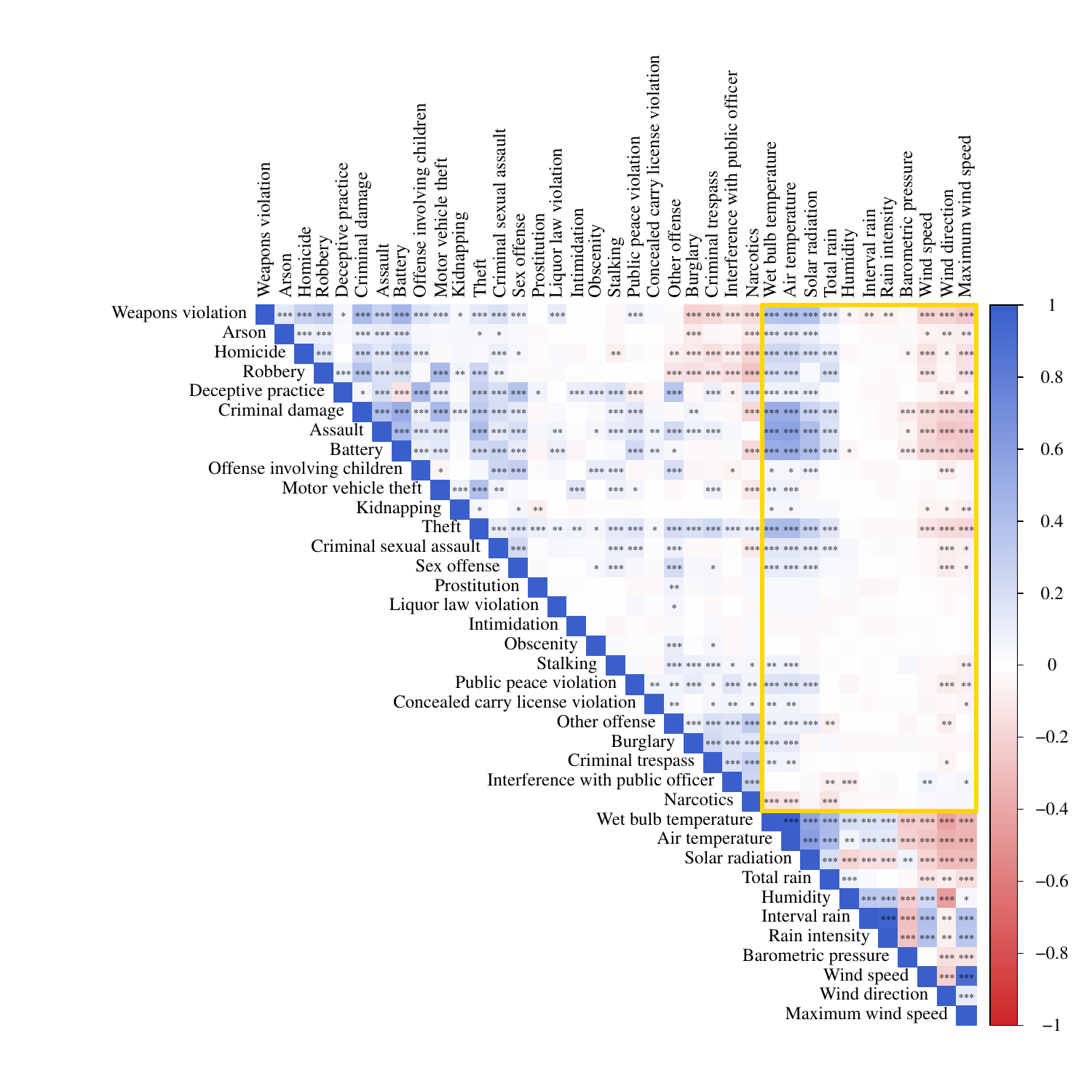}
		\caption{\label{fig:corr_crime_weather_PC}Pearson correlation matrix.}
	\end{subfigure}
    \caption{
    It is the full version of Figure
    \ref{fig:corr_crime_weather_submatrix}.
    Two types of estimated correlation matrices. The first $S_{\text{crime}}=26$ variables
    are crime-related, whereas the last $S_{\text{weather}}=11$ are weather-related variables.
    The ordering of the variables within each of these two categories is determined by the first principal components of
    the trend correlation matrix.
    The proposed trend correlation test and the Pearson's correlation test based on the R-function \texttt{cor.test}
    are performed.
    The cells are marked with `\texttt{*}', `\texttt{**}', and `\texttt{***}' if the corresponding $p$-value is less than $5\%$, $1\%$, and $0.1\%$,
    respectively.
    }
    \label{fig:corr_crime_weather}
\end{figure}

\begin{figure}[!t]
    \centering
    \includegraphics[width=\textwidth]{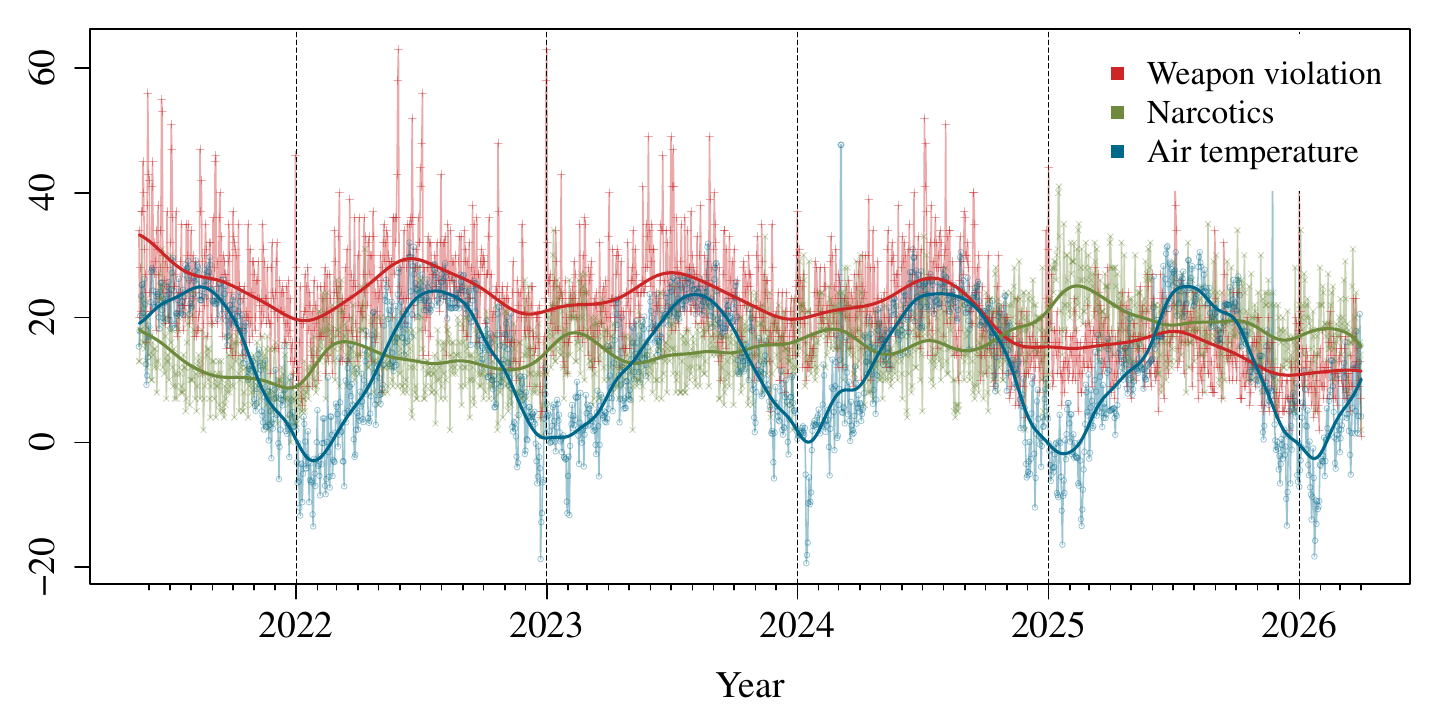}
    \caption{Time series plots of air temperature, weapon violation, and narcotics.
    The local linear regression estimates against time are also plotted for reference.}
    \label{fig:tcor_J01_crime_tsplot}
\end{figure}

The full correlation matrices are shown in
Figure \ref{fig:corr_crime_weather}.
We also show time-series plots of air temperature, weapon violation, and narcotics in Figure \ref{fig:tcor_J01_crime_tsplot}.
The figure shows that weapon violation follows the seasonality of air temperature very closely, whereas the local maxima usually appear around periods with low temperatures and remain fairly constant at other times.

\section{Proofs of Main Results}\label{sec:proofMain}

Throughout the proof, we omit the superscript ${}^{(m)}$ in $\widehat{\theta}^{(m)}$ to lighten the notation when there is no confusion.
Denote
\begin{align}\label{eqt:defM0M1}
	\mathcal{M}_0
		= \frac{\ell}{n}(\mathcal{C} + \mathcal{SJ})^2,
	\quad
	\mathcal{M}_1
		= \frac{\ell}{n}\left(\frac{\mathcal{C}^2\ell}{n} + \mathcal{S}^2\mathcal{J}\right),
	\quad	\text{and} \quad
	\mathcal{M}_{\mu Z} = \frac{\mathcal{C} + \mathcal{SJ}}{n^{3/2}} ,
\end{align}
which are the reminder terms appeared in Lemmas \ref{prop:B2} and \ref{lem:B8}.
As these remainder terms will appear frequently throughout Section \ref{sec:proofMain},
we summarize them above for easy reference.

\subsection{Proof of Theorem \ref{thm:4.3}}\label{sec:A1}
For part (i), we prove the result in five steps as follows:
\begin{itemize}[wide=0pt]
	\item \underline{\textbf{Step 1}: Cram\'{e}r--Wold device.}
		It suffices to derive the result for $S=4$, since it encompasses all possible cases.
		The general result can be obtained by symmetry based on the result for $S=4$.
        Let $c_{j,j'}\in \mathbb{R}$ for $(j,j')\in\mathbb{J}$ be any fixed constants, where
		\[ \mathbb{J} = \{(1,1),(2,2),(3,3),(4,4),(1,2),(2,3),(3,4)\}. \]
		Consider the linear combination of the unique elements in the matrix $\widehat{\theta}$:
		\begin{equation*}
			\widehat{L} = \sum_{(j,j')\in\mathbb{J}} c_{j,j'}\widehat{\theta}_{j,j'}.
		\end{equation*}
		To prove the joint limiting normality of
        the seven elements of $\widehat{\theta}$,
		it suffices to prove that $\widehat{L}$ is asymptotically  normal, after suitable normalization,
		according to the Cram\'{e}r--Wold device.
        The complete result for $\widehat{\theta}$
        follows by analogous arguments without additional technical difficulty.
        We therefore omit the full version for brevity.

	\item \underline{\textbf{Step 2}: Approximating $\widehat{L}$ by a quadratic form $\widetilde{L}$}:
		We approximate $\widehat{\theta}_{j_1,j_2}$ by the following quadratic form:
		{\small
	\begin{align}\label{eqt:tildetheta}
	    \widetilde{\theta}_{j_1,j_2} = \sum_{1 \leq i,i' \leq n} a_{n,i - i'}(X_{i,j_1} - \bar{X}_{n,j_1})(X_{i'j_2} - \bar{X}_{n,j_2}),
	\end{align}
	where
	\begin{align*}
		a_{n,k} = \frac{\{K(k/\ell) - K_{m}(k/\ell)\}\mathbb{1}(|k| \leq \ell + mh)}{n\ell \kappa_\ell}
		\qquad\text{and}\qquad
		K_{m}(t) = \sum^{\left \lfloor{\min\left(m, \frac{1-t}{\lambda}\right)}\right \rfloor}_{s = \left \lceil{\max\left(-m, \frac{-1-t}{\lambda}\right)}\right \rceil} \delta_{|s|}K(t + \lambda s).
	\end{align*}
	}
	In view of Proposition \ref{prop:hat_tilde_diff}, we have
\begin{equation}\label{eqt:boundAhatAtilde}
    \left\| \widehat{\theta}_{j_1,j_2} - \widetilde{\theta}_{j_1,j_2} \right\|
    	=  O\left( \frac{\ell^{1/2}}{n} +\mathcal{M}_0 + \frac{\ell}{n}\mathcal{M}_1 \right) .
\end{equation}
	Denote
	\begin{equation*}
		\widetilde{L} = \sum_{(j,j')\in\mathbb{J}} c_{j,j'}\widetilde{\theta}_{j,j'}.
	\end{equation*}
	which is a sum of seven terms.
	Applying (\ref{eqt:boundAhatAtilde}) seven times for different values of $j_1$ and $j_2$,
	we know that the bound in (\ref{eqt:boundAhatAtilde}) holds also for $\|\widehat{L} - \widetilde{L}\|$,
    i.e., $\|\widehat{L} - \widetilde{L}\|
    =  O\left(  \ell^{1/2}/{n} +\mathcal{M}_0 + {\ell}\mathcal{M}_1/{n} \right) $.
    The assumptions on $\mathcal{C}$, $\mathcal{S}$, and $\mathcal{J}$ ensure that
    $\mathcal{M}_0 =  o(1/\sqrt{n})$ and
    $\mathcal{M}_1 =  o(\sqrt{n}/\ell)$.
    Thus, we know that $\|\widehat{L} - \widetilde{L}\|= o(1/\sqrt{n})$.
    Hence,
    it suffices for us to establish the result for $\widetilde{L}$ instead of $\widehat{L}$.

	\item \underline{\textbf{Step 3}: Decomposing $\widetilde{L}$ into $\widetilde{L}^{(\mu,\mu)},\widetilde{L}^{(Z,Z)},\widetilde{L}^{(\mu,Z)},\widetilde{L}^{(\mu,Z)}$}:
Note that $X_{i,j} = \mu_{i,j} + Z_{i,j}$. So, for fixed $j_1,j_2$, we have
$\widetilde{\theta}_{j_1,j_2}
	    = \widetilde{\theta}_{j_1,j_2}^{(\mu,\mu)} + \widetilde{\theta}_{j_1,j_2}^{(Z,Z)} +
        \{\widetilde{\theta}_{j_1,j_2}^{(\mu,Z)}+ \widetilde{\theta}_{j_1,j_2}^{(Z,\mu)}\}$,
where
{\footnotesize
\begin{align}
    \widetilde{\theta}_{j_1,j_2}^{(\mu,\mu)}
    	&= \sum_{1 \leq i,i' \leq n} a_{n,i - i'} (\mu_{i,j_1} - \bar{\mu}_{n,j_1})(\mu_{i',j_2} - \bar{\mu}_{n,j_2}) , \label{eqt:tildetheta1}\\
	\widetilde{\theta}_{j_1,j_2}^{(Z,Z)}
    	&= \sum_{1 \leq i,i' \leq n} a_{n,i - i'} (Z_{i,j_1} - \bar{Z}_{n,j_1})(Z_{i',j_2} - \bar{Z}_{n,j_2}) ,\label{eqt:tildetheta2}\\
	\widetilde{\theta}_{j_1,j_2}^{(\mu,Z)}+\widetilde{\theta}_{j_1,j_2}^{(Z,\mu)}
	    & =   \sum_{1 \leq i,i' \leq n} a_{n,i - i'} \left\{(\mu_{i',j_2} - \bar{\mu}_{n,j_2})(Z_{i,j_1} - \bar{Z}_{n,j_1})
				+ (\mu_{i,j_1} - \bar{\mu}_{n,j_1})(Z_{i',j_2} - \bar{Z}_{n,j_2}) \right\}.\label{eqt:tildetheta3}
\end{align}}
Using the above decomposition, we can write $\widetilde{L}$ as
\begin{equation}\label{eqt:Tj1j2diffm_3terms}
    \widetilde{L} = \widetilde{L}^{(\mu,\mu)}
    	+ \widetilde{L}^{(Z,Z)}
        + \{\widetilde{L}^{(\mu,Z)}+\widetilde{L}^{(Z,\mu)}\},
\end{equation}
where $\widetilde{L}^{(\alpha,\beta)} =
		\sum_{(j,j')\in\mathbb{J}} c_{j,j'}\widetilde{\theta}_{j,j'}^{(\alpha,\beta)}$, for $\alpha, \beta\in\{\mu, Z\}$.
	\item \underline{\textbf{Step 4}: Handling the three terms in (\ref{eqt:Tj1j2diffm_3terms}) one by one}:
\begin{itemize}
	\item By Lemma \ref{lem:B4}, $\widetilde{L}^{(\mu,\mu)} = \sum_{(j,j')\in\mathbb{J}}^4 c_{j,j'}\theta_{j,j'}+ O(\mathcal{M}_0+\mathcal{M}_1)$.
	\item By Lemma \ref{lem:B6}, $\widetilde{L}^{(Z,Z)} = O_p(1/\sqrt{n\ell})$.
	\item By Lemma \ref{lem:B8}, $\sqrt{n\kappa_\ell^2}\{\widetilde{L}^{(\mu,Z)} +\widetilde{L}^{(Z,\mu)}\}\inD \Normal(0,\widetilde{v}_{\mu Z})$,
	 		where $\widetilde{v}_{\mu Z}$ is defined in (\ref{eqt:CLT_vtildemuZ}).
				provided that
				$\mathcal{M}_{\mu Z} = o(1/\sqrt{n})$, which is satisfied
				given the assumptions on $\mathcal{C}$, $\mathcal{S}$, and $\mathcal{J}$
                stated in the theorem.,
\end{itemize}
	Combining them,
	we have, as $n \rightarrow \infty$,
\begin{equation*}
    \sqrt{n\kappa_\ell^2}\left\{\widetilde{L} - \sum_{(j,j')\in\mathbb{J}}^4 c_{j,j'}\theta_{j,j'}\right\} \inD \Normal(0,\widetilde{v}_{\mu Z}).
\end{equation*}
	\item \underline{\textbf{Step 5}: Representation in matrix form}:
		Combining Steps 1--4 and the Cram\'{e}r--Wold device, we obtain
		\[
			\sqrt{n}
			\left[
				(\widehat{\theta}_{11},\widehat{\theta}_{22},\widehat{\theta}_{33},\widehat{\theta}_{44},\widehat{\theta}_{12},\widehat{\theta}_{23},\widehat{\theta}_{34})^{\T}
				-(\theta_{11},\theta_{22},\theta_{33},\theta_{44},\theta_{12},\theta_{23},\theta_{34})^{\T}
			\right]
			\inD \Normal_7\left( 0_7 , \Omega_{\theta}^{\#}
			\right),
		\]
		where the variance-covariance matrix
		\[
			\Omega_{\theta}^{\#} =
			\left[
			\begin{array}{c|c}
			\Omega_{\theta}^{\#11} & \Omega_{\theta}^{\#12} \\
			\hline
			\Omega_{\theta}^{\#21} & \Omega_{\theta}^{\#22}
			\end{array}
			\right]
		\]
		is partition as follows:
		\begin{gather*}
		\resizebox{\textwidth}{!}{$
		\Omega_{\theta}^{\#11} =
		4\begin{bmatrix}
			v_{11}\theta_{11} & v_{12}\theta_{12} & v_{13}\theta_{13} & v_{14}\theta_{14} \\
			v_{12}\theta_{12} & v_{22}\theta_{22} & v_{23}\theta_{23} & v_{24}\theta_{24} \\
			v_{13}\theta_{13} & v_{23}\theta_{23} & v_{33}\theta_{33} & v_{34}\theta_{34} \\
			v_{14}\theta_{14} & v_{24}\theta_{24} & v_{34}\theta_{34} & v_{44}\theta_{44} \\
		\end{bmatrix},
		\quad
		\Omega_{\theta}^{\#12} = (\Omega_{\theta}^{\#21})^{\T} =
		2\begin{bmatrix}
			v_{11}\theta_{12} + v_{12}\theta_{11} & v_{12}\theta_{13} + v_{13}\theta_{12} & v_{13}\theta_{14} + v_{14}\theta_{13}\\
			v_{22}\theta_{12} + v_{12}\theta_{22} & v_{22}\theta_{23} + v_{23}\theta_{22} & v_{23}\theta_{24} + v_{24}\theta_{23}\\
			v_{13}\theta_{23} + v_{23}\theta_{13} & v_{33}\theta_{23} + v_{23}\theta_{33} & v_{33}\theta_{34} + v_{34}\theta_{33}\\
			v_{14}\theta_{24} + v_{24}\theta_{14} & v_{24}\theta_{34} + v_{34}\theta_{24} & v_{44}\theta_{34} + v_{34}\theta_{44}\\		\end{bmatrix},  $}\\
			\resizebox{\textwidth}{!}{$
			\Omega_{\theta}^{\#22}
			 =
			 \begin{bmatrix}
			 2v_{14}\theta_{24} + 2v_{24}\theta_{14} & v_{11}\theta_{22} + v_{22}\theta_{11} + 2v_{12}\theta_{12} & v_{22}\theta_{13} + 2v_{12}\theta_{23} + 2v_{13}\theta_{22} + 2v_{23}\theta_{12} & 2v_{13}\theta_{24} + 2v_{14}\theta_{23} + 2v_{23}\theta_{14} + 2v_{24}\theta_{13}\\
2v_{24}\theta_{34} + 2v_{34}\theta_{24} & v_{22}\theta_{13} + 2v_{12}\theta_{23} + 2v_{13}\theta_{22} + 2v_{23}\theta_{12} & v_{22}\theta_{33} + v_{33}\theta_{22} + 2v_{23}\theta_{23} & v_{33}\theta_{24} + 2v_{23}\theta_{34} + 2v_{24}\theta_{33} + 2v_{34}\theta_{23}\\
2v_{44}\theta_{34} + 2v_{34}\theta_{44} & 2v_{13}\theta_{24} + 2v_{14}\theta_{23} + 2v_{23}\theta_{14} + 2v_{24}\theta_{13} & v_{33}\theta_{24} + 2v_{23}\theta_{34} + 2v_{24}\theta_{33} + 2v_{34}\theta_{23} & v_{33}\theta_{44} + v_{44}\theta_{33} + 2v_{34}\theta_{34}
			\end{bmatrix}$}.
		\end{gather*}
		It is interesting to note that the structure of $\Omega_{\theta}^{\#22}$ is not the same
		for $\Cov(\widehat{\theta}_{12},\widehat{\theta}_{23})$ and
		$\Cov(\widehat{\theta}_{12},\widehat{\theta}_{34})$.
		This is the reason why we need to consider at least $S=4$ for obtaining the most general result.
		Finally, representing it in a matrix form for general $S$, we obtain (\ref{eqt:CLT_thetaHat}).
		Thus, the desired result in part (i) follows.
\end{itemize}

For part (ii), it is a straightforward application of the Delta method.
Let
    \begin{equation*}
        g(x_1,x_2,x_3,x_4,x_5,x_6,x_7) =
        \begin{pmatrix}
            \dfrac{x_5}{\sqrt{x_1x_2}}, & \dfrac{x_6}{\sqrt{x_2x_3}}, & \dfrac{x_7}{\sqrt{x_3x_4}}
        \end{pmatrix}^{\T},
    \end{equation*}
    where $x_1,x_2,x_3,x_4 > 0$.
    For the rest of the proof, we abbreviate $\nabla g(\theta_{11},\theta_{22},\theta_{33},\theta_{44},\theta_{12},\theta_{23},\theta_{34})$ as $\nabla g$ for simplicity.
    Then
    \begin{align*}
        \nabla g
		=\begin{bmatrix}
            -\dfrac{\theta_{2 2}\theta_{1 2}}{2(\theta_{1 1}\theta_{2 2})^{3/2}} & 0 & 0 \\
            -\dfrac{\theta_{1 1}\theta_{1 2}}{2(\theta_{1 1}\theta_{2 2})^{3/2}} & -\dfrac{\theta_{3 3}\theta_{2 3}}{2(\theta_{2 2}\theta_{3 3})^{3/2}} & 0 \\
            0 & -\dfrac{\theta_{2 2}\theta_{2 3}}{2(\theta_{2 2}\theta_{3 3})^{3/2}} & -\dfrac{\theta_{4 4}\theta_{3 4}}{2(\theta_{3 3}\theta_{4 4})^{3/2}} \\
            0 & 0 & -\dfrac{\theta_{3 3}\theta_{3 4}}{2(\theta_{3 3}\theta_{4 4})^{3/2}} \\
            \dfrac{1}{\sqrt{\theta_{1 1}\theta_{2 2}}} & 0 & 0 \\
            0 & \dfrac{1}{\sqrt{\theta_{2 2}\theta_{3 3}}} & 0 \\
            0 & 0 & \dfrac{1}{\sqrt{\theta_{3 3}\theta_{4 4}}}
        \end{bmatrix}.
    \end{align*}
    By the Delta method, we have
    \[
      \sqrt{n}
      \left[
        (\widehat{\rho}_{12},\widehat{\rho}_{23},\widehat{\rho}_{34})^{\T}
        -(\rho_{12},\rho_{23},\rho_{34})^{\T}
      \right]
      \inD \Normal_3\left( 0_3 , \Omega_{\rho}^{\#}
      \right),
    \]
    where $\Omega_{\rho}^{\#} = \nabla g^{\T}  \Omega_{\theta}^{\#} \nabla g$.
	Note that it suffices to compute the first row of $\Omega_{\rho}^{\#}$.
	Other entries of $\Omega_{\rho}^{\#}$ can be obtained by symmetry.
	For any matrix $M$, denote $(M)_{i,\bullet}$ and $(M)_{\bullet, j}$ denote the $i$th row and $j$th column of $M$, respectively.
    By some simple algebras, we have
    \begin{align*}
         (\Omega_{\theta})_{1,\bullet} \left(\nabla g\right)_{\bullet,1} &= 2v_{12}\left(1 - \rho_{12}^2\right)\sqrt{\frac{\theta_{11}}{\theta_{22}}},\\
         (\Omega_{\theta})_{2,\bullet} \left(\nabla g\right)_{\bullet,1} &= 2v_{12}\left(1 - \rho_{12}^2\right)\sqrt{\frac{\theta_{22}}{\theta_{11}}},\\
         (\Omega_{\theta})_{3,\bullet} \left(\nabla g\right)_{\bullet,1} &= -2v_{13}\rho_{12}\frac{\theta_{13}}{\theta_{11}} - 2v_{23}\rho_{12}\frac{\theta_{23}}{\theta_{22}} + \frac{2v_{13}\theta_{23}}{\sqrt{\theta_{11}\theta_{22}}} + \frac{2v_{23}\theta_{13}}{\sqrt{\theta_{11}\theta_{22}}},\\
         (\Omega_{\theta})_{4,\bullet} \left(\nabla g\right)_{\bullet,1} &= -2v_{14}\rho_{12}\frac{\theta_{14}}{\theta_{11}} - 2v_{24}\rho_{12}\frac{\theta_{24}}{\theta_{22}} + \frac{2v_{14}\theta_{24}}{\sqrt{\theta_{11}\theta_{22}}} + \frac{2v_{24}\theta_{14}}{\sqrt{\theta_{11}\theta_{22}}},\\
         (\Omega_{\theta})_{5,\bullet} \left(\nabla g\right)_{\bullet,1} &= \left(1 - \rho_{12}^2\right)\left(v_{11}\sqrt{\frac{\theta_{22}}{\theta_{11}}} + v_{22}\sqrt{\frac{\theta_{11}}{\theta_{22}}}\right),\\
         (\Omega_{\theta})_{6,\bullet} \left(\nabla g\right)_{\bullet,1} &= -v_{12}\rho_{12}\frac{\theta_{13}}{\theta_{11}} - v_{13}\rho_{12}^2\sqrt{\frac{\theta_{22}}{\theta_{11}}} - v_{22}\rho_{12}\frac{\theta_{23}}{\theta_{22}} + \frac{v_{22}\theta_{13}}{\sqrt{\theta_{11}\theta_{22}}} + \frac{v_{12}\theta_{23}}{\sqrt{\theta_{11}\theta_{22}}} + v_{13}\sqrt{\frac{\theta_{22}}{\theta_{11}}},\\
         (\Omega_{\theta})_{7,\bullet} \left(\nabla g\right)_{\bullet,1} &= -v_{13}\rho_{12}\frac{\theta_{14}}{\theta_{11}} - v_{14}\rho_{12}\frac{\theta_{13}}{\theta_{11}} - v_{23}\rho_{12}\frac{\theta_{24}}{\theta_{22}} - v_{24}\rho_{12}\frac{\theta_{23}}{\theta_{22}}\\
        &\qquad + \frac{v_{13}\theta_{24}}{\sqrt{\theta_{11}\theta_{22}}} +
        \frac{v_{14}\theta_{23}}{\sqrt{\theta_{11}\theta_{22}}} + \frac{v_{23}\theta_{14}}{\sqrt{\theta_{11}\theta_{22}}} +
        \frac{v_{24}\theta_{13}}{\sqrt{\theta_{11}\theta_{22}}},
    \end{align*}
    where $\Omega_{\theta}$ is the asymptotic covariance matrix in Theorem \ref{thm:4.3}.
    Then,
    {\small
    \begin{align*}
        \left(\Omega_{\rho}^{\#}\right)_{11}
        &= \left(1 - \rho_{12}^2\right)\left(\frac{v_{11}}{\theta_{11}} + \frac{v_{22}}{\theta_{22}} - \frac{2v_{12}}{\sqrt{\theta_{11}\theta_{22}}}\rho_{12}\right),\\
        \left(\Omega_{\rho}^{\#}\right)_{12}
        &= \frac{v_{12}}{\sqrt{\theta_{11}\theta_{22}}}\left(\rho_{23}\rho_{12}^2 - \rho_{12}\rho_{13}\right) + \frac{v_{13}}{\sqrt{\theta_{11}\theta_{33}}}\left(\rho_{12}\rho_{13}\rho_{23} - \rho_{23}^2 - \rho_{12}^2 + 1\right)\\
        &\qquad + \frac{v_{23}}{\sqrt{\theta_{22}\theta_{33}}}\left\{\rho_{23}\left(\rho_{12}\rho_{23} - \rho_{13}\right)\right\} + \frac{v_{22}}{\theta_{22}}\left(\rho_{13} - \rho_{12}\rho_{23}\right),\\
        \left(\Omega_{\rho}^{\#}\right)_{13}
        &= \frac{v_{13}}{\sqrt{\theta_{11}\theta_{33}}}\left(\rho_{12}\rho_{13}\rho_{34} - \rho_{23}\rho_{34} - \rho_{12}\rho_{14} + \rho_{24}\right) \\
        &\qquad + \frac{v_{23}}{\sqrt{\theta_{22}\theta_{33}}}\left(\rho_{12}\rho_{23}\rho_{34} - \rho_{13}\rho_{34} - \rho_{12}\rho_{24} + \rho_{14}\right)\\
        &\qquad + \frac{v_{14}}{\sqrt{\theta_{11}\theta_{44}}}\left(\rho_{12}\rho_{14}\rho_{34} - \rho_{24}\rho_{34} - \rho_{12}\rho_{13} + \rho_{23}\right) \\
         &\qquad + \frac{v_{24}}{\sqrt{\theta_{22}\theta_{44}}}\left(\rho_{12}\rho_{24}\rho_{34} - \rho_{14}\rho_{34} - \rho_{12}\rho_{23} + \rho_{13}\right).
    \end{align*}}
    Representing it in a matrix form for general $S$, we obtain the desired result in part (ii).

\begin{remark}[Fisher's transform]\label{rem:fiskerTransform}
We can project $\rho_{j_1,j_2}\in[-1,1]$ to $\mathbb{R}$ by the Fisher transform. Define
			$\varrho_{j_1,j_2} =\atanh\rho_{j_1,j_2}$ and
			$\hat{\varrho}_{j_1,j_2}^{(m)}=\atanh \hat{\rho}_{j_1,j_2}^{(m)}$,
where
$\atanh(z) = \frac{1}{2}\log\frac{1 + z}{1 - z}$.
Denote them in matrix form as $\hat{\varrho}^{(m)}$ and $\varrho$,
			where the diagonal entries are $+\infty$.
			By Theorem \ref{thm:4.3} and the Delta method,
			$n^{1/2} \vectorized(\hat{\varrho}^{(m)} - \varrho )
					\inD \Normal(0, \Omega_{\varrho})$,
where
$\Omega_{\varrho} = \diag\{\vectorized(J_S-\rho\odot \rho)\}^{-1} \Omega_{\rho} \diag\{\vectorized(J_S-\rho\odot \rho)\}^{-1}$
			and $J_S$ is a $S\times S$ matrix of ones.
			The entry of $\Omega_{\varrho}$ corresponding to $\Cov(\hat{\varrho}_{j_1,j_2}^{(m)},\hat{\varrho}_{j_3,j_4}^{(m)})$ is
            $\Omega_{\varrho,(j_1,j_2),(j_3,j_4)}
            = \Omega_{\rho,(j_1,j_2),(j_3,j_4)} (1 - \rho_{j_1,j_2}^2)^{-1}(1 - \rho_{j_3,j_4}^2)^{-1}$.
To derive it,
we can use the Delta's method. Let $f(z_1,z_2,z_3) = (\atanh(z_1) , \atanh(z_2) , \atanh(z_3)^{\T}$
for $-1 < z_1,z_2,z_3 < 1$.
So,
    \begin{equation*}
        \nabla f\left(\rho_{j_1,j_2},\rho_{j_2,j_3},\rho_{j_3j_4}\right)^{\T} =
        \begin{pmatrix}
            \dfrac{1}{1 - \rho_{j_1,j_2}^2} & 0 & 0\\
            0 & \dfrac{1}{1 - \rho_{j_2,j_3}^2} & 0\\
            0 & 0 & \dfrac{1}{1 - \rho_{j_3j_4}^2}
        \end{pmatrix}.
    \end{equation*}
Then the desired result follows.
\end{remark}

\subsection{Proof of Theorem \ref{thm:5.3}}
The proof logic is similar to the proof of Theorem \ref{thm:4.3}.
We prove as follows:
\begin{itemize}
	\item \underline{\textbf{Step 1}: Approximation by quadratic form.}
		We approximate $\widehat{\theta}$ by the quadratic form $\widetilde{\theta}$, where  $\widetilde{\theta}$ is defined in (\ref{eqt:tildetheta}).
			In view of Proposition \ref{prop:hat_tilde_diff} with $\mathcal{M}_0=\mathcal{M}_1=0$, we have
			\begin{equation}\label{eqt:boundAhatAtilde2}
			    \left\| \widehat{\theta}_{j_1,j_2} - \widetilde{\theta}_{j_1,j_2} \right\| = O(\ell^{1/2}/n).
			\end{equation}
		Since $\ell \ll n^{1/2}$, we obtain $\left\| \widehat{\theta}_{j_1,j_2} - \widetilde{\theta}_{j_1,j_2} \right\| =o(1/\sqrt{n\ell})$.
		So, it suffices to consider $\widetilde{\theta}$.
	\item \underline{\textbf{Step 2}: Cram\'{e}r--Wold device.}
			Now we establish the result through the Cram\'{e}r--Wold device.
Let $c_{j_1,j_2} \in \mathbb{R}$, where $j_1,j_2 = 1,\ldots,S$.
Then we consider the linear combination:
\begin{align}
    \widetilde{L}^{\circ} = \sum_{j_1 = 1}^S \sum_{j_2 = j_1}^S c_{j_1,j_2} \widetilde{\theta}_{j_1,j_2},
    \quad \text{where} \quad
    \widetilde{\theta}_{j_1,j_2} = \sum_{1 \leq i,i' \leq n} a_{n,i - i'}(X_{i,j_1} - \bar{X}_{n,j_1})(X_{i'j_2} - \bar{X}_{n,j_2}). \label{eqt:tildeLcirc_tildeTheta}
\end{align}
	\item \underline{\textbf{Step 3}: Decomposing $\widetilde{L}^{\circ}$.}
		Using (\ref{eqt:tildetheta1})--(\ref{eqt:tildetheta3}), we have
		$\widetilde{\theta}_{j_1,j_2}
	    = \widetilde{\theta}_{j_1,j_2}^{(\mu,\mu)} + \widetilde{\theta}_{j_1,j_2}^{(Z,Z)} +
        \{\widetilde{\theta}_{j_1,j_2}^{(\mu,Z)}+\widetilde{\theta}_{j_1,j_2}^{(Z,\mu)}\}$.
	    Let $\widetilde{L}^{\circ(\alpha\beta)} = \sum_{j_1 = 1}^S \sum_{j_2 = j_1}^S c_{j_1,j_2} \widetilde{\theta}_{j_1,j_2}^{(\alpha,\beta)}$
	    for $\alpha, \beta\in\{\mu,Z\}$.
	    So, we have
	    \[
	    	\widetilde{L}^{\circ} = \widetilde{L}^{\circ(\mu\mu)} + \widetilde{L}^{\circ(ZZ)} +
            \{\widetilde{L}^{\circ(\mu Z)}+\widetilde{L}^{\circ(Z\mu)}\}.
		\]
	\item \underline{\textbf{Step 4}: Handling the three terms one by one}.
		\begin{itemize}
			\item We have $\widetilde{L}^{\circ(\mu\mu)} = 0$ since $\mu(\cdot)$ is a constant function.
			\item We also have $\widetilde{L}^{\circ(\mu Z)} = \widetilde{L}^{\circ(Z\mu)} = 0$ since $\mu(\cdot)$ is a constant function.
			\item Next we use Lemma \ref{lem:B11} to derive the limiting distribution of $\widetilde{L}^{\circ(ZZ)}$.
				In light of the calculations done in the proof of Lemma \ref{lem:B6},
				we only need to further show that $\sigma_n^2 \sim 4A(\Delta_m - \Delta_0)/(n\ell \kappa_\ell^2)$.
Note that
{\small
\begin{align*}
    \sigma_n^2
    &= 2\sum_{k = 1}^n \sum_{t = 1}^n b_{n,t - k}^2\\
    &= 2\sum_{k = 1}^n \sum_{t = 1}^n \frac{\mathbb{1}(|t - k| \leq \ell + mh)}{n^2 \ell^2 \kappa_\ell^2}  \left\{K^2\left(\frac{t - k}{\ell}\right) - 2K\left(\frac{t - k}{\ell}\right)K_{m}\left(\frac{t - k}{\ell}\right) + K_{m}^2\left(\frac{t - k}{\ell}\right)\right\}\\
    &= \frac{2}{n^2 \ell^2 \kappa_\ell^2}\sum_{|j'| \leq \ell + mh}\left\{K^2\left(\frac{j'}{\ell}\right) - 2K\left(\frac{j'}{\ell}\right)K_{m}\left(\frac{j'}{\ell}\right) + K_{m}^2\left(\frac{j'}{\ell}\right)\right\}(n - |j'|)\\
    &= \frac{2}{n \ell^2 \kappa_\ell^2}\sum_{|j'| \leq \ell + mh}\left\{K^2\left(\frac{j'}{\ell}\right) - 2K\left(\frac{j'}{\ell}\right)K_{m}\left(\frac{j'}{\ell}\right) + K_{m}^2\left(\frac{j'}{\ell}\right)\right\} + O\left(\frac{1}{n^2 \ell \kappa_\ell^2}\right)\\
    &= \frac{2}{n \ell^2 \kappa_\ell^2}\sum_{|j'| \leq \ell} K^2\left(\frac{j'}{\ell}\right) - \frac{2}{n \ell^2 \kappa_\ell^2}\sum_{|j'| \leq \ell} 2K\left(\frac{j'}{\ell}\right)K_{m}\left(\frac{j'}{\ell}\right) \\
    &\qquad  + \frac{2}{n \ell^2 \kappa_\ell^2}\sum_{|j'| \leq \ell} K_{m}^2\left(\frac{j'}{\ell}\right) + o\left(\frac{1}{n\ell}\right)\\
    &= \frac{2}{n \ell^2 \kappa_\ell^2}\sum_{|j'| \leq \ell} K^2\left(\frac{j'}{\ell}\right) - \frac{2}{n \ell^2 \kappa_\ell^2}\sum_{|j'| \leq \ell} 2K\left(\frac{j'}{\ell}\right)K\left(\frac{j'}{\ell}\right)\delta_0 \\
    &\qquad + \frac{2}{n \ell^2 \kappa_\ell^2}\sum_{|j| \leq m} \sum_{|j''| \leq \ell + j\lambda \ell} K^2\left(\frac{j''}{\ell} - j\lambda\right)\delta_{j}^2 + o\left(\frac{1}{n\ell}\right)\\
    &\sim \frac{4}{n\ell \kappa_\ell^2}A - \frac{8}{n\ell \kappa_\ell^2}A + \frac{4}{n\ell \kappa_\ell^2}A\Delta_m\\
    &= \frac{4A(\Delta_m - \Delta_0)}{n\ell \kappa_\ell^2}.
\end{align*}}
			Hence, we have
			\[
				\sqrt{n\ell\kappa_\ell^2} \widetilde{L}^{\circ(ZZ)}
				\inD \Normal\left( 0,
				\sum_{r = 1}^S \sum_{s = r}^S \sum_{r' = 1}^S \sum_{s' = r'}^S c_{rs}c_{r's'}
					(v_{rr'}v_{ss'} + v_{rs'}v_{sr'})
				\right).
			\]
		\end{itemize}
	\item \underline{\textbf{Step 5}: Representation in matrix form}:
		Combining Steps 1--4 and the Cram\'{e}r--Wold device, we obtain the desired result.
\end{itemize}

\subsection{Proof of Corollary \ref{cor:improvedVersion}}
We prove (\ref{eqt:limitDist_general_theta}), and the proof of (\ref{eqt:limitDist_general_rho}) is similar.
Note that
\[
	\bar{\theta}_{j_1, j_2} = \widehat{\theta}_{j_1, j_2} \mathbb{1}(|\widehat{\theta}_{j_1, j_1}|\geq \epsilon_{j_1})\mathbb{1}(|\widehat{\theta}_{j_2, j_2}|\geq \epsilon_{j_2}).
\]
We consider two cases:
\begin{itemize}
	\item \underline{\textbf{Case 1}: at least one of $\theta_{j_1, j_1}$ and $\theta_{j_2, j_2}$ is zero}.
            Without loss of generality, we assume $\theta_{j_1, j_1}=0$.
			In this case, we have that $\theta_{j_1,j_2} =0$ by definition.
			Then
			\begin{align*}
				\sqrt{n} (\bar{\theta}_{j_1, j_2} -\theta_{j_1,j_2})
					&= \left( \sqrt{n} \widehat{\theta}_{j_1, j_2} \right)
						\mathbb{1}(\sqrt{n\ell}|\widehat{\theta}_{j_1, j_1}|\geq \sqrt{n\ell}\epsilon_{j_1})\mathbb{1}(|\widehat{\theta}_{j_2, j_2}|\geq \epsilon_{j_2}) \\
                    &= O_p(1) o_p(1) O_p(1)
                    \inP 0 ,
			\end{align*}
			where we have used the results that
            $\sqrt{n} \widehat{\theta}_{j_1, j_2} = O_p(1)$
            by Theorem \ref{thm:4.3};
			$\sqrt{n\ell} \widehat{\theta}_{j_1, j_1} = O_p(1)$
            by the univariate special case of
            Theorem \ref{thm:5.3}; and
            the assumption that $\sqrt{n\ell}\epsilon_{j_1} \inP \infty$.
	\item \underline{\textbf{Case 2}: Neither of $\theta_{j_1, j_1}$ and $\theta_{j_2, j_2}$ is zero}.
			First, we find the weak limit of $\mathbb{1}(|\widehat{\theta}_{j_1, j_1}|\geq \epsilon_{j_1})$.
			Note that $\theta_{j_1, j_1}>0$ by definition and the assumption in this case.
			Then we consider
			\begin{align*}
				\pr(|\widehat{\theta}_{j_1, j_1}| < \epsilon_{j_1})
                    &= \pr( \theta_{j_1,j_1}-|\widehat{\theta}_{j_1, j_1}| + \epsilon_{j_1} > \theta_{j_1,j_1}) \\
                    &\leq \pr( |\widehat{\theta}_{j_1, j_1}-\theta_{j_1,j_1}| + \epsilon_{j_1} > \theta_{j_1,j_1}) \\
                    &\leq
                    \pr(|\widehat{\theta}_{j_1, j_1}-\theta_{j_1, j_1}| > \theta_{j_1,j_1}/2)
                    + \pr(\epsilon_{j_1} >\theta_{j_1,j_1}/2),
			\end{align*}
			where the first term goes to zero as
            $\widehat{\theta}_{j_1, j_1}\inP\theta_{j_1, j_1}$
            by Theorem \ref{thm:4.3}, and
            the second term goes to zero as $\epsilon_{j_1}\inP 0$ by assumption.
            Hence, $\pr(|\widehat{\theta}_{j_1, j_1}| < \epsilon_{j_1}) \rightarrow 0$, i.e., $\mathbb{1}(|\widehat{\theta}_{j_1, j_1}|\geq \epsilon_{j_1}) \inP 1$.
			Similarly, we have $\mathbb{1}(|\widehat{\theta}_{j_2, j_2}|\geq \epsilon_{j_2}) \inP 1$.
			Finally, we obtain that
			\begin{align*}
				\sqrt{n} (\bar{\theta}_{j_1, j_2} -\theta_{j_1,j_2})
					&= \left\{ \sqrt{n} (\widehat{\theta}_{j_1, j_2} -\theta_{j_1,j_2})\right\}
						\mathbb{1}(|\widehat{\theta}_{j_1, j_1}|\geq \epsilon_{j_1})\mathbb{1}(|\widehat{\theta}_{j_2, j_2}|\geq \epsilon_{j_2})\\
					&\inD \Normal(0, \Omega_{\theta, (j_1, j_2), (j_1,j_2)})
			\end{align*}
			by Theorem \ref{thm:4.3} and Slutsky's Theorem.

\end{itemize}

\section{Auxiliary Results}\label{sec:auxillary}

\subsection{Proof of (\ref{eqt:betaWrongLimit})}\label{sec:proofOfReg}
The result follows directly from the calculations below.
\begin{align*}
     \widehat{\beta}_{1}
    &= \frac{\sum_{i = 1}^n (X_{i,j_1} - \bar{X}_{n,j_1})(X_{i,j_2} - \bar{X}_{n,j_2})}{\sum_{i = 1}^n (X_{i,j_1} - \bar{X}_{n,j_1})^2}\\
    &= \frac{\sum_{i = 1}^n (X_{i,j_1} - \bar{X}_{n,j_1})(X_{i,j_2} - \bar{X}_{n,j_2})}{\sqrt{\sum_{i = 1}^n (X_{i,j_1} - \bar{X}_{n,j_1})^2}\sqrt{\sum_{i = 1}^n (X_{i,j_2} - \bar{X}_{n,j_2})^2}}\frac{\sqrt{\sum_{i = 1}^n (X_{i,j_2} - \bar{X}_{n,j_2})^2}}{\sqrt{\sum_{i = 1}^n (X_{i,j_1} - \bar{X}_{n,j_1})^2}}\\
    &= \frac{(1/n)\sum_{i = 1}^n (X_{i,j_1} - \bar{X}_{n,j_1})(X_{i,j_2} - \bar{X}_{n,j_2})}{\sqrt{(1/n)\sum_{i = 1}^n (X_{i,j_1} - \bar{X}_{n,j_1})^2}\sqrt{(1/n)\sum_{i = 1}^n (X_{i,j_2} - \bar{X}_{n,j_2})^2}}\frac{\sqrt{(1/n)\sum_{i = 1}^n (X_{i,j_2} - \bar{X}_{n,j_2})^2}}{\sqrt{(1/n)\sum_{i = 1}^n (X_{i,j_1} - \bar{X}_{n,j_1})^2}}\\
    &= \frac{(\widehat{\gamma}_0^{(0)})_{j_1,j_2}}{\sqrt{(\widehat{\gamma}_0^{(0)})_{j_1,j_1}(\widehat{\gamma}_0^{(0)})_{j_2,j_2}}}\frac{\sqrt{(\widehat{\gamma}_0^{(0)})_{j_2,j_2}}}{\sqrt{(\widehat{\gamma}_0^{(0)})_{j_1,j_1}}}\\
    &\inP \frac{(\gamma_0)_{j_1,j_2} + \theta_{j_1,j_2}}{\sqrt{((\gamma_0)_{j_1,j_1} + \theta_{j_1,j_1})((\gamma_0)_{j_2,j_2} + \theta_{j_2,j_2})}}\frac{\sqrt{(\gamma_0)_{j_2,j_2} + \theta_{j_2,j_2}}}{\sqrt{(\gamma_0)_{j_1,j_1} + \theta_{j_1,j_1}}}\\
    &= \beta_{1,\mix},
\end{align*}
where we use Proposition \ref{prop:B3} when we take $n \rightarrow \infty$.

\subsection{Proposition \ref{prop:hat_tilde_diff}}
Recall the definition of the quadratic form $\widetilde{\theta}_{j_1,j_2}$ in (\ref{eqt:tildetheta})
as an approximation of $\widehat{\theta}_{j_1,j_2}$.

\begin{proposition}\label{prop:hat_tilde_diff}
	Let $m \geq 1$ be fixed,
	$u_q := \max_{j = 1,\ldots,S} \sum_{k \in \mathbb{Z}} |k|^q |(\gamma_k)_{jj}| < \infty$ for some $q \in \mathbb{N}$, $h=2\ell$
	and
	$n^{1/(2 + 2q)} \lesssim \ell \ll \sqrt{n}$.
	Suppose Assumptions \ref{assum:jumps}, \ref{assum:weak dependence} and \ref{assum:kernel} hold.
Then, for any $j_1,j_2\in\{1, \ldots, S\}$,
\[
	\left\| \widehat{\theta}_{j_1,j_2} - \widetilde{\theta}_{j_1,j_2} \right\|
	= O\left( \frac{\ell^{1/2}}{n} +\mathcal{M}_0 + \frac{\ell}{n}\mathcal{M}_1 \right).
\]
\end{proposition}
\begin{proof}[Proof of Proposition \ref{prop:hat_tilde_diff}]
    Recall that
    \begin{eqnarray*}
        \widehat{\theta}_{j_1,j_2} &=& \frac{\sum_{|k| \leq \ell} K(k/\ell)\{(\widehat{\gamma}_k^{(0)})_{j_1,j_2} - (\widehat{\gamma}_k^{(m)})_{j_1,j_2}\}}{\sum_{|k| \leq \ell} K(k/\ell)},\\
        \widetilde{\theta}_{j_1,j_2} &=& \sum_{1 \leq i,i' \leq n} a_{n,i - i'}(X_{i,j_1} - \bar{X}_{n,j_1})(X_{i'j_2} - \bar{X}_{n,j_2}) ,
    \end{eqnarray*}
    where $(\widehat{\gamma}_k^{(m)})_{j_1,j_2}$ is as defined in Proposition \ref{prop:B3}, and
    \begin{equation*}
        a_{n,k} = \frac{\{K(k/\ell) - K_{m}(k/\ell)\}\mathbb{1}(|k| \leq \ell + mh)}{n\ell \kappa_\ell}
		\quad\text{and}\quad
		K_{m}(t) = \sum^{\left \lfloor{\min\left(m, \frac{1-t}{\lambda}\right)}\right \rfloor}_{s = \left \lceil{\max\left(-m, \frac{-1-t}{\lambda}\right)}\right \rceil} \delta_{|s|}K(t + \lambda s).
    \end{equation*}
    Using the property $\sum_{j = 0}^m d_j = 0$, we have
\begin{align*}
    \widehat{\theta}_{j_1,j_2}
    &= \frac{1}{\ell \kappa_\ell}
    	\sum_{|k| \leq \ell} K\left( \frac{k}{\ell} \right)
		\left\{
			(\widehat{\gamma}_k^{(0)})_{j_1,j_2}
			- (\widehat{\gamma}_k^{(m)})_{j_1,j_2}
		\right\}\\
    &= \frac{1}{\ell \kappa_\ell}\sum_{|k| \leq \ell}
    	K\left(\frac{k}{\ell}\right)
		\left\{
			(\widehat{\gamma}_k^{(0)})_{j_1,j_2} -
			\sum_{|s| \leq m}\delta_s (\widehat{\gamma}^{(0)}_{k + sh})_{j_1,j_2}
		\right\}
		+ R_1\\
    &= \frac{1}{\ell \kappa_\ell}
    	\left\{  \sum_{|k| \leq \ell} K\left( \frac{k}{\ell} \right)  (\widehat{\gamma}_k^{(0)})_{j_1,j_2}
				- \sum_{|s| \leq m} \delta_s
				\sum_{k = -\ell}^{\ell} K\left( \frac{k}{\ell} \right)  (\widehat{\gamma}^{(0)}_{k + sh})_{j_1,j_2}
		\right\} + R_1\\
    &= \frac{1}{\ell \kappa_\ell}\left\{\sum_{|k| \leq \ell}
    		K\left(\frac{k}{\ell}\right)  (\widehat{\gamma}_k^{(0)})_{j_1,j_2}
			- \sum_{|s| \leq m} \delta_s
			\sum_{k = -\ell + sh}^{\ell + sh} K\left( \frac{k - sh}{\ell} \right)  (\widehat{\gamma}_k^{(0)})_{j_1,j_2}
		\right\}
		+ R_1\\
    &= \frac{1}{\ell \kappa_\ell}
    	\sum_{|k| \leq \ell + mh}
		\left\{ K\left(\frac{k}{\ell}\right)
			- \sum_{|s| \leq m} \delta_s K\left( \frac{k - sh}{\ell} \right)
		\right\}
		(\widehat{\gamma}_k^{(0)})_{j_1,j_2}
		+ R_1\\
    &= \frac{1}{\ell \kappa_\ell}
    	\sum_{|k| \leq \ell + mh}
		\left\{
			K\left(\frac{k}{\ell}\right) - K_m\left( \frac{k}{\ell} \right)
		\right\}
		(\widehat{\gamma}_k^{(0)})_{j_1,j_2}
		+ R_1 \\
    &= \widetilde{\theta}_{j_1,j_2} + R_1,
\end{align*}
where the reminder term $R_1$ is defined as
\begin{align*}
    R_1 &= -\frac{1}{\ell \kappa_\ell}
    	\sum_{|k| \leq \ell} K\left(\frac{k}{\ell}\right)
		\left\{(\widehat{\gamma}_k^{(m)})_{j_1,j_2}
		- \sum_{|s| \leq m} \delta_s
			(\widehat{\gamma}^{(0)}_{k + sh})_{j_1,j_2}\right\} .
\end{align*}
The difference $(\widehat{\gamma}_k^{(m)})_{j_1,j_2} - \sum_{|s| \leq m} \delta_s (\widehat{\gamma}^{(0)}_{k + sh})_{j_1,j_2}$
is contributed by the boundary indices.
To see it, we expand them one by one:
\begin{align*}
	(\widehat{\gamma}_k^{(m)})_{j_1,j_2}
		&=\frac{1}{n} \sum_{j=0}^m \sum_{j'=0}^m d_j d_{j'}
			\sum_{i = (m-j)h + k^+ + 1}^{n - jh -k^-} D_{i,j_1}^{(0)} D_{i-k-(j'-j)h, j_2}^{(0)}, \\
	\sum_{|s| \leq m} \delta_s (\widehat{\gamma}^{(0)}_{k + sh})_{j_1,j_2}
		&=\frac{1}{n} \sum_{j=0}^m \sum_{j'=0}^m d_j d_{j'}
				\sum_{i = \{k+(j'-j)h\}^++1}^{n-(k+(j'-j)h)^-} D_{i,j_1}^{(0)}  D_{i-k-(j'-j)h,j_2}^{(0)} ,
\end{align*}
where $x_- = \max(-x,0)$ and
$X_+ = \max(x,0)$.
Hence, $R_1$ can be simplified as
\begin{align*}
    R_1 &= \frac{1}{ \kappa_\ell n}
    		\sum_{|k|\leq\ell}
			K\left( \frac{k}{\ell} \right)
			\sum_{j=0}^m \sum_{j'=0}^m d_j d_{j'} (\bar{\gamma}^{(0)}_{k+(j'-j)h})_{j_1,j_2}
\end{align*}
where
\[
	(\bar{\gamma}^{(0)}_{k+(j'-j)h})_{j_1,j_2}
	=\frac{1}{\ell}\left(
					\sum_{i=\{k+(j'-j)h\}^++1}^{(m-j)h+k^+}
					+ \sum_{i=n-jh+1-k^-}^{n-\{k+(j'-j)h\}^-}
				\right)
				D^{(0)}_{i,j_1}D^{(0)}_{i-k-(j'-j)h,j_2}.
\]
    Applying Proposition \ref{prop:B3} and the property $\theta_{j_1,j_2} = O\{(\mathcal{C}+\mathcal{S}\mathcal{J})^2\} = O(\mathcal{M}_0n/\ell)$, we have
    \begin{align*}
    	\E\{ (\bar{\gamma}^{(0)}_{k+(j'-j)h})_{j_1,j_2}\}
    		&= O\{ (\gamma_{k+(j'-j)h})_{j_1,j_2} + \theta_{j_1,j_2} + \mathcal{M}_0 + \mathcal{M}_1\}\\
			&= O\{ (\gamma_{k+(j'-j)h})_{j_1,j_2} + \mathcal{M}_0 n /\ell + \mathcal{M}_1\} ,  \\
		\Var\{ (\bar{\gamma}^{(0)}_{k+(j'-j)h})_{j_1,j_2}\}
			&=O\{ (\mathcal{M}_0 + \mathcal{M}_1 +1 )/\ell\}.
	\end{align*}
	Consequently, using the fact that $\kappa_\ell = O(1)$ and $\sum_{k = -\infty}^\infty |(\gamma_k)_{j_1,j_2}| <\infty$,
	we have
    \begin{align*}
        \E(R_1) &= O\left\{ \frac{1+\ell(\mathcal{M}_0n/\ell + \mathcal{M}_1)}{n} \right\}
        		= O\left( \frac{1}{n} +\mathcal{M}_0 + \frac{\ell}{n}\mathcal{M}_1 \right), \\
        \|R_1-\E(R_1)\| &= O\left(\frac{\ell}{n}\right) O\left\{ \left( \frac{ \mathcal{M}_0 + \mathcal{M}_1 +1 }{\ell} \right)^{1/2} \right\}
			= O\left\{ \frac{\ell^{1/2} (\mathcal{M}_0^{1/2} + \mathcal{M}_1^{1/2} + 1)}{n}  \right\} ,
    \end{align*}
    where the last line follows from the inequality $(a+b)^{1/2} \leq a^{1/2} + b^{1/2}$.
	By Minkowski inequality, we have
	\[
		\|R_1\| \leq  |\E(R_1)|+ \|R_1-\E(R_1)\|
		= O\left( \frac{\ell^{1/2}}{n} +\mathcal{M}_0 + \frac{\ell}{n}\mathcal{M}_1 \right).
	\]
    Hence the desired result follows.
\end{proof}

\subsection{Proposition \ref{prop:B1}}
\begin{proposition}\label{prop:B1}
For any $x_{1,j_1},\ldots,x_{n,j_1},x_{1,j_2},\ldots,x_{n,j_2}\in\mathbb{R}$,
$k \in \{0, \pm 1,  \ldots, \pm(n-1)\}$,
and $n\in\mathbb{N}$,
define
$\bar{x}_{j} = \sum_{i = 1}^n x_{i,j}/n$,
$k_- = \max(-k,0)$,
$k_+ = \max(k,0)$, and
\[
    (\widehat{\gamma}^x_k)_{j_1,j_2} = \frac{1}{n}\sum_{i = 1+k_+}^{n-k_-} (x_{i,j_1} - \bar{x}_{j_1})(x_{i - k,j_2} - \bar{x}_{j_2}).
\]
Then the following identity holds:
\begin{align*}
    (\widehat{\gamma}^x_k)_{j_1,j_2} &= (\widehat{\gamma}^x_0)_{j_1,j_2} - \frac{1}{n}\sum_{i = 1}^{|k|} (x_{i,j_1} - \bar{x}_{j_1})(x_{i,j_2} - \bar{x}_{j_2}) \\
    &\qquad - \frac{1}{n}\sum_{i = |k| + 1}^n \left\{ (x_{i,j_1} - x_{i - k_-, j_1})(x_{i,j_2} - \bar{x}_{j_2}) + (x_{i - k_-, j_1} - \bar{x}_{j_1})(x_{i,j_2} - x_{i - k_+, j_2}) \right\},
\end{align*}
\end{proposition}

\begin{proof}[Proof of Proposition \ref{prop:B1}]
Let the centered variables be $x_{i,j}^{\circ}=x_{i,j} - \bar{x}_j$.
By definition, we have $(\widehat{\gamma}^x_0)_{j_1,j_2}=\sum_{i=1}^n x_{i,j_1}^{\circ} x_{i,j_2}^{\circ}/ n$.
We consider two cases: $k\geq 0$ and $k<0$.
For $k\geq 0$, we have
\begin{align*}
    (\widehat{\gamma}^x_k)_{j_1,j_2}
    	&= \frac{1}{n}\sum_{i=k+1}^n x_{i,j_1}^\circ \left\{x_{i,j_2}^\circ-(x_{i,j_2}^\circ - x_{i - k,j_2}^\circ) \right\} \\
		&= \left\{\frac{1}{n}\sum_{i=k+1}^n x_{i,j_1}^\circ x_{i,j_2}^\circ\right\}
			- \frac{1}{n}\sum_{i = k + 1}^n x_{i,j_1}^\circ (x_{i,j_2}^\circ - x_{i - k,j_2}^\circ) \\
		&= \left\{ (\widehat{\gamma}^x_0)_{j_1,j_2} - \frac{1}{n}\sum_{i = 1}^k x_{i,j_1}^\circ x_{i,j_2}^\circ  \right\}
			- \frac{1}{n}\sum_{i = k + 1}^n x_{i,j_1}^\circ (x_{i,j_2}^\circ - x_{i - k,j_2}^\circ),
\end{align*}
which is the desired expression.
The derivation for $k<0$ is similar.
Thus, the desired result follows.
\end{proof}

\subsection{Proposition \ref{prop:B2}}
\begin{proposition}\label{prop:B2}
    Let $m \geq 0$ be fixed, $1/\ell + \ell/n = o(1)$ and $h/\ell = \lambda \in [2,\infty)$.
    Let Assumptions \ref{assum:jumps} and \ref{assum:weak dependence} hold.
    For $|k| \leq \ell$, define
    \begin{equation*}
        D^{(\mu)}_{i,s} = \sum_{j = 0}^m d_{j} (\mu_{i - hj,s} - \bar{\mu}_{n,s}),
    \end{equation*}
    where $s = 1,\ldots,S$.
    Denote
    \[
    	\mathcal{M}_0 = \frac{\ell}{n}(\mathcal{C} + \mathcal{SJ})^2
		\qquad \text{and} \qquad
    	\mathcal{M}_1 = \frac{\ell}{n}\left(\frac{\mathcal{C}^2\ell}{n} + \mathcal{S}^2\mathcal{J}\right).
    \]
    Then for $j_1,j_2 = 1,\ldots,S$,
    \begin{equation*}
        \frac{1}{n}\sum_{i = 1+mh + k^+}^{n-k^-} D^{(\mu)}_{i,j_1}D^{(\mu)}_{i - k, j_2} =
        \begin{cases}
            \displaystyle O(\mathcal{M}_1), ~ &\text{if} ~ m > 0,\\
            \displaystyle \theta_{j_1,j_2} + O(\mathcal{M}_0), ~ &\text{if} ~ m = 0.
        \end{cases}
    \end{equation*}
\end{proposition}
\begin{proof}[Proof of Proposition \ref{prop:B2}]
    We first prove the case $m > 0$.
    Recall that $\mu_{i,j_1} = \mathfrak{c}_{i,j_1} + \mathfrak{s}_{i,j_1}$, where $\mathfrak{c}_{i,j_1} = \mathfrak{c}_{j_1}(i/n)$ and $\mathfrak{s}_{i,j_1} = \mathfrak{s}_{j_1}(i/n) = \sum_{j = 0}^\mathcal{J} \xi_{j,j_1}\mathbb{1}\left(T_{j} \leq i < T_{j + 1}\right)$ for each $i$. Define
	\begin{equation*}
		D^{(\mathfrak{c})}_{i,j_1} = \sum_{j = 0}^m d_{j}\mathfrak{c}_{i - hj,j_1}
		\quad \text{and} \quad
		D^{(\mathfrak{s})}_{i,j_1} = \sum_{j = 0}^m d_{j}\mathfrak{s}_{i - hj,j_1} \;
	\end{equation*}
	for each $i$.
	Then using the inequality $|xy| \leq x^2 + y^2$, we have
	\begin{align*}
		\left|D_{i,j_1}^{(\mu)}D_{i - k,j_2}^{(\mu)}\right|
			&= \left|D_{i,j_1}^{(\mathfrak{c})}D_{i - k,j_2}^{(\mathfrak{c})} + D_{i,j_1}^{(\mathfrak{c})}D_{i - k,j_2}^{(\mathfrak{s})} + D_{i,j_1}^{(\mathfrak{s})}D_{i - k,j_2}^{(\mathfrak{c})} + D_{i,j_1}^{(\mathfrak{s})}D_{i - k,j_2}^{(\mathfrak{s})}\right|\\
			&\leq 3\left(\left|D_{i,j_1}^{(\mathfrak{c})}\right|^2 + \left|D_{i - k,j_2}^{(\mathfrak{c})}\right|^2\right) + 3\left(\left|D_{i,j_1}^{(\mathfrak{s})}\right|^2 + \left|D_{i - k,j_2}^{(\mathfrak{s})}\right|^2\right).
	\end{align*}
	Hence we have
	\begin{align*}
		\left|\frac{1}{n}\sum_{i = 1+mh + k^+}^{n-k^-} D_{i,j_1}^{(\mu)}D_{i - k,j_2}^{(\mu)}\right|
		\leq \frac{3}{n}\sum_{i = 1+mh + k^+}^{n-k^-} \left\{\left(\left|D_{i,j_1}^{(\mathfrak{c})}\right|^2 + \left|D_{i - k,j_2}^{(\mathfrak{c})}\right|^2\right) + \left(\left|D_{i,j_1}^{(\mathfrak{s})}\right|^2 + \left|D_{i - k,j_2}^{(\mathfrak{s})}\right|^2\right)\right\}.
	\end{align*}
    Note that $D_{i,j_1}^{(\mathfrak{s})} = 0$ if there is no change point within the time interval $i - mh,\ldots,i$ for any $i$ and $j_1 = 1,\ldots,S$.
	Since there are $\mathcal{J}$ change points, at most $(mh + 1)\mathcal{J}$ elements in $\{D_{i,j_1}^{(\mathfrak{s})}\}_{i = mh + 1}^n$ are non-zero.
    By assumption, the non-zero $D_{i,j_1}^{(\mathfrak{s})}$ satisfies $\left|D_{i,j_1}^{(\mathfrak{s})}\right| \leq \sum_{j = 0}^m \left|d_{j}\right|\mathcal{S} = O(\mathcal{S})$ since $m < \infty$.
    Hence
	\begin{equation}\label{eqt:B3}
		\frac{3}{n}\sum_{i = 1+mh + k^+}^{n-k^-} \left(\left|D_{i,j_1}^{(\mathfrak{s})}\right|^2 + \left|D_{i - k,j_2}^{(\mathfrak{s})}\right|^2\right) \leq O\left(\frac{\mathcal{S}^2\mathcal{J}\ell}{n}\right).
	\end{equation}
	On the other hand, we have $\left|D_{i,j_1}^{(\mathfrak{c})}\right| \leq \mathcal{C}(mh + 1)/n = O(\mathcal{C}h/n)$ for any $i$. Hence
	\begin{equation}\label{eqt:B4}
		\frac{3}{n}\sum_{i = 1+mh + k^+}^{n-k^-} \left(\left|D_{i,j_1}^{(\mathfrak{c})}\right|^2 + \left|D_{i - k,j_2}^{(\mathfrak{c})}\right|^2\right) \leq O\left(\frac{\mathcal{C}^2 \ell^2}{n^2}\right).
	\end{equation}
	By combining (\ref{eqt:B3}) and (\ref{eqt:B4}), we have
	\begin{equation*}
		\left|\frac{1}{n}\sum_{i = 1+mh + k^+}^{n-k^-} D_{i,j_1}^{(\mu)}D_{i - k,j_2}^{(\mu)}\right| \leq O\left(\frac{\mathcal{S}^2\mathcal{J}\ell}{n}\right) + O\left(\frac{\mathcal{C}^2 \ell^2}{n^2}\right) = O\left\{\frac{\ell}{n}\left(\frac{\mathcal{C}^2 \ell}{n} + \mathcal{S}^2\mathcal{J}\right)\right\}.
	\end{equation*}
	Now we prove the case $m = 0$.
    By Proposition \ref{prop:B1}, we have
	\begin{align*}
		& \frac{1}{n}\sum_{i = 1+mh + k^+}^{n-k^-} D_{i,j_1}^{(\mu)}D_{i - k,j_2}^{(\mu)} \\
		&\qquad= \frac{1}{n}\sum_{i = 1+|k|}^{n} \left(\mu_{i-k^-,j_1} - \bar{\mu}_{n,j_1}\right)\left(\mu_{i - k^+,j_2} - \bar{\mu}_{n,j_2}\right)\\
		&\qquad= \frac{1}{n}\sum_{i = 1}^n \left(\mu_{i,j_1} - \bar{\mu}_{n,j_1}\right)\left(\mu_{i,j_2} - \bar{\mu}_{n,j_2}\right)\\
        &\qquad - \frac{1}{n}\sum_{i = 1+|k|}^{n} \left\{ (\mu_{i,j_1} - \mu_{i - k^-, j_1})(\mu_{i,j_2} - \bar{\mu}_{n,j_2}) + (\mu_{i - k^-, j_1} - \bar{\mu}_{n,j_1})(\mu_{i,j_2} - \mu_{i - k^+, j_2}) \right\}
        - r_k,
	\end{align*}
	where $r_0 = 0$, and for $k=1,\ldots,\ell$
	\begin{align*}
        r_k = r_{-k} &= \frac{1}{n}\sum_{i = 1}^{|k|} (\mu_{i,j_1} - \bar{\mu}_{n,j_1})(\mu_{i,j_2} - \bar{\mu}_{n,j_2})
        = O\left\{\frac{\ell}{n}(\mathcal{C} + \mathcal{SJ})^2\right\}.
	\end{align*}
	For $k = 1,\ldots,\ell$,
	\begin{align*}
        \left\vert \frac{1}{n}\sum_{i = |k| + 1}^n (\mu_{i, j_1} - \bar{\mu}_{n,j_1})(\mu_{i,j_2} - \mu_{i - k, j_2}) \right\vert
		&\leq \max_{1\leq i\leq n}|\mu_{i,j_1}-\bar{\mu}_{n,j_1}| \frac{1}{n}\sum_{i=k+1}^n |\mu_{i,j_2}-\mu_{i-k,j_2}| \\
		&\leq (\mathcal{C}+\mathcal{S}\mathcal{J}) \frac{k}{n}\sum_{i=2}^n |\mu_{i,j_2}-\mu_{i-1,j_2}| \\
		&\leq \frac{k}{n}(\mathcal{C}+\mathcal{S}\mathcal{J})^2\leq \mathcal{M}_0.
	\end{align*}
	On the other hand, we have $\sum_{i = 1}^n (\mu_{i,j_1} - \bar{\mu}_{n,j_1})(\mu_{i,j_2} - \bar{\mu}_{n,j_2})/n = \theta_{j_1,j_2} + O\{(\mathcal{C} + \mathcal{SJ})^2/n\}$.
    Hence,
	\begin{align*}
		\frac{1}{n}\sum_{i = 1+k^+}^{n-k^-} \left(\mu_{i,j_1} - \bar{\mu}_{n,j_1}\right)\left(\mu_{i - k,j_2} - \bar{\mu}_{n,j_2}\right)
			&= \theta_{j_1,j_2} + O\left\{\frac{(\mathcal{C} + \mathcal{SJ})^2}{n}+ \mathcal{M}_0\right\} \\
			&= \theta_{j_1,j_2} + O\left(\mathcal{M}_0 /\ell + \mathcal{M}_0\right) \\
			&= \theta_{j_1,j_2} + O\left(\mathcal{M}_0 \right)
	\end{align*}
	and the desired result follows.
\end{proof}

\subsection{Proposition \ref{prop:B3}}
\begin{proposition}\label{prop:B3}
    Let $m \geq 0$ be fixed, $|k| \lesssim \ell$ and $h \asymp \ell$, where $\ell$ satisfies $1/\ell + \ell/n = o(1)$.
    Let Assumptions \ref{assum:jumps} and \ref{assum:weak dependence} hold.
    For $j_1, j_2 = 1,\ldots,S$, let
    \begin{equation*}
        \left(\widehat{\gamma}^{(m)}_k\right)_{j_1,j_2}
        = \frac{1}{n}\sum_{i = mh + 1 + k_+}^{n-k_-} D_{i,j_1}D_{i - k,j_2},
    \end{equation*}
    where we recall that $k_- = \max(-k,0)$ and
    $k_+ = \max(k,0)$.
    Then, as $n \rightarrow \infty$,
    \begin{equation*}
        \E\left\{\left(\widehat{\gamma}^{(m)}_k\right)_{j_1,j_2}\right\} = \left(\gamma_k\right)_{j_1,j_2} + \theta_{j_1,j_2} \mathbb{1}(m = 0) + R_{1,1}, \qquad \Var\left\{\left(\widehat{\gamma}^{(m)}_k\right)_{j_1,j_2}\right\} = R_{1,2},
    \end{equation*}
    where
    \begin{align*}
        R_{1,1} &= -\frac{|k| + mh}{n}\left(\gamma_k\right)_{j_1,j_2} + \left(1 - \frac{|k| + mh}{n}\right)\sum_{s = 1}^m \delta_s\left\{\left(\gamma_{k + sh}\right)_{j_1,j_2} + \left(\gamma_{k - sh}\right)_{j_1,j_2}\right\} \\
       &\qquad +  O\left\{ \mathcal{M}_0\mathbb{1}(m = 0) +  \mathcal{M}_1\right\}, \\
        R_{1,2} &= {\color{black} O\left\{  \frac{\mathcal{M}_0\mathbb{1}(m=0)+\mathcal{M}_1+1}{n} \right\}}.
    \end{align*}
\end{proposition}
\begin{proof}[Proof of Proposition \ref{prop:B3}]
    We first prove the case for $m = 0$.
    Without loss of generality, we consider $k\geq 0$.
    For $k<0$, the results follow immediately as
    $(\widehat{\gamma}^{(m)}_{-k})_{j_1,j_2} = (\widehat{\gamma}^{(m)}_k)_{j_2,j_1}$.
	Notice that
	{\small
	\begin{align*}
		\left(\widehat{\gamma}_k^{(0)}\right)_{j_1,j_2}
			&= \frac{1}{n}\sum_{i = k + 1}^n \left(X_{i,j_1} - \bar{X}_{n,j_1}\right)\left(X_{i - k, j_2} - \bar{X}_{n,j_2}\right)\\
			&= \left\{\frac{1}{n}\sum_{i = k + 1}^n \left(\mu_{i, j_1} - \bar{\mu}_{n,j_1}\right)\left(\mu_{i - k, j_2} - \bar{\mu}_{n,j_2}\right) \right\}
				- \left\{\frac{1}{n}\sum_{i = k + 1}^n \left(\mu_{i, j_1} - \bar{\mu}_{n,j_1}\right)\left(Z_{i - k, j_2} - \bar{Z}_{n,j_2}\right) \right\}\\
				&\quad - \left\{\frac{1}{n}\sum_{i = k + 1}^n \left(Z_{i,j_1} - \bar{Z}_{n,j_1}\right)\left(\mu_{i - k, j_2} - \bar{\mu}_{n,j_2}\right)\right\}
				+ \left\{\frac{1}{n}\sum_{i = k + 1}^n \left(Z_{i,j_1} - \bar{Z}_{n,j_1}\right)\left(Z_{i - k, j_2} - \bar{Z}_{n,j_2}\right) \right\}\\
			&=: B_1 - B_2 - B_3 + B_4,
	\end{align*}}
	where $\bar{\mu}_{n,j_1} = \sum_{i = 1}^n \mu_{i, j_1}/n$ and $\bar{Z}_{n,j_1} = \sum_{i = 1}^n Z_{i,j_1}/n$.
	Since $\{Z_{i,j_1}\}$ is mean-zero for each $j_1 = 1,\ldots,S$, $\E\left(B_2\right) = \E\left(B_3\right) = 0$.
	Since $B_1$ is non-stochastic, $\E\left(B_1\right) = B_1$.
	So
	\begin{equation}\label{eqt:B5}
		\E\left\{\left(\widehat{\gamma}_k^{(0)}\right)_{j_1,j_2}\right\} = B_1 + \E\left(B_4\right).
	\end{equation}
	Note that
	\begin{align*}
		B_1
			&= \frac{1}{n}\sum_{i = k + 1}^n \left\{ \left(\mu_{i, j_1} - \bar{\mu}_{j_1}\right) + \left(\bar{\mu}_{j_1} - \bar{\mu}_{n,j_1}\right)\right\} \left\{\left(\mu_{i - k, j_2} - \bar{\mu}_{j_2}\right) + \left(\bar{\mu}_{j_2} - \bar{\mu}_{n,j_2}\right)\right\}\\
			&= \left\{ \frac{1}{n}\sum_{i = k + 1}^n \left(\mu_{i, j_1} - \bar{\mu}_{j_1}\right)\left(\mu_{i - k, j_2} - \bar{\mu}_{j_2}\right) \right\}
				+ \left\{\frac{1}{n}\sum_{i = k + 1}^n \left(\bar{\mu}_{j_1} - \bar{\mu}_{n,j_1}\right)\left(\mu_{i - k, j_2} - \bar{\mu}_{j_2}\right) \right\}\\
				&\qquad + \left\{\frac{1}{n}\sum_{i = k + 1}^n \left(\bar{\mu}_{j_2} - \bar{\mu}_{n,j_2}\right)\left(\mu_{i, j_1} - \bar{\mu}_{j_1}\right) \right\}
				+ \left\{ \frac{1}{n}\sum_{i = k + 1}^n \left(\bar{\mu}_{j_1} - \bar{\mu}_{n,j_1}\right)\left(\bar{\mu}_{j_2} - \bar{\mu}_{n,j_2}\right) \right\}\\
			&= B^{(1)}_1 + B^{(2)}_1 + B^{(3)}_1 + B^{(4)}_1.
	\end{align*}
	For sufficiently large $n$ and $j_1 = 1,\ldots,S$, we have
	\begin{align}
		|\bar{\mu}_{n,j_1} - \bar{\mu}_{j_1}|
			&\leq \frac{1}{n} \sum_{i = 1}^n \sup_{i - 1\leq nt \leq i} \left\vert \mu_{j_1}(t) - \mu_{j_1}(i/n)\right\vert \nonumber\\
			&\leq \frac{1}{n} \sum_{i = 1}^n \left\{ \sup_{i - 1\leq nt \leq i} \left\vert \mathfrak{c}_{j_1}(t) - \mathfrak{c}_{j_1}(i/n)\right\vert
						+ \sup_{i - 1\leq nt \leq i} \left\vert \mathfrak{s}_{j_1}(t) - \mathfrak{s}_{j_1}(i/n)\right\vert \right\}
			\leq \frac{\mathcal{C} + \mathcal{S}\mathcal{J}}{n}, \label{eqt:B6}
	\end{align}
	where we have used
	Assumption \ref{assum:jumps} to argue that
	the time interval $(i - 1)/n \leq t \leq i/n$ contains at most one jump.
	Similarly,
	\begin{align}\label{eqt:diff_incomplete_meanMu}
		\left\vert \frac{1}{n}\sum_{i = k + 1}^n \mu_{i - k, j_1}  - \bar{\mu}_{j_1}\right\vert \leq (\mathcal{C} + \mathcal{S}\mathcal{J})/n
	\end{align}
	for all $k\leq \ell = o(n)$.
	Besides, $\bar{\mu}_{j_1} = \int_0^1 \mathfrak{c}_{j_1}(t)\, \dd t + \int_0^1 \mathfrak{s}_{j_1}(t)\, \dd t = O(\mathcal{C} + \mathcal{S}\mathcal{J})$.
	So,
	\begin{align*}
		B^{(2)}_1
			=  \left(\bar{\mu}_{j_1} - \bar{\mu}_{n,j_1}\right)
				\left\{\left(  \frac{1}{n}\sum_{i = k + 1}^n \mu_{i - k, j_2} - \bar{\mu}_{j_2}\right) + \frac{k}{n}\bar{\mu}_{j_2} \right\}
			= O\left\{ \frac{\ell(\mathcal{C}+\mathcal{S}\mathcal{J})^2}{n^2} \right\}
			= O(\mathcal{M}_0/n).
	\end{align*}
	Similarly, $B^{(3)}_1 = O(\mathcal{M}_0/n)$.
	Using (\ref{eqt:B6}) again, we have
	\[
		B^{(4)}_1
			= O\{ (\mathcal{C} + \mathcal{S}\mathcal{J})^2/n^2 \}
			= O\{\mathcal{M}_0/(n\ell)\}.
	\]
	By Proposition \ref{prop:B2},
	$
		B^{(1)}_1 = \theta_{j_1,j_2} + O(\mathcal{M}_0+\mathcal{M}_1).
	$
	Hence, we have
	\begin{equation}\label{eqt:B7}
		B_1 = \theta_{j_1,j_2} + O(\mathcal{M}_0+\mathcal{M}_1),
	\end{equation}
	since $B_1^{(2)}$, $B_1^{(3)}$ and $B_1^{(4)}$ are of order $o(\mathcal{M}_0)$.
	It remains to calculate $\E\left(B_4\right)$.
	Note that
	\begin{align*}
		\E\left(B_4\right)
			&= \frac{1}{n}\sum_{i = k + 1}^n \E\left\{\left(Z_{i,j_1} - \bar{Z}_{n,j_1}\right)\left(Z_{i - k, j_2} - \bar{Z}_{n,j_2}\right)\right\}\\
			&= \frac{1}{n}\sum_{i = k + 1}^n \Cov\left(Z_{i,j_1} - \bar{Z}_{n,j_1},Z_{i - k, j_2} - \bar{Z}_{n,j_2}\right)\\
			&= \left\{ \frac{1}{n}\sum_{i = k + 1}^n \Cov\left(Z_{i,j_1},Z_{i - k, j_2}\right) \right\}
				- \left\{ \frac{1}{n}\sum_{i = k + 1}^n \Cov\left(Z_{i,j_1},\bar{Z}_{n,j_2}\right) \right\}\\
			&\qquad	- \left\{ \frac{1}{n}\sum_{i = k + 1}^n \Cov\left(\bar{Z}_{n,j_1},Z_{i - k, j_2}\right) \right\}
				+ \left\{ \frac{1}{n}\sum_{i = k + 1}^n \Cov\left(\bar{Z}_{n,j_1},\bar{Z}_{n,j_2}\right) \right\}\\
			&=: B^{(1)}_4 - B^{(2)}_4 - B^{(3)}_4 + B^{(4)}_4.
	\end{align*}
	Since Assumption \ref{assum:weak dependence} implies $\sum_{h \in \mathbb{Z}} |(\gamma_h)_{j_1,j_2}| < \infty$,
	we have
	\begin{align}\label{eqt:B8}
		B^{(2)}_4
		&= \frac{1}{n}\sum_{i = k + 1}^n \Cov\left(Z_{i,j_1},\frac{1}{n}\sum_{i' = 1}^n Z_{i',j_2}\right) \nonumber\\
		&= \frac{1}{n^2}\sum_{i = k + 1}^n \sum_{i' = 1}^n \Cov(Z_{i,j_1},Z_{i',j_2})
		= \frac{1}{n^2}\sum_{i = k + 1}^n \sum_{i' = 1}^n \left(\gamma_{i' - i}\right)_{j_1,j_2} \nonumber \\
		&\leq \frac{1}{n^2}\sum_{i = 1}^n \sum_{i' = 1}^n \left|\left(\gamma_{i' - i}\right)_{j_1,j_2}\right|
		= \frac{1}{n^2}\sum_{|h| \leq n - 1}(n - h)\left|\left(\gamma_h\right)_{j_1,j_2}\right|
		= \frac{1}{n}\sum_{|h| \leq n - 1} \frac{n - h}{n}\left|\left(\gamma_h\right)_{j_1,j_2}\right| \nonumber \\
		&\leq \frac{1}{n}\sum_{h \in \mathbb{Z}} \left|\left(\gamma_h\right)_{j_1,j_2}\right|
		= O\left(\frac{1}{n}\right).
	\end{align}
	Similarly, $B^{(3)}_4 = O(1/n)$ and $B^{(4)}_4 = O\{(n - k)/n^2\} = O(1/n)$.
	On the other hand, $B^{(1)}_4 = (1 - k/n)(\gamma_k)_{j_1,j_2}$.
	Hence
	\begin{equation}\label{eqt:B9}
		\E\left(B_4\right) = \left(1 - \frac{k}{n}\right)\left(\gamma_k\right)_{j_1,j_2} + O\left(\frac{1}{n}\right).
	\end{equation}
	Hence, putting (\ref{eqt:B7}) and (\ref{eqt:B9}) into (\ref{eqt:B5}), we have
	\begin{equation}\label{eqt:B10}
		\E\left\{\left(\widehat{\gamma}_k^{(0)}\right)_{j_1,j_2}\right\} = \left(1 - \frac{k}{n}\right)\left(\gamma_k\right)_{j_1,j_2} + \theta_{j_1,j_2} + O(\mathcal{M}_0 + \mathcal{M}_1).
	\end{equation}
	Now, we derive $\Var\{(\widehat{\gamma}_k^{(0)})_{j_1,j_2}\}$.
	Since $B_1$ is non-stochastic, we know that $\Var(B_1) = 0$ and $\Cov(B_1,B_{j}) = 0$ for $j = 2,3,4$.
	So, $\Var\{(\widehat{\gamma}_k^{(0)})_{j_1,j_2}\}$ can be written as
	\begin{equation*}
		\Var\left\{\left(\widehat{\gamma}_k^{(0)}\right)_{j_1,j_2}\right\} = \sum_{i = 2}^4 \Var\left(B_{i}\right) + 2\left\{\Cov\left(B_2,B_3\right) + \Cov\left(B_2,B_4\right) + \Cov\left(B_3,B_4\right)\right\}.
	\end{equation*}
	{\color{black}By Proposition \ref{prop:B2} and
	the assumption $\sum_{k'\in\mathbb{Z}} |(\gamma_{k'})_{j_2,j_2}| <\infty$, we know that }
	\begin{align}
		\Var\left(B_2\right)
		&= \frac{1}{n^2}\sum_{i = k + 1}^n \sum_{i' = k + 1}^n (\mu_{i, j_1} - \bar{\mu}_{n,j_1})(\mu_{i',j_1} - \bar{\mu}_{n,j_1})\Cov\left(Z_{i - k, j_2} - \bar{Z}_{n,j_2},Z_{i' - k,j_2} - \bar{Z}_{n,j_2}\right) \nonumber\\
		&= \frac{1}{n^2}\sum_{i = k + 1}^n \sum_{i' = k + 1}^n (\mu_{i, j_1} - \bar{\mu}_{n,j_1})(\mu_{i',j_1} - \bar{\mu}_{n,j_1})\left\{\left(\gamma_{i' - i}\right)_{j_2,j_2} + O\left(\frac{1}{n}\right)\right\}\nonumber\\
		&=\frac{1}{n^2}\sum_{|k'| \leq n-(k+1)}  \sum_{i = k + 1}^n {(\mu_{i, j_1} - \bar{\mu}_{n,j_1}) (\mu_{i + |k'|,j_1} - \bar{\mu}_{n,j_1})}
					(\gamma_{k'})_{j_2,j_2} \nonumber\\
		&\leq\frac{1}{n^2} n\left\{\theta+O(\mathcal{M}_0+\mathcal{M}_1  )\right\}
				\sum_{|k'| \leq n-(k+1)} |(\gamma_{k'})_{j_2,j_2}|
			= {\color{black} O\{(\mathcal{M}_0+\mathcal{M}_1+1)/n)\}}. \label{eqt:orderVarB2}
	\end{align}
	Similarly, $\Var\left(B_3\right) = {\color{black} O\{(\mathcal{M}_0+\mathcal{M}_1+1)/n)\}}$.
	By inequality (A.5) of \citet{WuZaffaroni2018} and the finiteness of $\Omega_{j_1,4}$ and $\Omega_{j_2,4}$,
	\begin{align}\label{eqt:B11}
		\Var\left(B_4\right)
			= \left\lVert \sum_{i,i' = 1}^n c_{i,i'}\left\{\left(Z_{i,j_1} - \bar{Z}_{n,j_1}\right)\left(Z_{i',j_2} - \bar{Z}_{n,j_2}\right) - \left(\gamma_{i - i'}\right)_{j_1,j_2}\right\} \right\rVert_2^2
            = O\left(\frac{1}{n}\right),
	\end{align}
	where $c_{i,i'} = \mathbb{1}(i - i' = k)/n$.
	By Cauchy--Schwarz inequality, for $i,j \in\{2,3\}$ and $i \neq j$,
	\begin{align*}
		\Cov\left(B_{i},B_{j}\right) &= {\color{black} O\{(\mathcal{M}_0+\mathcal{M}_1+1)/n)\}}, \\
		\Cov(B_{i},B_4) &= {\color{black} O\{(\mathcal{M}_0+\mathcal{M}_1+1)^{1/2}/n)\}} .
	\end{align*}
	Hence,
	\[
		\Var\left\{\left(\widehat{\gamma}_k^{(0)}\right)_{j_1,j_2}\right\}
		= {\color{black}O\left( \frac{\mathcal{M}_0 +\mathcal{M}_1 +1}{n}\right)}.
	\]

	Now we consider the case $m > 0$.
	Define, for $j_1 = 1,\ldots,S$,
	\begin{equation*}
		D^{(\mu)}_{i,j_1} = \sum_{j = 0}^m d_{j}\mu_{i - hj}^{[j_1]}, \quad
		D^{(Z)}_{i,j_1} = \sum_{j = 0}^m d_{j}Z_{i - hj}^{[j_1]}, \quad
		D^{(\mathfrak{c})}_{i,j_1} = \sum_{j = 0}^m d_{j}\mathfrak{c}_{i - hj}^{[j_1]}, \quad
		D^{(\mathfrak{s})}_{i,j_1} = \sum_{j = 0}^m d_{j}\mathfrak{s}_{i - hj}^{[j_1]}.
	\end{equation*}
	for each $i$.
	Notice that
	\begin{align*}
		\left(\widehat{\gamma}_{k}^{(m)}\right)_{j_1,j_2}
			&= \frac{1}{n}\sum_{i = mh + k + 1}^n D_{i,j_1}D_{i - k,j_2}\\
			&= \left\{ \frac{1}{n}\sum_{i = mh + k + 1}^n D^{(\mu)}_{i,j_1}D^{(\mu)}_{i - k, j_2} \right\}
				+ \left\{ \frac{1}{n}\sum_{i = mh + k + 1}^n D^{(\mu)}_{i,j_1}D^{(Z)}_{i - k, j_2} \right\}\\
				&\qquad
				+ \left\{\frac{1}{n}\sum_{i = mh + k + 1}^n D^{(Z)}_{i,j_1}D^{(\mu)}_{i - k, j_2} \right\}
				+ \left\{ \frac{1}{n}\sum_{i = mh + k + 1}^n D^{(Z)}_{i,j_1}D^{(Z)}_{i - k, j_2} \right\}\\
			&=: B_5 + B_6 + B_7 + B_8.
	\end{align*}
	Since $\{Z_{i}\}$ is mean-zero $\E\left(B_6\right) = \E\left(B_7\right) = 0$.
	Since $B_5$ is non-stochastic, $\E\left(B_5\right) = B_5$.
	So
	\begin{align}\label{eqt:B12}
		\E\left\{\left(\widehat{\gamma}_k^{(m)}\right)_{j_1,j_2}\right\} = B_5 + \E\left(B_8\right).
	\end{align}
	Note that
	\begin{align}\label{eqt:B13}
		\E\left(B_8\right)
			&= \frac{1}{n}\sum_{i = mh + k + 1}^n \E\left\{\left(\sum_{j = 0}^m d_{j}Z_{i - hj,j_1}\right)\left(\sum_{j' = 0}^m d_{j'}Z_{i - k - hj',j_2}\right)\right\} \nonumber\\
			&= \frac{1}{n}\sum_{i = mh + k + 1}^n \sum_{j = 0}^m \sum_{j' = 0}^m d_{j}d_{j'}\E\left(Z_{i - hj,j_1}Z_{i - k - hj',j_2}\right) \nonumber\\
			&= \frac{1}{n}\sum_{i = mh + k + 1}^n \sum_{j = 0}^m \sum_{j' = 0}^m d_{j}d_{j'}\left(\gamma_{h(j - j') - k}\right)_{j_1,j_2} \nonumber\\
			&= \frac{n - mh - k}{n}\sum_{s = -m}^m \delta_s\left(\gamma_{sh - k}\right)_{j_1,j_2}.
	\end{align}
	By Proposition \ref{prop:B2} for $m>0$,
	\begin{equation}\label{eqt:B14}
		B_5 = {\color{black}O(\mathcal{M}_1)}.
	\end{equation}
	Hence, putting (\ref{eqt:B13}) and (\ref{eqt:B14}) into (\ref{eqt:B12}),
	we obtain
	\begin{equation}\label{eqt:B15}
		\E\left\{\left(\widehat{\gamma}_k^{(m)}\right)_{j_1,j_2}\right\}
			= {\color{black}\left(1 - \frac{k + mh}{n}\right)\sum_{s = -m}^m \delta_s \left(\gamma_{k + sh}\right)_{j_1,j_2} + O(\mathcal{M}_1)}.
	\end{equation}
	Next we derive the value of $\Var\{(\widehat{\gamma}_k^{(m)})_{j_1,j_2}\}$ for $m > 0$.
	Since $B_5$ is non-stochastic, we know that $\Var\left(B_5\right) = 0$ and $\Cov\left(B_5,B_{j}\right) = 0$ for $j = 6,7,8$.
	So, we have
	\begin{equation*}
		\Var\left\{\left(\widehat{\gamma}_k^{(m)}\right)_{j_1,j_2}\right\} = \sum_{i = 6}^8 \Var\left(B_{i}\right) + 2\left\{\Cov\left(B_6,B_7\right) + \Cov\left(B_6,B_8\right) + \Cov\left(B_7,B_8\right)\right\}.
	\end{equation*}
	{\color{black}Similar to the derivation in (\ref{eqt:orderVarB2}), we have }
	\begin{align*}
		\Var\left(B_6\right)
			&= \frac{1}{n^2}\sum_{i = mh + k + 1}^n \sum_{i' = mh + k + 1}^n D_{i,j_1}^{(\mu)}D_{i',j_1}^{(\mu)}\Cov\left(D_{i - k, j_2}^{(Z)},D_{i' - k, j_2}^{(Z)}\right) \\
			&= \frac{1}{n^2}\sum_{i = mh + k + 1}^n \sum_{i' = mh + k + 1}^n D_{i,j_1}^{(\mu)}D_{i',j_1}^{(\mu)} (\gamma_{i'-i})_{j_2,j_2} \\
			&=\frac{1}{n^2}\sum_{|k'| \leq n-(mh+k+1)}  \sum_{i = mh + k + 1}^n D_{i,j_1}^{(\mu)}D_{i+|k'|,j_1}^{(\mu)}
					(\gamma_{k'})_{j_2,j_2}  \\
			&\leq\frac{1}{n^2} O(n\mathcal{M}_1) \sum_{|k'| \leq n-(mh+k+1)} |(\gamma_{k'})_{j_2,j_2}|
			= {\color{black} O(\mathcal{M}_1/n)}.
	\end{align*}
	where the last line is followed from Proposition \ref{prop:B2} and
	the assumption $\sum_{k'\in\mathbb{Z}} |(\gamma_{k'})_{j_2,j_2}| <\infty$.
	Similarly, $\Var\left(B_7\right) = {\color{black} O(\mathcal{M}_1/n)}$.
	By a similar argument as in (\ref{eqt:B11}), we have $\Var\left(B_8\right) = O(1/n)$.
	By Cauchy--Schwarz inequality, $\Cov\left(B_{i},B_{j}\right) = {\color{black} O(\mathcal{M}_1/n)}$ and
	$\Cov(B_{i},B_8) = {\color{black}O(\mathcal{M}_1^{1/2}/n)}$ for $i,j \in\{6,7\}$ and $i \neq j$.
	Hence,
	\[
		\Var\left\{\left(\widehat{\gamma}_k^{(m)}\right)_{j_1,j_2}\right\}
			= {\color{black}O\left( \frac{\mathcal{M}_1 +1}{n}\right)}.
	\]
	Thus, the results follow.
\end{proof}

\subsection{Lemma \ref{lem:B4}}
Recall the definition of $\widetilde{\theta}_{j_1,j_2}^{(\mu,\mu)}$ in (\ref{eqt:tildetheta1}):
\[
	\widetilde{\theta}_{j_1,j_2}^{(\mu,\mu)}
    	= \sum_{1 \leq i,i' \leq n} a_{n,i - i'} (\mu_{i,j_1} - \bar{\mu}_{n,j_1})(\mu_{i',j_2} - \bar{\mu}_{n,j_2}) .
\]

\begin{lemma}\label{lem:B4}
    Suppose the conditions in Theorem \ref{thm:4.3} hold.
    Let
    $j_1, j_2\in \{ 1,\ldots,S \}$.
	Then
	$
		\color{black}
		\widetilde{\theta}_{j_1,j_2}^{(\mu,\mu)} = \theta_{j_1,j_2} + O(\mathcal{M}_0)
	$
	as $n \rightarrow \infty$.
\end{lemma}
\begin{proof}[Proof of Lemma \ref{lem:B4}]
	Recall the definition of $a_{n,k}$ in (\ref{eqt:tildetheta}).
    Note that {\footnotesize
	\begin{align*}
		\widetilde{\theta}_{j_1,j_2}^{(\mu,\mu)}
		&=\frac{1}{n\ell \kappa_\ell}\sum_{|k| \leq \ell + mh} \left\{K\left(\frac{k}{\ell}\right) - K_{m}\left(\frac{k}{\ell}\right)\right\}\sum_{i = |k| + 1}^n \left(\mu_{i - |k|, j_1} - \bar{\mu}_{n,j_1}\right)\left(\mu_{i,j_2} - \bar{\mu}_{n,j_2}\right)\\
		&= \frac{1}{n\ell \kappa_\ell}\sum_{|k| \leq \ell + mh} \left\{K\left(\frac{k}{\ell}\right) - K_{m}\left(\frac{k}{\ell}\right)\right\}\sum_{i = |k| + 1}^n \left(\mu_{i,j_2} - \bar{\mu}_{n,j_2}\right)\left\{\left(\mu_{i, j_1} - \bar{\mu}_{n,j_1}\right) + \left(\mu_{i - |k|, j_1} - \mu_{i, j_1}\right)\right\}\\
		&= \bigg[
			\frac{1}{n\ell \kappa_\ell}\sum_{|k| \leq \ell + mh} \left\{K\left(\frac{k}{\ell}\right) - K_{m}\left(\frac{k}{\ell}\right)\right\} \\
		& \qquad\qquad\qquad\qquad \times\sum_{i = 1}^n \left\{\left(\mu_{i, j_1} - \bar{\mu}_{n,j_1}\right)\left(\mu_{i,j_2} - \bar{\mu}_{n,j_2}\right) + \left(\mu_{i,j_2} - \bar{\mu}_{n,j_2}\right)\left(\mu_{i - |k|, j_1} - \mu_{i, j_1}\right)\right\}
			\bigg]\\
		&\quad - \bigg[\frac{1}{n\ell \kappa_\ell}\sum_{|k| \leq \ell + mh} \left\{K\left(\frac{k}{\ell}\right) - K_{m}\left(\frac{k}{\ell}\right)\right\}\\
		&\qquad\qquad\qquad\qquad \times \sum_{i = 1}^{|k|} \left\{\left(\mu_{i, j_1} - \bar{\mu}_{n,j_1}\right)\left(\mu_{i,j_2} - \bar{\mu}_{n,j_2}\right) + \left(\mu_{i,j_2} - \bar{\mu}_{n,j_2}\right)\left(\mu_{i - |k|, j_1} - \mu_{i, j_1}\right)\right\}\bigg]\\
		&=: B_9 - B_{10}.
	\end{align*}}
	For $B_9$, we have
    {\small
	\begin{align*}
		B_9 &= \frac{1}{n\ell \kappa_\ell}\sum_{|k| \leq \ell + mh} \left\{K\left(\frac{k}{\ell}\right) - K_{m}\left(\frac{k}{\ell}\right)\right\}\\
		&\qquad \times \sum_{i = 1}^n \left\{\left(\mu_{i, j_1} - \bar{\mu}_{n,j_1}\right)\left(\mu_{i,j_2} - \bar{\mu}_{n,j_2}\right) + \left(\mu_{i,j_2} - \bar{\mu}_{n,j_2}\right)\left(\mu_{i - |k|, j_1} - \mu_{i, j_1}\right)\right\}\\
		&= \left\{\frac{1}{n}\sum_{i = 1}^n \left(\mu_{i, j_1} - \bar{\mu}_{j_1}\right)\left(\mu_{i,j_2} - \bar{\mu}_{j_2}\right) \right\}
			- \left(\bar{\mu}_{j_1} - \bar{\mu}_{n,j_1}\right)\left(\bar{\mu}_{j_2} - \bar{\mu}_{n,j_2}\right) \\
		&\qquad 	+ \left[ \frac{1}{n\ell \kappa_\ell}\sum_{|k| \leq \ell + mh} \left\{K\left(\frac{k}{\ell}\right) - K_{m}\left(\frac{k}{\ell}\right)\right\}\sum_{i = 1}^n \left(\mu_{i,j_2} - \bar{\mu}_{n,j_2}\right)\left(\mu_{i - |k|, j_1} - \mu_{i, j_1}\right) \right]\\
		&=: B_9^{(1)} - B_9^{(2)} + B_9^{(3)}.
	\end{align*}}
	By Proposition \ref{prop:B2} and (\ref{eqt:B6}), we have
	\begin{align*}
		B_9^{(1)} = \theta_{j_1,j_2} + {\color{black}O(\mathcal{M}_0) }
		\qquad \text{and} \qquad
		B_9^{(2)} = O\left\{\left(\frac{\mathcal{C} + \mathcal{SJ}}{n}\right)^2\right\}
		= \color{black}O\left( \frac{\mathcal{M}_0}{n\ell} \right).
	\end{align*}
	For $B_9^{(3)}$, we have
	\begin{align*}
		\left|B_9^{(3)}\right| &\leq \frac{1}{n\ell \kappa_\ell}\sum_{|k| \leq \ell + mh}
			\left\vert
				K\left(\frac{k}{\ell}\right) - K_{m}\left(\frac{k}{\ell}\right)
			\right\vert \\
            &\qquad
			\times \sum_{i = 1}^n \left\vert\mu_{i,j_2} - \bar{\mu}_{n,j_2}\right\vert
			\left\{
				\frac{\mathcal{C}|k|}{n}
				+ \sum_{j = 1}^\mathcal{J} \left|\xi_{j} - \xi_{j - 1}\right|_\infty
				\mathbb{1}\left( i - |k|\leq T_{j} \leq i\right)
			\right\} \\
		&\leq  \frac{\left(2\ell + 1\right)
			K_{\mathsf{max}}}{n\ell \kappa_\ell}\sum_{i = 1}^n
			\left\vert\mu_{i,j_2} - \bar{\mu}_{n,j_2}\right\vert\left\{\frac{\mathcal{C}\ell}{n}
			+ \sum_{j = 1}^\mathcal{J} \left|\xi_{j} - \xi_{j - 1}\right|_\infty
				\mathbb{1}\left( i - |\ell + mh|\leq T_{j} \leq i\right)
			\right\} \\
		&= O(\mathcal{M}_0),
	\end{align*}
	where $K_{\mathsf{max}} = \max_{|k| \leq \ell + mh} \left|K(k/\ell) - K_{m}(k/\ell)\right| <\infty$.
	On the other hand, following similar calculations as above, we have
	$
		B_{10} = O(\mathcal{M}_0).
	$
	The desired result then follows.
\end{proof}

\subsection{Lemma \ref{lem:B6}}
Recall the definition of $\widetilde{\theta}_{j_1,j_2}^{(Z,Z)}$ in (\ref{eqt:tildetheta2}):
\[
	\widetilde{\theta}_{j_1,j_2}^{(Z,Z)}
    	= \sum_{1 \leq i,i' \leq n} a_{n,i - i'} (Z_{i,j_1} - \bar{Z}_{n,j_1})(Z_{i',j_2} - \bar{Z}_{n,j_2}) .
\]
\begin{lemma}\label{lem:B6}
    Suppose the conditions in Theorem \ref{thm:4.3} hold.
    Let $j_1, j_2\in \{ 1,\ldots,S \}$.
    Then $\widetilde{\theta}_{j_1,j_2}^{(Z,Z)} = O_p(1/\sqrt{n\ell})$ as $n\rightarrow\infty$.
\end{lemma}
\begin{proof}[Proof of Lemma \ref{lem:B6}]
	First, when $X_i = Z_i$ for all $i$, we have
	$\widehat{\theta}_{j_1,j_2}^{(Z,Z)} = \widehat{\theta}_{j_1,j_2}$ and $\theta_{j_1,j_2}=0$.
	By Theorem \ref{thm:B9}, we obtain
	\begin{align}\label{eqt:hatthetaZZ_exp}
		\E\left(\widehat{\theta}_{j_1,j_2}^{ZZ}\right)
		=
		o\left(\frac{1}{\ell^{q + 1}}\right) .
	\end{align}
	Next, we plan to use Lemma \ref{lem:B11} and verify the conditions needed.
    Once again, recall the definition of $a_{n,k}$ in (\ref{eqt:tildetheta}).
    So, in this case,
    \begin{align}
		\zeta_n^2
		&= \sum_{k=1}^n a_{n,k}^2
		= O\left(\frac{1}{n^2\ell \kappa_\ell^2}\right), \nonumber\\
		\sigma_n^2
		&= 2\sum_{i=1}^n\sum_{i'=1}^n a_{n,i'-i}^2
		= 2\left\{na_{n,0}^2 + 2\sum_{i'=1}^{n-1} (n-i')a_{n,i'}^2\right\}
		= O\left(\frac{n\ell}{n^2\ell^2 \kappa_\ell^2}\right)
        = O\left(\frac{1}{n\ell}\right), \label{eqt:orderSigman}
	\end{align}
	where we have used the fact that $\kappa_\ell =  \kappa + o(1)$, where $\kappa\in\mathbb{R}$.
	Therefore, we have
	\begin{align*}
		\max_{0 \leq k \leq n} a_{n,k}^2
			&= \frac{1}{n^2\ell^2 \kappa_\ell^2} = o\left(\frac{1}{n^2\ell \kappa_\ell^2}\right) = o(\zeta_n^2);\\
		n\zeta_n^2
			&= O\left(\frac{1}{n\ell \kappa_\ell^2}\right) = O(\sigma_n^2); \\
		\sum_{k=1}^n\sum_{t=1}^{k-1}\left(\sum_{i=k+1}^n a_{n,k-i}a_{n,t-i}\right)^2
			&= \sum_{k=1}^n\sum_{i,i'=k+1}^n\sum_{t=1}^{k-1} a_{n,k-i}a_{n,k-i'}a_{n,t-i}a_{n,t-i'} \nonumber\\
			&= \sum_{k=1}^n\sum_{i,i'=1}^{n-k}\sum_{t=1-k}^{-1} a_{n,-i}a_{n,-i'}a_{n,t-i}a_{n,t-i'}  \nonumber \\
			&= \sum_{k=1}^n\sum_{i,i'=1}^{n-k} a_{n,-i}a_{n,-i'}O\left(\frac{1}{n^2\ell \kappa_\ell^2}\right) \nonumber\\
			&= O\left\{n\left(\frac{1}{n^2\ell \kappa_\ell^2}\right)^2  \right\}
			= O\left(\frac{1}{n^3\ell^2 \kappa_\ell^4}\right)
			= o\left(\frac{1}{n^2\ell^2 \kappa_\ell^4}\right)
			= o(\sigma_n^4);\\
		\sum_{i=1}^n (a_{n,i} - a_{n,i-1})^2
			&= O\left(\frac{1}{n^2\ell^2 \kappa_\ell^2}\right) = o\left(\frac{1}{n^2\ell \kappa_\ell^2}\right) = o(\zeta_n^2).
		\end{align*}
    Hence, by Lemma \ref{lem:B11},
    $
    	\{\widetilde{\theta}_{j_1,j_2}^{(Z,Z)}- \E(\widetilde{\theta}_{j_1,j_2}^{(Z,Z)})\}/\sigma_n = O_p(1).
    $
    Together with (\ref{eqt:hatthetaZZ_exp}) and (\ref{eqt:orderSigman}),
    we have  $\widetilde{\theta}_{j_1,j_2}^{(Z,Z)} = O_p(1/\sqrt{n\ell})$.
\end{proof}

\subsection{Lemma \ref{lem:B8}}
Recall the definition of
$\widetilde{\theta}_{j_1,j_2}^{(\mu,Z)}$
and
$\widetilde{\theta}_{j_1,j_2}^{(Z,\mu)}$
in (\ref{eqt:tildetheta3}):
\[
	\widetilde{\theta}_{j_1,j_2}^{(\mu,Z)}+\widetilde{\theta}_{j_1,j_2}^{(Z,\mu)}
	     =   \sum_{1 \leq i,i' \leq n} a_{n,i - i'} \left\{(\mu_{i',j_2} - \bar{\mu}_{n,j_2})(Z_{i,j_1} - \bar{Z}_{n,j_1})
				+ (\mu_{i,j_1} - \bar{\mu}_{n,j_1})(Z_{i',j_2} - \bar{Z}_{n,j_2}) \right\},
\]
and the definition of $\widetilde{L}^{(\mu,Z)}+\widetilde{L}^{(Z,\mu)}$ in (\ref{eqt:Tj1j2diffm_3terms}):
    \begin{align*}
		\widetilde{L}^{(\mu,Z)}+\widetilde{L}^{(Z \mu)}
		= \sum_{(j_1,j_2)\in\mathbb{J}} c_{j_1,j_2} \widetilde{\theta}_{j_1,j_2}^{(\mu,Z)} ,
    \end{align*}
where $\mathbb{J} = \{(1,1), (2,2), (3,3), (4,4), (1,2), (2,3), (3,4)\}$.

\begin{lemma}\label{lem:B8}
    Suppose the conditions in Theorem \ref{thm:4.3} hold.
    Let $c_{j,j'}\in \mathbb{R}$ for $(j,j')\in\mathbb{J}$ be any fixed constants.
	If $\mathcal{M}_{\mu Z} = o(1/\sqrt{n})$, where $\mathcal{M}_{\mu Z}$ is defined in (\ref{eqt:defM0M1}),
    then as $n \rightarrow \infty$
    \begin{align}\label{eqt:CLT_vtildemuZ}
        \sqrt{n}\{\widetilde{L}^{(\mu,Z)} +\widetilde{L}^{(Z,\mu)} \}
        	\inD \Normal\left(0,\widetilde{v}_{\mu Z}\right),
    \quad \text{where} \quad
    	\widetilde{v}_{\mu Z} \sum_{(j_1,j_2)\in\mathbb{J}}\sum_{(j_3,j_4)\in\mathbb{J}} c_{j_1,j_2} c_{j_3,j_4} \Omega_{\theta,(j_1,j_2),(j_3,j_4)},
    \end{align}
    and
    $\Omega_{\theta,(j_1,j_2),(j_3,j_4)}$ is defined in (\ref{eqt:Omegathetaj1234}).
\end{lemma}
\begin{proof}[Proof of Lemma \ref{lem:B8}]
    Let
    \[
    	\widetilde{c}_{11} = c_{11}, \quad
		\widetilde{c}_{22} = c_{22}/2, \quad
		\widetilde{c}_{33} = c_{33}/2, \quad
    	\widetilde{c}_{44} = c_{44}, \quad
		\widetilde{c}_{12} = c_{12}, \quad
		\widetilde{c}_{23} = c_{23}, \quad
		\widetilde{c}_{34} = c_{34}.
	\]
    Note that we can write
    {\small
    \begin{align}
        &\widetilde{L}^{(\mu,Z)}+\widetilde{L}^{(Z,\mu)}\\
        &\qquad= 2\widetilde{c}_{11}\sum_{1 \leq i,i' \leq n} a_{n,i - i'}(\mu_{i,1} - \bar{\mu}_{n,1})(Z_{i',1} - \bar{Z}_{n,1}) \nonumber\\
        &\qquad\quad+ 2\widetilde{c}_{44}\sum_{1 \leq i,i' \leq n} a_{n,i - i'}(\mu_{i,4} - \bar{\mu}_{n,4})(Z_{i',4} - \bar{Z}_{n,4})\nonumber\\
        &\qquad\quad + 4\sum_{r = 2}^3 \widetilde{c}_{rr}\sum_{1 \leq i,i' \leq n} a_{n,i - i'}(\mu_{i,r} - \bar{\mu}_{n,r})(Z_{i',r} - \bar{Z}_{n,r})\nonumber\\
        &\qquad\quad + \sum_{r = 1}^3 \widetilde{c}_{r,r + 1}\sum_{1 \leq i,i' \leq n} a_{n,i - i'}\{(\mu_{i,r} - \bar{\mu}_{n,r})(Z_{i',r + 1} - \bar{Z}_{n,r + 1}) + (\mu_{i,r + 1} - \bar{\mu}_{n,r + 1})(Z_{i',r} - \bar{Z}_{n,r})\}\nonumber\\
        &\qquad= \sum_{w = 1}^3 \sum_{r = w}^{w + 1} \sum_{s = r}^{w + 1} \widetilde{c}_{rs}\frac{1}{n\ell \kappa_\ell}\sum_{|k| \leq \ell + mh} \left\{K\left(\frac{k}{\ell}\right) - K_{m}\left(\frac{k}{\ell}\right)\right\}\sum_{i = |k| + 1}^n \left(\mu_{i - |k|,r} - \bar{\mu}_{n,r}\right)\left(Z_{i,s} - \bar{Z}_{n,s}\right)\nonumber\\
        &\qquad\quad + \sum_{w = 1}^3 \sum_{r = w}^{w + 1} \sum_{s = r}^{w + 1} \widetilde{c}_{rs}\frac{1}{n\ell \kappa_\ell}\sum_{|k| \leq \ell + mh} \left\{K\left(\frac{k}{\ell}\right) - K_{m}\left(\frac{k}{\ell}\right)\right\}\sum_{i = |k| + 1}^n \left(\mu_{i - |k|,s} - \bar{\mu}_{n,s}\right)\left(Z_{i,r} - \bar{Z}_{n,r}\right)\nonumber\\
        &\qquad= \sum_{w = 1}^3 \sum_{r = w}^{w + 1} \sum_{s = r}^{w + 1} \widetilde{c}_{rs}\left\{\frac{-2}{n\ell \kappa_\ell}\sum_{k = \ell + 1}^{\ell + mh} K_{m}\left(\frac{k}{\ell}\right) \sum_{i = |k| + 1}^n Z_{i,s}\left(\mu_{i - |k|,r} - \bar{\mu}_{n,r}\right)\right\}\nonumber\\
        &\qquad\quad + \sum_{w = 1}^3 \sum_{r = w}^{w + 1} \sum_{s = r}^{w + 1} \widetilde{c}_{rs}\left\{\frac{-2}{n\ell \kappa_\ell}\sum_{k = \ell + 1}^{\ell + mh} K_{m}\left(\frac{k}{\ell}\right) \sum_{i = |k| + 1}^n Z_{i,r}\left(\mu_{i - |k|,s} - \bar{\mu}_{n,s}\right)\right\} + O_p\left\{\frac{(\mathcal{C} + \mathcal{SJ})}{n^{3/2}}\right\} \label{eqt:OpCSJ_remidnderTerm}\\
        &\qquad=: B_{12}^{(1)} + B_{12}^{(2)} +
        O_p\left(\frac{\mathcal{C} + \mathcal{SJ}}{n^{3/2}}\right), \nonumber
    \end{align}
    }
    where we have used the property that $\bar{Z}_{n,j} = O_p(1/\sqrt{n})$ for each $j$,
    the bounds in (\ref{eqt:B6})--(\ref{eqt:diff_incomplete_meanMu}).
    The detailed calculation of
    the $O_p$-term in (\ref{eqt:OpCSJ_remidnderTerm})
    is straightforward but technical.
    We defer it to Remark \ref{OpCSJ_remidnderTerm}
    at the end of this proof.

    Using the Riemann approximation
    $\bar{\mu}_{n,s} - \bar{\mu}_{s} = O\{(\mathcal{C} + \mathcal{SJ})/n\}$ (see (\ref{eqt:B6}) for the proof),
    we have
    \begin{align*}
        B_{12}^{(1)} &= \sum_{w = 1}^3 \sum_{r = w}^{w + 1} \sum_{s = r}^{w + 1} \widetilde{c}_{rs}\left\{\frac{-2}{n\ell \kappa_\ell}\sum_{i = 1}^n Z_{i,s}\sum_{k = \ell + 1}^{(\ell + mh) \wedge (i - 1)} K_{m}\left(\frac{k}{\ell}\right)\left(\mu_{i - |k|,r} - \bar{\mu}_{r}\right)\right\} + O_p\left(\frac{\mathcal{C} + \mathcal{SJ}}{n^{3/2}}\right),
        \\
        B_{12}^{(2)} &= \sum_{w = 1}^3 \sum_{r = w}^{w + 1} \sum_{s = r}^{w + 1} \widetilde{c}_{rs}\left\{\frac{-2}{n\ell \kappa_\ell}\sum_{i = 1}^n Z_{i,r}\sum_{k = \ell + 1}^{(\ell + mh) \wedge (i - 1)} K_{m}\left(\frac{k}{\ell}\right)\left(\mu_{i - |k|,s} - \bar{\mu}_{s}\right)\right\} + O_p\left(\frac{\mathcal{C} + \mathcal{SJ}}{n^{3/2}}\right),
    \end{align*}
    where the $O_p$ terms in the above two equations
    follow from the calculations below:
    \begin{align*}
        &\left| \frac{-2}{n\ell \kappa_\ell}\sum_{i = 1}^n Z_{i,s}\sum_{k = \ell + 1}^{(\ell + mh) \wedge (i - 1)} K_{m}\left(\frac{k}{\ell}\right)\left(\bar{\mu}_{n,r} - \bar{\mu}_{r}\right) \right|\\
        &\qquad\leq 2\left| \bar{\mu}_{n,r} - \bar{\mu}_{r} \right| \left| \frac{1}{n}\sum_{i = 1}^n Z_{i,s} \right| \sup_{i = 1,\ldots,n}\left| \frac{1}{\ell \kappa_\ell}\sum_{k = \ell + 1}^{(\ell + mh) \wedge (i - 1)} K_{m}\left(\frac{k}{\ell}\right) \right|\\
        &\qquad= O\left(\frac{\mathcal{C} + \mathcal{SJ}}{n}\right) \times O_p\left(\frac{1}{\sqrt{n}}\right) \times O(1)\\
        &\qquad= O_p\left(\frac{\mathcal{C} + \mathcal{SJ}}{n^{3/2}}\right),
    \end{align*}
    since $\bar{Z}_{n,s} = O_p(1/\sqrt{n})$, $\ell \kappa_\ell \asymp \ell$ and $K_m$ is bounded with support of order $\ell$, the inner sum is $O(\ell)$ uniformly in $i$, implying the supremum is $O(1)$.
    Denote $Q_{i} = (Z_{i,1},Z_{i,2},Z_{i,3},Z_{i,4})^{\T}$ and $b_i = (b_{1i},b_{2i},b_{3i},b_{4i})^{\T}$,
    where
    {\small
	\begin{align*}
	        b_{1i} &= \frac{2\widetilde{c}_{11}}{\ell \kappa_\ell}\sum_{k = \ell + 1}^{(\ell + mh) \wedge (i - 1)} -K_{m}\left(\frac{k}{\ell}\right)\left(\mu_{i - |k|,1} - \bar{\mu}_{1}\right) + \frac{2}{\ell \kappa_\ell}\sum_{s = 1}^2 \widetilde{c}_{1s}\sum_{k = \ell + 1}^{(\ell + mh) \wedge (i - 1)} -K_{m}\left(\frac{k}{\ell}\right)\left(\mu_{i - |k|,s} - \bar{\mu}_{s}\right),\\
	        b_{2i} &= \frac{2\widetilde{c}_{22}}{\ell \kappa_\ell}\sum_{k = \ell + 1}^{(\ell + mh) \wedge (i - 1)} -K_{m}\left(\frac{k}{\ell}\right)\left(\mu_{i - |k|,2} - \bar{\mu}_{2}\right) + \frac{2}{\ell \kappa_\ell}\sum_{r = 1}^2 \widetilde{c}_{r2}\sum_{k = \ell + 1}^{(\ell + mh) \wedge (i - 1)} -K_{m}\left(\frac{k}{\ell}\right)\left(\mu_{i - |k|,r} - \bar{\mu}_{r}\right)\\
	        &\quad + \frac{2\widetilde{c}_{22}}{\ell \kappa_\ell}\sum_{k = \ell + 1}^{(\ell + mh) \wedge (i - 1)} -K_{m}\left(\frac{k}{\ell}\right)\left(\mu_{i - |k|,2} - \bar{\mu}_{2}\right) + \frac{2}{\ell \kappa_\ell}\sum_{s = 2}^3 \widetilde{c}_{2s}\sum_{k = \ell + 1}^{(\ell + mh) \wedge (i - 1)} -K_{m}\left(\frac{k}{\ell}\right)\left(\mu_{i - |k|,s} - \bar{\mu}_{s}\right)\\
	        b_{3i} &= \frac{2\widetilde{c}_{33}}{\ell \kappa_\ell}\sum_{k = \ell + 1}^{(\ell + mh) \wedge (i - 1)} -K_{m}\left(\frac{k}{\ell}\right)\left(\mu_{i - |k|,3} - \bar{\mu}_{3}\right) + \frac{2}{\ell \kappa_\ell}\sum_{r = 2}^3 \widetilde{c}_{r3}\sum_{k = \ell + 1}^{(\ell + mh) \wedge (i - 1)} -K_{m}\left(\frac{k}{\ell}\right)\left(\mu_{i - |k|,r} - \bar{\mu}_{r}\right)\\
	        &\quad + \frac{2\widetilde{c}_{33}}{\ell \kappa_\ell}\sum_{k = \ell + 1}^{(\ell + mh) \wedge (i - 1)} -K_{m}\left(\frac{k}{\ell}\right)\left(\mu_{i - |k|,3} - \bar{\mu}_{3}\right) + \frac{2}{\ell \kappa_\ell}\sum_{s = 3}^4 \widetilde{c}_{3s}\sum_{k = \ell + 1}^{(\ell + mh) \wedge (i - 1)} -K_{m}\left(\frac{k}{\ell}\right)\left(\mu_{i - |k|,s} - \bar{\mu}_{s}\right)\\
	        b_{4i} &= \frac{2\widetilde{c}_{44}}{\ell \kappa_\ell}\sum_{k = \ell + 1}^{(\ell + mh) \wedge (i - 1)} -K_{m}\left(\frac{k}{\ell}\right)\left(\mu_{i - |k|,4} - \bar{\mu}_{4}\right) + \frac{2}{\ell \kappa_\ell}\sum_{r = 3}^4 \widetilde{c}_{r4}\sum_{k = \ell + 1}^{(\ell + mh) \wedge (i - 1)} -K_{m}\left(\frac{k}{\ell}\right)\left(\mu_{i - |k|,r} - \bar{\mu}_{r}\right).
	    \end{align*}
	    }
    So, we can now write
    \begin{equation*}
        \widetilde{L}^{(\mu,Z)}+\widetilde{L}^{(Z,\mu)}
        = \frac{1}{n}\sum_{i = 1}^n b_i^{\T} Q_i +  {O_p\left\{\frac{(\mathcal{C} + \mathcal{SJ})}{n^{3/2}}\right\}}
    \end{equation*}
    Let $S_{i} = \sum_{i' = 1}^i Q_{i'}$ for $i \in\mathbb{Z}$.
	By Theorem 2.2 of \cite{KarmakarWu2020}
    there exist independent four-dimensional Gaussian vectors
    $\mathbb{Q}_1,\mathbb{Q}_2, \ldots$ such that
	\begin{equation}\label{eqt:B28}
		\max_{1 \leq i \leq n}
        |S_{i} - G_{i}| = o_p(n^{1/p})
	\end{equation}
    in a richer probability space,
	for $p>2$,
    where
    $G_i = \sum_{i'=1}^i \mathbb{Q}_{i'}$.
	On the other hand, by Theorem 3 of \citet{wu2011}, the invariance principle for $t \mapsto \sum_{i = 1}^{\lfloor nt \rfloor} Q_i$ holds.
	By the Cram\'{e}r--Wold device, as $n \rightarrow \infty$,
	\begin{equation}\label{eqt:B29}
		\left\{\frac{1}{\sqrt{n}}\sum_{i = 1}^{\lfloor nt \rfloor} Q_{i}: t \in [0,1]\right\} \inD \left\{\Sigma^{1/2}
			\mathbb{B}(t) : t \in [0,1]\right\},
	\end{equation}
	where $\Sigma = \lim_{n \rightarrow \infty} \Var(\sum_{i = 1}^n Q_{i}/\sqrt{n})$,
    which is invertible by Assumption \ref{assum:weak dependence}.
	Now observe that for each $s = 1,\ldots,4$, since $\mu_{i - |k|,s} - \bar{\mu}_{s} = O(\mathcal{C} + \mathcal{SJ})$, $\ell \kappa_\ell \asymp \ell$, and $\sum_{k = \ell + 1}^{(\ell + mh) \wedge (i - 1)} -K_{m}(k/\ell) = O(\ell)$,
    \begin{equation*}
        \left|\frac{1}{\ell \kappa_\ell}\sum_{k = \ell + 1}^{(\ell + mh) \wedge (i - 1)} -K_{m}\left(\frac{k}{\ell}\right)\left(\mu_{i - |k|,s} - \bar{\mu}_{s}\right)\right| =  O(\mathcal{C} + \mathcal{SJ}),
    \end{equation*}
    so
    \begin{equation}\label{eqt:orderSumAbsDiffOfb1}
    |b_{s,n}| = O(\mathcal{C} + \mathcal{SJ}).
    \end{equation}
    On the other hand, for each $s = 1,\ldots,4$ and $i = 1,\ldots,n$, since $b_{s,i + 1} - b_{s,i}$ contributes to at most one additional term of magnitude $O\{(\mathcal{C} + \mathcal{SJ})/\ell\}$ and the number of contributions is $O(\ell)$, we have
    \begin{equation}\label{eqt:orderSumAbsDiffOfb2}
        \sum_{i = 0}^{n - 1} |b_{s,i + 1} - b_{s,i}|
        = O(\mathcal{C} + \mathcal{SJ}).
    \end{equation}
    Hence, in view of (\ref{eqt:orderSumAbsDiffOfb1})--(\ref{eqt:orderSumAbsDiffOfb2}), we have
    \begin{equation*}
        |b_{s,n}| + \sum_{i = 0}^{n - 1} |b_{s,i + 1} - b_{s,i}|
        = O(\mathcal{C} + \mathcal{SJ}),
        \qquad s = 1,\ldots,4.
    \end{equation*}
    By (\ref{eqt:B28}), Lemma B.7 of \citet{ToChan2024} and the above, we have
    \begin{align}\label{eqt:B30}
       	\left|\frac{1}{\sqrt{n}}\sum_{i = 1}^n
        b_i^{\T}
        (Q_{i} - \mathbb{Q}_{i})\right|
        &= \frac{1}{\sqrt{n}}\left|
        \left\{ b_i^{\T} \sum_{i = 1}^n (Q_{i} - \mathbb{Q}_{i}) \right\}
        + \sum_{i' = 0}^{n - 1}
        (b_{i'+1}-b_{i'})^{\T}
        \sum_{i = 0}^{i'} (Q_{i} - \mathbb{Q}_{i})\right|\\
        &\leq o_p\left(n^{1/p}\right)\frac{1}{\sqrt{n}}\left\{\sum_{s = 1}^4 |b_{s,n}| + \sum_{i = 0}^{n - 1} \sum_{s = 1}^4 \left(|b_{s,i + 1} - b_{s,i}|\right)\right\} \nonumber\\
        &= o_p\left\{(\mathcal{C} + \mathcal{SJ})n^{1/p - 1/2}\right\}.
    \end{align}
    Applying (\ref{eqt:B30}), we can write $\sqrt{n}\{\widetilde{L}^{(\mu,Z)}+\widetilde{L}^{(Z \mu)}\}$ as
    \begin{align}\label{eqt:B31}
        \sqrt{n}\{\widetilde{L}^{(\mu,Z)} +\widetilde{L}^{(Z,\mu)}\}
        &= \frac{1}{\sqrt{n}}\sum_{i = 1}^n
		b_i^{\T}
        \mathbb{Q}_{i} + O_p\left\{\frac{(\mathcal{C} + \mathcal{SJ})}{n}\right\}
        + o_p\left\{(\mathcal{C} + \mathcal{SJ})n^{1/p - 1/2}\right\}\nonumber\\
        &= \frac{1}{\sqrt{n}}\sum_{i = 1}^n
		b_i^{\T}
        \mathbb{Q}_{i} + o_p(1) ,
    \end{align}
    where the last line is followed from the assumption made on
    $\mathcal{C} + \mathcal{SJ}$.
    Using (\ref{eqt:B29}), the limit of (\ref{eqt:B31}) as $n \rightarrow \infty$ can be expressed as follows:
    \begin{equation*}
        \sqrt{n}\{\widetilde{L}^{(\mu,Z)}+\widetilde{L}^{(Z,\mu)}\}
        \inD \int_0^1
        b(t)^{\T} \, \dd \Sigma^{1/2}
        \mathbb{B}(t),
    \end{equation*}
    where $b(t) = [b_1(t), \ldots, b_4(t)]^{\T}$ and
    \begin{align*}
        b_1(t) &= \lim_{n \rightarrow \infty} a_{1,\lfloor nt \rfloor} = 2\widetilde{c}_{11}(\mu_{1}(t) - \bar{\mu}_{1}) + \widetilde{c}_{12}(\mu_{2}(t) - \bar{\mu}_{2}),\\
        b_2(t) &= \lim_{n \rightarrow \infty} a_{2,\lfloor nt \rfloor} = 4\widetilde{c}_{22}(\mu_{2}(t) - \bar{\mu}_{2}) + \widetilde{c}_{12}(\mu_{1}(t) - \bar{\mu}_{1}) + \widetilde{c}_{23}(\mu_{3}(t) - \bar{\mu}_{3}),\\
        b_3(t) &= \lim_{n \rightarrow \infty} a_{3,\lfloor nt \rfloor} = 4\widetilde{c}_{33}(\mu_{3}(t) - \bar{\mu}_{3}) + \widetilde{c}_{23}(\mu_{2}(t) - \bar{\mu}_{2}) + \widetilde{c}_{34}(\mu_{4}(t) - \bar{\mu}_{4}),\\
        b_4(t) &= \lim_{n \rightarrow \infty} a_{4,\lfloor nt \rfloor} = 2\widetilde{c}_{44}(\mu_{4}(t) - \bar{\mu}_{4}) + \widetilde{c}_{34}(\mu_{3}(t) - \bar{\mu}_{3}).
    \end{align*}
    Let
    \begin{equation*}
        \Sigma^{1/2} =
        \begin{pmatrix}
            \ell_{11} & 0 & 0 & 0\\
            \ell_{21} & \ell_{22} & 0 & 0\\
            \ell_{31} & \ell_{32} & \ell_{33} & 0\\
            \ell_{41} & \ell_{42} & \ell_{43} & \ell_{44}
        \end{pmatrix},
    \end{equation*}
    be the Cholesky decomposition representation.
    Then, using the It\^{o} isometry and the independence of $\mathbb{B}_{s}(t)$ across $s = 1,\ldots,4$, the above weak limit can be expressed as
    \begin{align*}
        & \Var\left\{\int_0^1
        A(t)^{\T}\, \dd \Sigma^{1/2}
        \mathbb{B}(t) \right\}\\
        &= \int_0^1 \left(\sum_{s = 1}^4 \ell_{s1}b_{s}(t)\right)^2 \, \dd t + \int_0^1 \left(\sum_{s = 2}^4 \ell_{s2}a_{s}(t)\right)^2 \, \dd t + \int_0^1 \left(\sum_{s = 3}^4 \ell_{s3}b_{s}(t)\right)^2 \, \dd t + \int_0^1 \ell_{44}^2a_4^2(t) \, \dd t.
    \end{align*}
    Now, we have
    \begin{align*}
        b_1^2(t) &= 4c_{11}^2\left(\mu_{1}(t) - \bar{\mu}_{1}\right)^2 + 4c_{11}c_{12}\left(\mu_{1}(t) - \bar{\mu}_{1}\right)\left(\mu_{2}(t) - \bar{\mu}_{2}\right) + c_{12}^2\left(\mu_{2}(t) - \bar{\mu}_{2}\right)^2,\\
        b_2^2(t) &= 4c_{22}^2\left(\mu_{2}(t) - \bar{\mu}_{2}\right)^2 + c_{12}^2\left(\mu_{1}(t) - \bar{\mu}_{1}\right)^2 + c_{23}^2\left(\mu_{3}(t) - \bar{\mu}_{3}\right)^2\\
        &\qquad + 4c_{22}c_{12}\left(\mu_{1}(t) - \bar{\mu}_{1}\right)\left(\mu_{2}(t) - \bar{\mu}_{2}\right) + 4c_{22}c_{23}\left(\mu_{2}(t) - \bar{\mu}_{2}\right)\left(\mu_{3}(t) - \bar{\mu}_{3}\right)\\
        &\qquad + 2c_{12}c_{23}\left(\mu_{1}(t) - \bar{\mu}_{1}\right)\left(\mu_{3}(t) - \bar{\mu}_{3}\right),\\
        b_3^2(t) &= 4c_{33}^2\left(\mu_{3}(t) - \bar{\mu}_{3}\right)^2 + c_{23}^2\left(\mu_{2}(t) - \bar{\mu}_{2}\right)^2 + c_{34}^2\left(\mu_{4}(t) - \bar{\mu}_{4}\right)^2\\
        &\qquad + 4c_{33}c_{23}\left(\mu_{2}(t) - \bar{\mu}_{2}\right)\left(\mu_{3}(t) - \bar{\mu}_{3}\right) + 4c_{33}c_{34}\left(\mu_{3}(t) - \bar{\mu}_{3}\right)\left(\mu_{4}(t) - \bar{\mu}_{4}\right)\\
        &\qquad + 2c_{23}c_{34}\left(\mu_{2}(t) - \bar{\mu}_{2}\right)\left(\mu_{4}(t) - \bar{\mu}_{4}\right),\\
        b_4^2(t) &= 4c_{44}^2\left(\mu_{4}(t) - \bar{\mu}_{4}\right)^2 + 4c_{44}c_{34}\left(\mu_{3}(t) - \bar{\mu}_{3}\right)\left(\mu_{4}(t) - \bar{\mu}_{4}\right) + c_{34}^2\left(\mu_{3}(t) - \bar{\mu}_{3}\right)^2,\\
        b_1(t)b_2(t) &= 4c_{11}c_{22}\left(\mu_{1}(t) - \bar{\mu}_{1}\right)\left(\mu_{2}(t) - \bar{\mu}_{2}\right) + 2c_{11}c_{12}\left(\mu_{1}(t) - \bar{\mu}_{1}\right)^2\\
        &\qquad + 2c_{11}c_{23}\left(\mu_{1}(t) - \bar{\mu}_{1}\right)\left(\mu_{3}(t) - \bar{\mu}_{3}\right) + 2c_{22}c_{12}\left(\mu_{2}(t) - \bar{\mu}_{2}\right)^2\\
        &\qquad + c_{12}^2\left(\mu_{1}(t) - \bar{\mu}_{1}\right)\left(\mu_{2}(t) - \bar{\mu}_{2}\right) + c_{12}c_{23}\left(\mu_{2}(t) - \bar{\mu}_{2}\right)\left(\mu_{3}(t) - \bar{\mu}_{3}\right),\\
        b_1(t)b_3(t) &= 4c_{11}c_{33}\left(\mu_{1}(t) - \bar{\mu}_{1}\right)\left(\mu_{3}(t) - \bar{\mu}_{3}\right) + 2c_{11}c_{23}\left(\mu_{1}(t) - \bar{\mu}_{1}\right)\left(\mu_{2}(t) - \bar{\mu}_{2}\right)\\
        &\qquad + 2c_{11}c_{34}\left(\mu_{1}(t) - \bar{\mu}_{1}\right)\left(\mu_{4}(t) - \bar{\mu}_{4}\right) + 2c_{33}c_{12}\left(\mu_{2}(t) - \bar{\mu}_{2}\right)\left(\mu_{3}(t) - \bar{\mu}_{3}\right)\\
        &\qquad + c_{12}c_{23}\left(\mu_{2}(t) - \bar{\mu}_{2}\right)^2 + c_{12}c_{34}\left(\mu_{2}(t) - \bar{\mu}_{2}\right)\left(\mu_{4}(t) - \bar{\mu}_{4}\right),\\
        b_1(t)b_4(t) &= 4c_{11}c_{44}\left(\mu_{1}(t) - \bar{\mu}_{1}\right)\left(\mu_{4}(t) - \bar{\mu}_{4}\right) + 2c_{11}c_{34}\left(\mu_{1}(t) - \bar{\mu}_{1}\right)\left(\mu_{3}(t) - \bar{\mu}_{3}\right)\\
        &\qquad + 2c_{44}c_{12}\left(\mu_{2}(t) - \bar{\mu}_{2}\right)\left(\mu_{4}(t) - \bar{\mu}_{4}\right) + c_{12}c_{34}\left(\mu_{2}(t) - \bar{\mu}_{2}\right)\left(\mu_{3}(t) - \bar{\mu}_{3}\right),\\
        b_2(t)b_3(t) &= 4c_{22}c_{33}\left(\mu_{2}(t) - \bar{\mu}_{2}\right)\left(\mu_{3}(t) - \bar{\mu}_{3}\right) + 2c_{22}c_{23}\left(\mu_{2}(t) - \bar{\mu}_{2}\right)^2\\
        &\qquad + 2c_{22}c_{34}\left(\mu_{2}(t) - \bar{\mu}_{2}\right)\left(\mu_{4}(t) - \bar{\mu}_{4}\right) + 2c_{33}c_{12}\left(\mu_{1}(t) - \bar{\mu}_{1}\right)\left(\mu_{3}(t) - \bar{\mu}_{3}\right)\\
        &\qquad + c_{12}c_{23}\left(\mu_{1}(t) - \bar{\mu}_{1}\right)\left(\mu_{2}(t) - \bar{\mu}_{2}\right) + c_{12}c_{34}\left(\mu_{1}(t) - \bar{\mu}_{1}\right)\left(\mu_{4}(t) - \bar{\mu}_{4}\right)\\
        &\qquad + 2c_{33}c_{23}\left(\mu_{3}(t) - \bar{\mu}_{3}\right)^2 + c_{23}^2\left(\mu_{2}(t) - \bar{\mu}_{2}\right)\left(\mu_{3}(t) - \bar{\mu}_{3}\right)\\
        &\qquad + c_{23}c_{34}\left(\mu_{3}(t) - \bar{\mu}_{3}\right)\left(\mu_{4}(t) - \bar{\mu}_{4}\right),\\
        b_2(t)b_4(t) &= 4c_{22}c_{44}\left(\mu_{2}(t) - \bar{\mu}_{2}\right)\left(\mu_{4}(t) - \bar{\mu}_{4}\right) + 2c_{22}c_{34}\left(\mu_{2}(t) - \bar{\mu}_{2}\right)\left(\mu_{3}(t) - \bar{\mu}_{3}\right)\\
        &\qquad + 2c_{44}c_{12}\left(\mu_{1}(t) - \bar{\mu}_{1}\right)\left(\mu_{4}(t) - \bar{\mu}_{4}\right) + c_{12}c_{34}\left(\mu_{1}(t) - \bar{\mu}_{1}\right)\left(\mu_{3}(t) - \bar{\mu}_{3}\right)\\
        &\qquad + 2c_{44}c_{23}\left(\mu_{3}(t) - \bar{\mu}_{3}\right)\left(\mu_{4}(t) - \bar{\mu}_{4}\right) + c_{23}c_{34}\left(\mu_{3}(t) - \bar{\mu}_{3}\right)^2,\\
        b_3(t)b_4(t) &= 4c_{33}c_{44}\left(\mu_{3}(t) - \bar{\mu}_{3}\right)\left(\mu_{4}(t) - \bar{\mu}_{4}\right) + 2c_{33}c_{34}\left(\mu_{3}(t) - \bar{\mu}_{3}\right)^2\\
        &\qquad + 2c_{44}c_{23}\left(\mu_{2}(t) - \bar{\mu}_{2}\right)\left(\mu_{4}(t) - \bar{\mu}_{4}\right) + c_{23}c_{34}\left(\mu_{2}(t) - \bar{\mu}_{2}\right)\left(\mu_{3}(t) - \bar{\mu}_{3}\right)\\
        &\qquad + 2c_{44}c_{34}\left(\mu_{4}(t) - \bar{\mu}_{4}\right)^2 + c_{34}^2\left(\mu_{3}(t) - \bar{\mu}_{3}\right)\left(\mu_{4}(t) - \bar{\mu}_{4}\right).
    \end{align*}
    By observing the relationships
    \begin{gather*}
        \ell_{11}^2 = v_{11}, \qquad \ell_{21}^2 + \ell_{22}^2 = v_{22}, \qquad \ell_{31}^2 + \ell_{32}^2 + \ell_{33}^2 = v_{33},\\
        \ell_{41}^2 + \ell_{42}^2 + \ell_{43}^2 + \ell_{44}^2 = v_{44}, \qquad \ell_{11}\ell_{21} = v_{12}, \qquad \ell_{11}\ell_{31} = v_{13}, \qquad \ell_{11}\ell_{41} = v_{14},\\
        \ell_{21}\ell_{31} + \ell_{22}\ell_{32} = v_{23}, \qquad \ell_{21}\ell_{41} + \ell_{22}\ell_{42} = v_{24}, \qquad \ell_{31}\ell_{41} + \ell_{32}\ell_{42} + \ell_{33}\ell_{43} = v_{34},
    \end{gather*}
    the desired result can be obtained through straightforward algebra.
\end{proof}

    \begin{remark}\label{OpCSJ_remidnderTerm}
    The $O_p$-term in (\ref{eqt:OpCSJ_remidnderTerm})
    is found by noticing the following calculations:
    \begin{eqnarray*}
        &\phantom{=}& \left| \frac{\bar{Z}_{n,s}}{n\ell \kappa_\ell}\sum_{|k| \leq \ell + mh} \left\{K\left(\frac{k}{\ell}\right) - K_{m}\left(\frac{k}{\ell}\right)\right\}\sum_{i = |k| + 1}^n \left(\mu_{i - |k|,r} - \bar{\mu}_{n,r}\right) \right|\\
        &=& \left| \frac{\bar{Z}_{n,s}}{\ell \kappa_\ell}\sum_{|k| \leq \ell + mh} \left\{K\left(\frac{k}{\ell}\right) - \sum_{s = 0}^m \delta_s K\left(\left|\frac{k}{\ell}\right| - s\lambda\right)\right\} \frac{1}{n}\sum_{i = |k| + 1}^n \left(\mu_{i - |k|,r} - \bar{\mu}_{n,r}\right) \right|\\
        &=& \left| \frac{\bar{Z}_{n,s}}{\ell \kappa_\ell}\sum_{|k| \leq \ell + mh} \left\{-\sum_{s = 1}^m \delta_s K\left(\left|\frac{k}{\ell}\right| - s\lambda\right)\right\} \frac{1}{n}\sum_{i = |k| + 1}^n \left(\mu_{i - |k|,r} - \bar{\mu}_{n,r}\right) \right|\\
        &\leq& \left| \bar{Z}_{n,s} \right| \sum_{|k| \leq \ell + mh} \left| \frac{1}{\ell \kappa_\ell}\left\{-\sum_{s = 1}^m \delta_s K\left(\left|\frac{k}{\ell}\right| - s\lambda\right)\right\}\right| \left| \frac{1}{n}\sum_{i = |k| + 1}^n \left(\mu_{i - |k|,r} - \bar{\mu}_{n,r}\right)\right|\\
        &=& O_p\left(\frac{1}{\sqrt{n}}\right) \sum_{|k| \leq \ell + mh} \left| \frac{1}{\ell \kappa_\ell}\left\{-\sum_{s = 1}^m \delta_s K\left(\left|\frac{k}{\ell}\right| - s\lambda\right)\right\}\right|O\left(\frac{\mathcal{C} + \mathcal{SJ}}{n}\right)\\
        &=& O_p\left(\frac{\mathcal{C} + \mathcal{SJ}}{n^{3/2}}\right),
    \end{eqnarray*}
    where in the second equality, we use the facts $\delta_0 = 1$ and $K(\cdot)$ is even, and in the last equality, we invoke the facts that $\ell \kappa_\ell \asymp \ell$ and $\sum_{|k| \leq \ell + mh} \left| \left\{-\sum_{s = 1}^m \delta_s K\left(\left|k/\ell\right| - s\lambda\right)\right\}\right| = O(\ell)$, so that their ratio becomes $O(1)$.
    \end{remark}

\subsection{Lemma \ref{lem:B11}}
Recall the definition of $\widetilde{L}^{\circ}$ in (\ref{eqt:tildeLcirc_tildeTheta}):
\begin{align*}
    \widetilde{L}^{\circ} = \sum_{j_1 = 1}^S \sum_{j_2 = j_1}^S c_{j_1,j_2} \widetilde{\theta}_{j_1,j_2},
    \qquad \text{where} \qquad
    \widetilde{\theta}_{j_1,j_2} = \sum_{1 \leq i,i' \leq n} a_{n,i - i'} X_{i,j_1}X_{i'j_2}.
\end{align*}
\begin{lemma}\label{lem:B11}
    Let
    \begin{equation*}
        \sigma_n^2 = 2\sum_{i = 1}^n \sum_{i' = 1}^n a_{n,i' - i}^2
        \qquad \text{and} \qquad
        \zeta_n^2 = \sum_{i = 1}^n a_{n,i}^2.
    \end{equation*}
    Assume that $\E(X_0) = 0$, $\| X_0 \|_4 = \{\sum_{j = 1}^S \E(|X_{0,j}|^4)\}^{1/4} < \infty$,
    $\Omega_4 = \max_{1 \leq j \leq S} \Omega_{4,j} < \infty$, and
    \begin{align}
        \max_{0 \leq k \leq n} a_{n,k}^2 &= o\left(\zeta_n^2\right);\label{eqt:CLT_cond1}\\
        n\zeta_n^2 &= O\left(\sigma_n^2\right); \label{eqt:CLT_cond2}\\
        \sum_{k = 1}^n \sum_{t = 1}^{k - 1}\left|\sum_{j = 1 + k}^n a_{n,k - j}a_{n,t - j}\right|^2 &= o\left(\sigma_n^4\right); \label{eqt:CLT_cond3}\\
        \sum_{k = 1}^n \left|a_{n,k} - a_{n,k - 1}\right|^2 &= o\left(\zeta_n^2\right). \label{eqt:CLT_cond4}
    \end{align}
    Let $c_{j_1,j_2} \in \mathbb{R}$, where $j_1,j_2 = 1,\ldots,S$.
    Then as $n \rightarrow \infty$,
    \begin{equation*}
        \frac{\widetilde{L}^{\circ} - \E\left(\widetilde{L}^{\circ}\right)}{\sigma_n}
        	\inD \Normal\left(0, \sum_{j_1 = 1}^S \sum_{j_2 = j_1}^S \sum_{j_3 = 1}^S \sum_{j_4 = j_3}^S c_{j_1, j_2}c_{j_3,j_4} \left( v_{j_1j_3}v_{j_2,j_4}  + v_{j_1,j_4}v_{j_2,j_3} \right)\right).
    \end{equation*}
\end{lemma}
\begin{proof}[Proof of Lemma \ref{lem:B11}]
    We follow the proofs of Theorem 6 in \citet{LiuWu2010} and Lemma A.6 in \citet{WuZaffaroni2018}.
    Define
    \begin{align*}
        \widetilde{L}^{\circ}_{\text{half}} &= \sum_{j_1 = 1}^S \sum_{j_2 = j_1}^S c_{j_1, j_2} \sum_{1 \leq i < i' \leq n} a_{n,i - i'}X_{i,j_1}X_{i',j_2}.
    \end{align*}
    Note that $\widetilde{L}^{\circ} = 2\widetilde{L}^{\circ}_{\text{half}} + a_{n,0} \sum_{j_1 = 1}^S \sum_{j_2 = j_1}^S c_{j_1, j_2} \sum_{i = 1}^n X_{i,j_1}X_{i,j_2}$.
    Because $\|X_0\|_4 < \infty$ and $\Omega_4 < \infty$, by the Minkowski inequality, we have for some constant $0<C<\infty$ that
    \[
    	\left\| \sum_{j_1 = 1}^S \sum_{j_2 = j_1}^S c_{j_1, j_2} \sum_{i = 1}^n \{ X_{i,j_1}X_{i,j_2} - (\gamma_0)_{j_1,j_2}\} \right\| \leq C\sqrt{n}\|X_0\|_4 \Omega_4 = O(\sqrt{n}).
	\]
	Note that the assumptions (\ref{eqt:CLT_cond1})--(\ref{eqt:CLT_cond2}) imply that $na_{n,0}^2 = o(\sigma_n^2)$.
	Hence, we obtain
	\[
		\frac{\widetilde{L}^{\circ} - \E\left(\widetilde{L}^{\circ}\right)}{\sigma_n}
			= \frac{2\widetilde{L}^{\circ}_{\text{half}} - \E\left(2\widetilde{L}^{\circ}_{\text{half}}\right)}{\sigma_n} + o_p(1).
	\]
	Let $m\geq 0$. Define $\widetilde{X}_{t,j} = \E(X_{t,j} \mid \varepsilon_{t - m},\ldots,\varepsilon_t) = \E(X_{t,j} \mid \mathcal{F}_{t - m,t}),m \geq 0$ as the $m$-dependent approximating sequence.
	Also let
	$D_{k,j} = A_{k,j} - \E\left(A_{k,j} \mid \mathcal{F}_{k - 1}\right)$, and
    $A_{k,j} = \sum_{t = 0}^\infty \E(\widetilde{X}_{t + k,j} \mid \mathcal{F}_k)$.
    By Propositions 1 and 2 of \citet{LiuWu2010},
    \[
    	\frac{2\widetilde{L}^{\circ}_{\text{half}} - \E\left(2\widetilde{L}^{\circ}_{\text{half}}\right)}{\sigma_n}
			= \frac{M_{n,>} + M_{n,<}}{\sigma_n} + o_p(1),
    \]
    where
   	\begin{align*}
		M_{n,<} &= \sum_{j_1 = 1}^S \sum_{j_2 = j_1}^S c_{j_1, j_2} \sum_{t = 1}^n D_{t,j_1} \sum_{i = 1}^{t - 1} a_{n,i - 1}D_{i,j_2}, \\
		M_{n,>} &= \sum_{j_2 = 1}^S \sum_{j_1 = j_2}^S c_{j_2, j_1} \sum_{t = 1}^n D_{t,j_1} \sum_{i = 1}^{t - 1} a_{n,i - 1}D_{i,j_2}.
	\end{align*}
    Consequently, it suffices to show that
    \begin{equation*}
        \frac{M_{n,>} + M_{n,<}}{\sigma_n}
        \inD \Normal\left(0, \sum_{j_1 = 1}^S \sum_{j_2 = j_1}^S \sum_{j_3 = 1}^S \sum_{j_4 = j_3}^S c_{j_1, j_2}c_{j_3,j_4} \left(\widetilde{v}_{j_1j_3}\widetilde{v}_{j_2,j_4} + \widetilde{v}_{j_1,j_4}\widetilde{v}_{j_2,j_3}\right)\right),
    \end{equation*}
    as $n \rightarrow \infty$ holds for every $m$, where $\widetilde{v}_{j_1,j_2} = \sum_{k = -m}^m \E(\widetilde{X}_{0,j_1}\widetilde{X}_{k,j_2})$, which satisfies that $\widetilde{v}_{j_1,j_2} \rightarrow v_{j_1,j_2}$ as $m\rightarrow\infty$.
	Now we decompose the most inner summation in $M_{n,<}$ as follows:
	\[
		\sum_{i = 1}^{t - 1} a_{n,i - 1}D_{i,j}
			= \underbrace{\left( \sum_{i = 1}^{t - 4m} a_{n,i - t}D_{i,j} \right)}_{U_{t,j}}
				+ \underbrace{\left(\sum_{i = (t - 4m + 1) \vee 1}^{t - 1} a_{n,i - t}D_{i,j} \right)}_{U_{t,j}^{\text{tail}}} .
	\]
    By Minkowski inequality, we have
    \[
    	\left\| \sum_{j_1 = 1}^S \sum_{j_2 = j_1}^S c_{j_1, j_2} \sum_{i = 1}^n D_{i,j_1}U_{i,j_2}^{\text{tail}} \right\|
			\leq \sum_{j_1 = 1}^S \sum_{j_2 = j_1}^S |c_{j_1, j_2}| \left\|\sum_{i = 1}^n D_{i,j_1}U_{i,j_2}^{\text{tail}}\right\|
			= O(\sqrt{n})\max_{1 \leq k \leq n} |a_{n,k}|
			= o(\sigma_n),
	\]
	where the last line follows from the assumptions (\ref{eqt:CLT_cond1})--(\ref{eqt:CLT_cond2}).
    So, it remains to show that
    \begin{align}\label{eqt:B35}
        & \frac{1}{\sigma_n}\sum_{i = 1 + 4m}^n \sum_{j_1 = 1}^S \sum_{j_2 = j_1}^S c_{j_1, j_2}\left(D_{i,j_1}U_{i,j_2} + D_{i,j_2}U_{i,j_1}\right)\nonumber\\
        	&\qquad \inD \Normal\left(0, \sum_{j_1 = 1}^S \sum_{j_2 = j_1}^S \sum_{j_3 = 1}^S \sum_{j_4 = j_3}^S c_{j_1, j_2}c_{j_3,j_4} \left(\widetilde{v}_{j_1j_3}\widetilde{v}_{j_2,j_4} + \widetilde{v}_{j_1,j_4}\widetilde{v}_{j_2,j_3}\right)\right),
    \end{align}
    as $n \rightarrow \infty$.
    Because
    \begin{align*}
        \left\| \sum_{j_1 = 1}^S \sum_{j_2 = j_1}^S c_{j_1, j_2} D_{i,j_1}U_{i,j_2} \right\|_4^4
        &= \E\left\{\left|\sum_{j_1 = 1}^S\sum_{j_2 = 1}^S c_{j_1, j_2} D_{i,j_1}U_{i,j_2}\right|^4\right\}\\
        &\leq \left\{\frac{S(S - 1)}{2}\right\}^4 \left( \max_{1 \leq j_1, j_2 \leq S} \left|c_{j_1, j_2}\right|^4 \right) \sum_{j_1 = 1}^S \sum_{j_2 = j_1}^S \left\|D_{i,j_1}U_{i,j_2}\right\|_4^4,
    \end{align*}
    we have $\sum_{i = 1 + 4m}^n \| \sum_{j_1 = 1}^S \sum_{j_2 = j_1}^S c_{j_1, j_2} D_{i,j_1}U_{i,j_2}\|_4^4 = O(n)\zeta_n^4 = o(\sigma_n^4)$.
    Then it follows that the Lindeberg condition is satisfied.
    By the martingale Central Limit Theorem, (\ref{eqt:B35}) holds if, as $n \rightarrow \infty$,
    \begin{align}\label{eqt:B36}
        & \frac{1}{\sigma_n^2} \sum_{i = 1 + 4m}^n \E\left[\left\{\sum_{j_1 = 1}^S \sum_{j_2 = j_1}^S c_{j_1, j_2}\left(D_{i,j_1}U_{i,j_2} + D_{i,j_2}U_{i,j_1}\right)\right\}^2 \mid \mathcal{F}_{i - 1}\right]  \nonumber \\
        &\qquad \inP  \sum_{j_1 = 1}^S \sum_{j_2 = j_1}^S \sum_{j_3 = 1}^S \sum_{j_4 = j_3}^S c_{j_1, j_2}c_{j_3,j_4} \left(\widetilde{v}_{j_1j_3} \widetilde{v}_{j_2,j_4}  + \widetilde{v}_{j_1,j_4} \widetilde{v}_{j_2,j_3}\right).
    \end{align}
    For the rest of the proof, we shall verify (\ref{eqt:B36}).
    Note that $\E(\cdot \mid \mathcal{F}_{i - 1}) = \sum_{g = 1}^m \{\E(\cdot \mid \mathcal{F}_{i - g}) - \E(\cdot \mid \mathcal{F}_{i - g - 1})\} + \E(\cdot \mid \mathcal{F}_{i - m - 1})$.
    For $-m \leq g \leq m - 1$, by the Minkowski inequality,
    {\footnotesize
    \begin{align*}
        & \Bigg\|\sum_{i = 1 + 4m}^n \Bigg(\E\left[\left\{\sum_{j_1 = 1}^S \sum_{j_2 = j_1}^S c_{j_1, j_2}\left(D_{i,j_1}U_{i,j_2} + D_{i,j_2}U_{i,j_1}\right)\right\}^2 \mid \mathcal{F}_{i - g}\right] \\
       &\qquad  - \E\left[\left\{\sum_{j_1 = 1}^S \sum_{j_2 = j_1}^S c_{j_1, j_2}\left(D_{i,j_1}U_{i,j_2} + D_{i,j_2}U_{i,j_1}\right)\right\}^2 \mid \mathcal{F}_{i - g - 1}\right]\Bigg)\Bigg\|^2\\
        &\qquad\qquad\leq 4\sum_{j_1 = 1}^S \sum_{j_2 = j_1}^S \sum_{j_3 = 1}^S \sum_{j_4 = j_3}^S c_{j_1, j_2}^2 c_{j_3,j_4}^2 \sum_{i = 1 + 4m}^n \left\|D_{i,j_1}\right\|_4^4 \left\|U_{i,j_2}\right\|_4^4
        \leq Cn\zeta_n^4
        = o\left(\sigma_n^4\right).
    \end{align*}}
    Because $D_{i,j_1}$ are $\mathcal{F}_{i - m,i}$-measurable and $U_{i,j_2}$ are $\mathcal{F}_{i - 4m}$-measurable,
    \[
    	\E\{(D_{i,j_1})^2(U_{i,j_2})^2 \mid \mathcal{F}_{i - m,i}\} = (U_{i,j_2})^2\E\{(D_{i,j_1})^2\}
	\]
	for all $j_1,j_2 = 1,\ldots,S$.
    So, (\ref{eqt:B36}) is reduced to
    {\small
    \begin{align}\label{eqt:B37}
        & \frac{1}{\sigma_n^2}\sum_{i = 1 + 4m}^n \Bigg[\sum_{j_1 = 1}^S \sum_{j_2 = j_1}^S \sum_{j_3 = 1}^S \sum_{j_4 = j_3}^S c_{j_1, j_2}c_{j_3,j_4}
        (U_{i,j_2}U_{i,j_4})\E(D_{i,j_1}D_{i,j_3}) +
        (U_{i,j_1}U_{i,j_3})\E(D_{i,j_2}D_{i,j_4}) + \nonumber\\
        &\qquad (U_{i,j_2}U_{i,j_3})\E(D_{i,j_1}D_{i,j_4}) + (U_{i,j_1}U_{i,j_4})\E(D_{i,j_2}D_{i,j_3}) \Bigg]  \nonumber\\
        &\qquad \inP \sum_{j_1 = 1}^S \sum_{j_2 = j_1}^S \sum_{j_3 = 1}^S \sum_{j_4 = j_3}^S c_{j_1, j_2}c_{j_3,j_4} \left\{\E\left(D_{i,j_1}D_{i,j_3}\right)\E\left(D_{i,j_2}D_{i,j_4}\right) + \E\left(D_{i,j_1}D_{i,j_4}\right)\E\left(D_{i,j_2}D_{i,j_3}\right)\right\},
    \end{align}
    }
    since $\E(D_{i,j_1}D_{i,j_2}) =  \widetilde{v}_{j_1,j_2}$.
    Now, since $\|\sum_{i = 1 + 4m}^n \{(U_{i,j_1}U_{i,j_2}) - \E(U_{i,j_1}U_{i,j_2})\}\| = o(\sigma_n^2)$, (\ref{eqt:B37}) is further reduced to
    {\small
    \begin{align}\label{eqt:B38}
        & \frac{1}{\sigma_n^2}\sum_{i = 1 + 4m}^n \Bigg[\sum_{j_1 = 1}^S \sum_{j_2 = j_1}^S \sum_{j_3 = 1}^S \sum_{j_4 = j_3}^S c_{j_1, j_2}c_{j_3,j_4}
        \E(U_{i,j_2}U_{i,j_4})\E(D_{i,j_1}D_{i,j_3}) +\E(U_{i,j_1}U_{i,j_3})\E(D_{i,j_2}D_{i,j_4})
        + \nonumber\\
        &\qquad
        \E(U_{i,j_2}U_{i,j_3})\E(D_{i,j_1}D_{i,j_4}) +
        \E(U_{i,j_1}U_{i,j_4})\E(D_{i,j_2}D_{i,j_3})\Bigg]  \nonumber\\
        &\qquad \inP\sum_{j_1 = 1}^S \sum_{j_2 = j_1}^S \sum_{j_3 = 1}^S \sum_{j_4 = j_3}^S c_{j_1, j_2}c_{j_3,j_4} \Bigg\{\E\left(D_{i,j_1}D_{i,j_3}\right)\E\left(D_{i,j_2}D_{i,j_4}\right) +  \E\left(D_{i,j_1}D_{i,j_4}\right)\E\left(D_{i,j_2}D_{i,j_3}\right)\Bigg\}. \nonumber\\
        &\qquad = \sum_{j_1 = 1}^S \sum_{j_2 = j_1}^S \sum_{j_3 = 1}^S \sum_{j_4 = j_3}^S c_{j_1, j_2}c_{j_3,j_4} \left(\widetilde{v}_{j_1j_3} \widetilde{v}_{j_2,j_4}  + \widetilde{v}_{j_1,j_4} \widetilde{v}_{j_2,j_3}\right).
    \end{align}
    }
    Hence, the result follows.
\end{proof}

\section{$\mathcal{L}^2$-consistency}
\subsection{Theorem \ref{thm:B9}}
\begin{theorem}\label{thm:B9}
    Let $m \geq 1$ be fixed, $1/\ell + \ell/n = o(1)$ and $h/\ell = \lambda \in [2,\infty)$.
    Let Assumptions \ref{assum:jumps}, \ref{assum:weak dependence} and \ref{assum:kernel} hold.
    Assume further that
    $\Omega_{0,4}<\infty$,
    $\kappa \neq 0$ and
    $u_{j_1j_2,q} = \sum_{k \in \mathbb{Z}} |k|^q |(\gamma_k)_{j_1,j_2}| < \infty$ for some $q \in \mathbb{N}$ and fixed $j_1,j_2 = 1,\ldots,S$.
    Denote $A = \int_0^1 K^2(u) \, \dd u$ and
    \[
    	\mathcal{M}
        = n^{1/2}\mathcal{M}_0^{1/2}+ \ell \mathcal{J} \mathcal{M}_0.
	\]
    Let
    \begin{equation*}
        \widehat{\theta}^{(m)}_{j_1,j_2} = \frac{\sum_{|k| \leq \ell} K(k/\ell)\{(\widehat{\gamma}_k^{(0)})_{j_1,j_2} - (\widehat{\gamma}_k^{(m)})_{j_1,j_2}\}}{\sum_{|k| \leq \ell} K(k/\ell)}.
    \end{equation*}
    Then, as $n \rightarrow \infty$,
    \begin{align*}
        \E\left(\widehat{\theta}_{j_1,j_2}^{(m)}\right) - \theta_{j_1,j_2}
        &= O\left( \mathcal{M}_0 + \mathcal{M}_1 +\frac{1}{n}\right) + o\left(\frac{1}{\ell^{q + 1}}\right),\\
        \Var\left(\widehat{\theta}_{j_1,j_2}^{(m)}\right) &= \frac{4w_{j_1,j_2}A(\Delta_m - \Delta_0)}{n\ell \kappa_\ell^2} + \frac{v_{j_1,j_1}\theta_{j_2,j_2} + v_{j_2,j_2}\theta_{j_1,j_1} + 2v_{j_1,j_2}\theta_{j_1,j_2}}{n} + R_2,
    \end{align*}
    where $w_{j_1,j_2} = (v_{j_1,j_1}v_{j_2,j_2} + v_{j_1,j_2}v_{j_1,j_2})/2$ and $R_2$ satisfies the following properties:
    (i) $R_2 = o(1/n\ell)$ if $\mathcal{M} = o(1)$; and
    (ii) $R_2 = o(1/n)$ if $\mathcal{M}/\ell = o(1)$ and $\mathcal{M}_0=o(1)$.
\end{theorem}
\begin{proof}[Proof of Theorem \ref{thm:B9}]
    We first establish the result for the bias of $\widehat{\theta}_{j_1,j_2}^{(m)}$.
	Following the bias calculation in Theorem 5.1 and Corollaries 5.5
	and 6.2 of \citet{Chan2020}, we have, for each fixed $m \geq 0$,
	\begin{align*}
	    \E_\mu\left(\widehat{v}_{j_1,j_2}^{(m)}\right)
	        - K(0)v_{j_1,j_2}
	    &= \frac{Bv_{j_1j_2,q}}{\ell^q}
	        + \left\{\kappa_\ell \ell \theta_{j_1,j_2}
	        + O(\ell\mathcal{M}_0)\right\}\mathbb{1}(m=0)
			+ O(\ell\mathcal{M}_1) + r_{\mathsf{bias}}^{(m)},
	\end{align*}
	where
	$r_{\mathsf{bias}}^{(m)}=o(1/\ell^q)+O(\ell/n)$.
		Note that $\widehat{\theta}_{j_1,j_2}^{(m)} = (\widehat{v}_{j_1,j_2}^{(0)} - \widehat{v}_{j_1,j_2}^{(m)})/(\ell \kappa_\ell)$, where $\widehat{v}_{j_1,j_2}^{(m)} = \sum_{|k| \leq \ell} K(k/\ell)(\widehat{\gamma}_k^{(m)})_{j_1,j_2}$.
	Since $\kappa_{\ell} \rightarrow \kappa \neq 0$,
	we have $(r_{\mathsf{bias}}^{(0)}-r_{\mathsf{bias}}^{(m)})/(\ell\kappa_\ell) =o(\ell^{-q-1})+O(1/n)$.
	Using the above results, we have
	\begin{equation*}
		\E\left(\widehat{\theta}_{j_1,j_2}^{(m)}\right) - \theta_{j_1,j_2}
		= O(\mathcal{M}_0) + O(\mathcal{M}_1) + O(1/n) + o(1/\ell^{q + 1} ).
	\end{equation*}
    Now we establish the result for the variance of $\widehat{\theta}_{j_1,j_2}^{(m)}$.
	Recall that $X_{i} = \mu_{i} + Z_{i}$, where $\E(Z_{i}) = 0$ for $i = 1,\ldots,n$.
    Note that $\widehat{\theta}_{j_1,j_2}^{(m)}$ can be approximated by
	\begin{equation}\label{eqt:B32}
	\widetilde{\theta}^{(m)}_{j_1,j_2}
		= \widetilde{\theta}^{(\mu,\mu)}_{j_1,j_2}
			+ \widetilde{\theta}^{(\mu,Z)}_{j_1,j_2}
			+ \widetilde{\theta}^{(Z \mu)}_{j_1,j_2}
			+ \widetilde{\theta}^{(Z,Z)}_{j_1,j_2},
	\end{equation}
	where
	\begin{equation}\label{eqt:B33}
	\widetilde{\theta}^{(\alpha \beta)}_{j_1,j_2} = \frac{1}{n\ell \kappa_\ell}\sum_{|k| \leq \ell + mh} \left\{K\left(\frac{k}{\ell}\right) - K_{m}\left(\frac{k}{\ell}\right)\right\}\sum_{i = 1+k^+}^{n-k^-} (\alpha_{i}^{[j_1]} - \bar{\alpha}^{[j_1]}_n)(\beta_{i - k}^{[j_2]} - \bar{\beta}^{[j_2]}_n),
	\end{equation}
	for $\alpha,\beta \in \{\mu,Z\}$.
    Since $\widetilde{\theta}^{(\mu,\mu)}_{j_1,j_2}$ is non-stochastic, $\Var (\widetilde{\theta}^{(\mu,\mu)}_{j_1,j_2} ) = 0$.
	On the other hand, by Corollary 6.2 and modified proof of Lemma B.7 in \citet{Chan2020}, we have
	\begin{equation*}
	\Var\left(\widetilde{\theta}^{(Z,Z)}_{j_1,j_2}\right) = \frac{4w_{j_1,j_2} A \left(\Delta_m - \Delta_0\right)}{n\ell \kappa_\ell^2} + o\left( \frac{1}{n\ell}\right).
	\end{equation*}
    By Lemma \ref{lem:B10},
	\begin{align*}
		&\Var\left(\widetilde{\theta}^{(Z,\mu)}_{j_1,j_2}\right)
			= \frac{v_{j_1,j_1}\theta_{j_2,j_2}}{n} + O(\mathcal{M}'),  \\
        &\Var\left(\widetilde{\theta}^{(\mu,Z)}_{j_1,j_2}\right)
        	= \frac{v_{j_2,j_2}\theta_{j_1,j_1}}{n} + O(\mathcal{M}'),  \\
        &\Cov\left(\widetilde{\theta}^{(Z,\mu)}_{j_1,j_2},\widetilde{\theta}^{(\mu,Z)}_{j_1,j_2}\right)
        	= \frac{v_{j_1,j_2}\theta_{j_1,j_2}}{n} + O(\mathcal{M}') ,
	\end{align*}
	where $\mathcal{M}'$ is defined in (\ref{eqt:Mp_orderOf_theta_muZ}) of Lemma \ref{lem:B10}.
	By Cauchy--Schwarz inequality and $\theta_{j_1,j_2} = O\{(\mathcal{C} + \mathcal{SJ})^2\}$,
	\begin{align*}
	\left|\Cov\left(\widetilde{\theta}^{(\mu,Z)}_{j_1,j_2},\widetilde{\theta}^{(Z,Z)}_{j_1,j_2}\right)\right|
	&\leq \sqrt{\Var\left(\widetilde{\theta}^{(\mu,Z)}_{j_1,j_2}\right)\Var\left(\widetilde{\theta}^{(Z,Z)}_{j_1,j_2}\right)}\\
	&=O\left(\sqrt{\frac{(\mathcal C+\mathcal{SJ})^2}{n}\cdot\frac{1}{n\ell}}\right)
	= O\left( \frac{\mathcal{C} + \mathcal{SJ}}{n\ell^{1/2}}\right) .
	\end{align*}
	The same upper bound holds for $\Cov(\widetilde{\theta}^{(Z \mu)}_{j_1,j_2},\widetilde{\theta}^{(Z,Z)}_{j_1,j_2})$.
	Hence, the result then follows with
	\begin{align*}
		R_2 &= o\left(\frac{1}{n\ell}\right)
				+ O(\mathcal{M}')
				+ O\left(\frac{\mathcal{M}_0}{n}\right)
 				+ O\left(\frac{\mathcal{C}+\mathcal{S}\mathcal{J}}{n\ell^{1/2}}\right) \\
		&= o\left( \frac{1}{n\ell} \right)
			+ O\left[  \frac{(\mathcal{C} + \mathcal{SJ})^2}{n} \left\{\frac{(\mathcal{J} + 1)\ell}{n} + \frac{1}{\ell^q} \right\} +\frac{\mathcal{C} + \mathcal{SJ}}{n\ell^{1/2}} \right].
	\end{align*}
	After performing some direct calculations, we have the following results.
	\begin{itemize}
		\item If $\mathcal{M} = o(1)$,
				then $R_2 = o(1/n\ell)$.
		\item If $\mathcal{M}/\ell = o(1)$ and $\mathcal{M}_0=o(1)$,
				then $R_2 = o(1/n)$.
	\end{itemize}
	It completes the proof of the theorem.
\end{proof}

\subsection{Lemma \ref{lem:B10}}
\begin{lemma}\label{lem:B10}\
    Let $\widetilde{\theta}^{(Z,\mu)}_{j_1,j_2}$ and $\widetilde{\theta}^{(\mu,Z)}_{j_1,j_2}$ be defined as in (\ref{eqt:B33}).
	Suppose the conditions for Theorem \ref{thm:B9} hold.
	Let
	\begin{align}\label{eqt:Mp_orderOf_theta_muZ}
		\mathcal{M}'
		= O\left[  \frac{(\mathcal{C} + \mathcal{SJ})^2}{n} \left\{\frac{(\mathcal{J} + 1)\ell}{n} + \frac{1}{\ell^q} \right\} \right].
	\end{align}
	Then as $n \rightarrow \infty$,
	\begin{align*}
		&\Var\left(\widetilde{\theta}^{(Z,\mu)}_{j_1,j_2}\right)
			= \frac{v_{j_1,j_1}\theta_{j_2,j_2}}{n} + O(\mathcal{M}'),  \\
        &\Var\left(\widetilde{\theta}^{(\mu,Z)}_{j_1,j_2}\right)
        	= \frac{v_{j_2,j_2}\theta_{j_1,j_1}}{n} + O(\mathcal{M}'),  \\
        &\Cov\left(\widetilde{\theta}^{(Z,\mu)}_{j_1,j_2},\widetilde{\theta}^{(\mu,Z)}_{j_1,j_2}\right)
        	= \frac{v_{j_1,j_2}\theta_{j_1,j_2}}{n} + O(\mathcal{M}') .
	\end{align*}
\end{lemma}
\begin{proof}[Proof of Lemma \ref{lem:B10}]
    We only establish the result for $\Var(\widetilde{\theta}^{(Z,\mu)}_{j_1,j_2})$; the other results follow from similar calculations.
    Notice that
	\begin{align*}
	\widetilde{\theta}^{(Z,\mu)}_{j_1,j_2}
		&= \frac{1}{n\ell \kappa_\ell}\sum_{|k| \leq \ell + mh} \left\{K\left(\frac{k}{\ell}\right) - K_{m}\left(\frac{k}{\ell}\right)\right\}\sum_{i = 1+k^+}^{n-k^-} (Z_{i,j_1} - \bar{Z}_{n,j_1})(\mu_{i - k, j_2} - \bar{\mu}_{n,j_2})\\
		&= \frac{1}{n \kappa_\ell}\sum_{i = 1}^n
			\widetilde{A}_i (Z_{i,j_1} - \bar{Z}_{n,j_1}),
	\end{align*}
	where
	\[
		\widetilde{A}_i = \left[\frac{1}{\ell}\sum_{|k| \leq \ell + mh} \left\{K\left(\frac{k}{\ell}\right) - K_{m}\left(\frac{k}{\ell}\right)\right\}(\mu_{i - k, j_2} - \bar{\mu}_{n,j_2})\mathbb{1}(1\leq i-k\leq n)\right] .
	\]
	By change of variable and the property that $\sum_{s=-m}^m \delta_{|s|} = 0$, we have
	\[
		\frac{1}{\ell}\sum_{|k|\leq \ell+mh}\left\{K\left(\frac{k}{\ell}\right) - K_{m}\left(\frac{k}{\ell}\right)\right\} = \kappa_{\ell}.
	\]
	Consequently, we have $\max_{i=1, \ldots, n}|\widetilde{A}_i |= O(\mathcal{C}+\mathcal{SJ})$.
	Then
	\begin{align*}
	\Var\left(\widetilde{\theta}^{(Z,\mu)}_{j_1,j_2}\right)
		&= \left(\frac{1}{n^2 \kappa_\ell^2} \sum_{|k'| \leq \ell} \left(\gamma_{k'}\right)_{j_1,j_1} \sum_{i - j = k'} \widetilde{A}_{i}\widetilde{A}_{j} \right)
			+ \left(\frac{1}{n^2 \kappa_\ell^2} \sum_{\ell < |k'| < n} \left(\gamma_{k'}\right)_{j_1,j_1}  \sum_{i - j = k'} \widetilde{A}_{i}\widetilde{A}_{j} \right) \\
		&\qquad + \left\{ \frac{1}{n^2\kappa_{\ell}^2}\left( \sum_{i=1}^n \widetilde{A}_i \right)^2\Var(\bar{Z}_{n,j_1})
				- \frac{2}{n^2 \kappa_{\ell}^2} \left( \sum_{i=1}^n \widetilde{A}_i \right) \sum_{i=1}^n \widetilde{A}_i\Cov(Z_{i,j_1}, \bar{Z}_{n,j_1}) \right\}\\
		&=: B_{13} + B_{14} +  B_{15},
	\end{align*}
	where the reminder term $B_{15}$ exists because of the errors of centering $Z_{i,j_1}$ by $\bar{Z}_{n,j_1}$.
    Observe that
	\begin{align}\label{eqt:B34}
	B_{13}
		&= \frac{1}{n^2 \kappa_\ell^2} \sum_{|k'| \leq \ell} \left(\gamma_{k'}\right)_{j_1,j_1} \left(\sum_{\substack{i - j = k',\\i \in \mathbb{S}}} \widetilde{A}_{i}\widetilde{A}_{j}\right)
			+ \frac{1}{n^2 \kappa_\ell^2} \sum_{|k'| \leq \ell} \left(\gamma_{k'}\right)_{j_1,j_1} \left(\sum_{\substack{i - j = k',\\i \in \mathbb{S}^{\comp}}} \widetilde{A}_{i}\widetilde{A}_{j}\right) \nonumber \\
		&=: B_{13}^{(1)} + B_{13}^{(2)},
	\end{align}
	where $\mathbb{S} = \{1 \leq i \leq n: |i - T_{j}| > mh+2\ell ~ \text{for} ~ j = 0,\ldots,\mathcal{J}+1\}$.
	We start with $B_{13}^{(1)}$.
	We have
	\begin{align*}
	B_{13}^{(1)}
		&= \frac{1}{n^2 \kappa_\ell^2} \sum_{|k'| \leq \ell} \left(\gamma_{k'}\right)_{j_1,j_1} \left(\sum_{\substack{i - j = k',\\i \in \mathbb{S}}} \widetilde{A}_{i}^2\right)
			+ \frac{1}{n^2 \kappa_\ell^2} \sum_{|k'| \leq \ell} \left(\gamma_{k'}\right)_{j_1,j_1} \left\{\sum_{\substack{i - j = k',\\i \in \mathbb{S}}} \widetilde{A}_{i}\left(\widetilde{A}_{j} - \widetilde{A}_{i}\right)\right\} \\
		&=: B_{13}^{(1,1)} + B_{13}^{(1,2)}.
	\end{align*}
	Using $\sum_{k' \in \mathbb{Z}} |(\gamma_{k'})_{j_1,j_1}| < \infty$ and $|\widetilde{A}_{i} - \widetilde{A}_{j}| = O(\mathcal{C\ell}/n)$ for all $|i - j| = k'$ and $i \in \mathbb{S}$, we have
	\begin{align*}
		B_{13}^{(1,2)}
		= O\left\{\frac{\mathcal{C}(\mathcal{C}+ \mathcal{SJ})\ell}{n^2}\right\} .
	\end{align*}
	For $B_{13}^{(1,1)}$,
	by $\sum_{i = 1}^n (\mu_{i,j_2} - \bar{\mu}_{n,j_2})^2/n = \theta_{j_2,j_2} + O\{(\mathcal{C} + \mathcal{SJ})^2/n\}$ and (\ref{eqt:B6}),
	we have
	\begin{align*}
		B_{13}^{(1,1)}
		&= \frac{\theta_{j_2,j_2} }{n} \sum_{|k|\leq \ell}(\gamma_{k})_{j_1,j_1}
			+ O\left\{\frac{(\mathcal{J} + 1)\ell(\mathcal{C} + \mathcal{SJ})^2}{n^2}\right\} \\
		&= \frac{\theta_{j_2,j_2} }{n}
			\left\{ v_{j_1, j_1}+O\left( \frac{1}{\ell^q}\right) \right\}
			+ O\left\{\frac{(\mathcal{J} + 1)\ell(\mathcal{C} + \mathcal{SJ})^2}{n^2}\right\} \\
		&= \frac{\theta_{j_2,j_2} v_{j_1,j_1}}{n}
			+ O\left[  \frac{(\mathcal{C} + \mathcal{SJ})^2}{n} \left\{\frac{(\mathcal{J} + 1)\ell}{n} + \frac{1}{\ell^q} \right\} \right] .
	\end{align*}
	On the other hand, for the second term $B_{13}^{(2)}$ of (\ref{eqt:B34}), we have
	\begin{align*}
		|B_{13}^{(2)}|
		\leq \frac{1}{n^2\kappa_\ell^2}\sum_{|k'| \leq \ell} \left|\left(\gamma_{k'}\right)_{j_1,j_1}\right| \mathcal{J} \ell
		\max_{1\leq i\leq n} \widetilde{A}_{i}^2
		= O\left\{ \frac{(\mathcal{J} + 1)\ell(\mathcal{C} + \mathcal{SJ})^2}{n^2}\right\} .
	\end{align*}
	Hence,
	\begin{align*}
	B_{13}
		&= \frac{\theta_{j_2,j_2} v_{j_1,j_1}}{n}
			+ O\left[  \frac{(\mathcal{C} + \mathcal{SJ})^2}{n} \left\{\frac{(\mathcal{J} + 1)\ell}{n} + \frac{1}{\ell^q} \right\} \right].
	\end{align*}
	For $B_{14}$,
	using $|\widetilde{A}_{i}| = O(\mathcal{C} + \mathcal{SJ})$ for each $i$
	and $\sum_{k' \in \mathbb{Z}} |(\gamma_{k'})_{j_1,j_1}| < \infty$,
	we have
	\[
		B_{14} = O\left\{ \frac{(\mathcal{C} + \mathcal{SJ})^2}{n\ell^q} \right\}.
	\]
	For $B_{15}$, using
	$|\sum_{i=1}^n\widetilde{A}_{i}| = O\{\ell(\mathcal{C} + \mathcal{SJ})\}$,
	$\max_{1\leq i\leq n} |\Cov(Z_{i,j_1}, \bar{Z}_{n,j_1})| = O(1/n)$, and
	$\Var(\bar{Z}_{n,j_1}) = O(1/n)$, we have
	\[
		B_{15} = O\left\{ \frac{\ell(\mathcal{C} + \mathcal{SJ})^2}{n^2} \right\}.
	\]
	Hence
	\begin{align*}
	\Var\left(\widetilde{\theta}^{(Z,\mu)}_{j_1,j_2}\right)
		= \frac{\theta_{j_2,j_2} v_{j_1,j_1}}{n}
			+ O\left[  \frac{(\mathcal{C} + \mathcal{SJ})^2}{n} \left\{\frac{(\mathcal{J} + 1)\ell}{n} + \frac{1}{\ell^q} \right\} \right].
	\end{align*}
	Thus, the desired result follows.
\end{proof}

\bibliographystyle{rss}
	{\small
	\setstretch{1.0}
	\bibliography{myRef}
	}
\end{document}